\documentclass[10pt,twocolumn]{iopart}

\usepackage[latin1]{inputenc}
\usepackage{color}
\usepackage{xspace}
\usepackage{graphicx}
\usepackage{iopams}
\usepackage{xspace}
\usepackage{dcolumn}
\usepackage{bm}
\usepackage{color}
\usepackage{citesort}
\usepackage[extra]{tipa}
\usepackage{times}

\newcommand{\oh}{\mbox{$\frac{1}{2}$}}

\newcommand{\text}[1]{\mathrm{#1}}

\newcommand{\las}[0]{\langle}
\newcommand{\ras}[0]{\rangle}
\newcommand{\llas}[0]{\langle\langle}
\newcommand{\rras}[0]{\rangle\rangle}
\newcommand{\la}[0]{\left\las}
\newcommand{\ra}[0]{\right\ras}
\newcommand{\ket}[1]{\left|#1\ra}
\newcommand{\bra}[1]{\la#1\right|}

\renewcommand{\tilde}[1]{\widetilde{#1}}

\newcommand{\Sxy}{S_\text{AF}^{xy}}
\newcommand{\Szz}{S_\text{AF}^{zz}}
\newcommand{\lso}[0]{\lambda_\text{SO}}
\newcommand{\lr}[0]{\lambda_\text{R}}

\newcommand{\Uc}[0]{U_\text{c}}

\newcommand{\ie}[0]{i.e.\@\xspace}
\newcommand{\eg}[0]{e.g.\@\xspace}
\newcommand{\UP}[0]{\uparrow}
\newcommand{\DO}[0]{\downarrow}

\newcommand{\om}[0]{\omega}

\newcommand{\kF}{k_\mathrm{F}}
\newcommand{\vF}{v_\mathrm{F}}

\newcommand{\nag}{{\phantom{\dag}}}

\newcommand{\on}[0]{\widehat{n}}

\newfont{\tensy}{cmsy10}
\newcommand{\chem}[1]{{$\fontdimen16\tensy=3.0pt
    \fontdimen17\tensy=3.0pt \mathrm{#1}$}}

\begin{document}

\topical[]{Correlation effects in two-dimensional topological insulators}

\author{M Hohenadler and F F Assaad}

\address{%
Institut f\"ur Theoretische Physik und Astrophysik,
Universit\"at W\"urzburg, \newline Am Hubland, 97074 W\"urzburg, DE}

\ead{mhohenadler@physik.uni-wuerzburg.de}

\begin{abstract}
  Topological insulators have become one of the most active research areas in
  condensed matter physics. This article reviews progress on the topic of
  electronic correlations effects in the two-dimensional case, with a focus
  on systems with intrinsic spin-orbit coupling and numerical results. Topics
  addressed include an introduction to the noninteracting case, an overview
  of theoretical models, correlated topological band insulators,
  interaction-driven phase transitions, topological Mott insulators and
  fractional topological states, correlation effects on helical edge states,
  and topological invariants of interacting systems.
\end{abstract}

\tableofcontents

\setlength\hoffset{-0.5in}\setlength\voffset{-0.5in}\setlength\textwidth{6.75in}
\setlength\columnsep{0.2in}\setlength\textheight{9.25in}\mathindent=0.in\twocolumn

\section{Introduction}\label{sec:intro}

The characterization of states of matter in terms of spontaneously broken
symmetries and the corresponding order parameters, known as Ginzburg-Landau
theory \cite{GinzburgLandau}, is one of the most important concepts in condensed matter
theory. Perhaps the best known and studied example is that of magnetic order,
for instance originating from strong electronic interactions. Remarkably,
there also exist insulating phases of matter which do not break any
symmetries, yet are distinct from simple band insulators. The first such
state to be discovered was the integer {\it quantum Hall state} of a
two-dimensional electron gas in a strong magnetic field
\cite{PhysRevLett.45.494}, see \cite{RevModPhys.80.1083} for an
introduction. Its quantized Hall conductivity (in two dimensions,
conductance and conductivity are identical) can be explained by the
existence of a topological invariant (the Chern number) calculated from a
single-particle Hamiltonian
\cite{PhysRevLett.49.405,PhysRevLett.51.51,Kohmoto1985343}. Because this
number cannot change under adiabatic transformations of the Hamiltonian, the
integer quantum Hall state is topologically different from a trivial band
insulator. In particular, a transition between these two states can only
occur via a closing of the band gap. As a result, metallic states emerge at
the edge of quantum Hall samples \cite{PhysRevB.25.2185}. Moreover, while
atomic insulators may be described as product states of completely local
electronic wave functions, topological states show a distinct response to
boundary conditions \cite{FuKa07}. Haldane \cite{Haldane98} showed that a
quantum Hall state with broken time-reversal symmetry can not only be
achieved with an external magnetic field, but also with a periodic magnetic
flux.

{\it Topological insulators} (a name apparently first coined in
\cite{PhysRevB.75.121306}) represent another class of states that are
topologically distinct from simple band insulators. They are symmetric under
time reversal, and can therefore be realized experimentally without magnetic
fields. Topological insulators were first discovered in a theoretical model
for graphene \cite{KaMe05a,KaMe05b}, in the context of the search for an
intrinsic quantum spin Hall effect. Kane and Mele \cite{KaMe05a,KaMe05b}
showed that the intrinsic spin-orbit coupling gives rise to an insulating
state with a topological band structure, and a nonzero spin Hall response
related to metallic edge states that are protected against perturbations 
by time-reversal symmetry. Topological insulators in three
dimensions were predicted soon after \cite{PhysRevLett.98.106803}, and are by
now well confirmed by experiments \cite{RevModPhys.83.1057}. Two-dimensional
topological insulators, or {\it quantum spin Hall insulators}, have close
conceptual relations to integer quantum Hall states: in the simplest case,
the theoretical model of Kane and Mele \cite{KaMe05a,KaMe05b} is completely
equivalent to two decoupled copies (with opposite chirality for the two spin
directions) of the Haldane model for the integer quantum Hall effect
\cite{KaMe05a}.

While the spin-orbit gap in graphene is too small to permit experimental
observation (about $10^{-3}$ meV, see for example \cite{PhysRevB.74.165310};
an experimentally observable spin-orbit gap has been predicted for
silicene \cite{PhysRevLett.107.076802}, see
\cite{PhysRevLett.108.155501,PhysRevLett.108.245501,PhysRevLett.109.056804})
for experimental progress with this honeycomb material),
the quantum spin Hall state has been observed in CdTe/HgTe/CdTe
heterostructures \cite{Koenig07,Roth09,Br.Ro.Bu.Ha.Mo.Ma.Qi.Zh.12}. As first
predicted theoretically \cite{BeHuZh06}, the low-energy physics of this
system is determined by the gap at the $\Gamma$ point. If the HgTe quantum
well has a thickness below a critical value $d_\text{c}$ of 6.3 nm, the band structure
is dominated by the surrounding CdTe layers, and the normal regime, with the
s ($\Gamma_6$) bands lying energetically above the p ($\Gamma_8$) bands, is
realized. If the thickness exceeds the critical value, the band structure
becomes dominated by that of the quantum well, and band inversion takes
place. Bernevig \etal \cite{BeHuZh06} predicted samples with such an inverted
band structure to be quantum spin Hall insulators with a band gap of tens of
meV, which was soon verified experimentally
\cite{Koenig07,Roth09}.

In the absence of interactions, topological insulators can be understood in
terms of single-particle Hamiltonians, allowing for a full classification
based on time-reversal, particle-hole, and chiral symmetry
\cite{PhysRevB.78.195125,Kitaev.09,Bu.Tr.12}. Recently, an extension to
include crystal symmetries has been given \cite{Sl.Me.Ju.Za.12}. This review
is primarily concerned with correlation effects, which have been explored in
the context of topological insulators in several different settings.  The
most direct way to theoretically explore correlation effects is to consider
the impact of electron-electron interactions on a noninteracting quantum spin
Hall state by adding interactions to models with intrinsic spin-orbit
coupling. Remarkably, some of these models permit the application of exact
numerical methods. With regard to experiment, such studies are relevant for
materials with significant spin-orbit and electron-electron
interactions. This situation is expected to occur in, for example, the
transition-metal oxide \chem{Na_2IrO_3} \cite{Irridates-Nagaosa}. The
interplay of strong correlations and spin-orbit coupling in this and other
materials may allow for realizations of complex spin models
\cite{PhysRevB.84.100406,Chaloupka10,Re.Th.Ra.12,Ka.La.Fi.12,arXiv:1210.2290}
and hence may provide a route to quantum spin liquids \cite{Balents10}.

Even in materials with rather weak intrinsic spin-orbit coupling, robust
quantum spin Hall states can emerge from strong electronic correlations 
via a dynamically generated spin-orbit coupling that is the result of
spontaneous symmetry breaking \cite{RaQiHo08}. The resulting {\it topological Mott insulators},
states with a band gap generated by interactions and edge states protected by
time-reversal symmetry, have so far been explored only at the mean-field
level. The term topological Mott insulator is also used in the literature to
refer to exotic states in which the interplay of spin-orbit coupling and
electronic correlations leads to spin-charge separation, gapped charge
excitations, and a topological band structure carried by low-energy spinon
excitations \cite{PeBa10}. A novel Mott state arising from strong
spin-orbit interaction has been found experimentally in \chem{Sr_2IrO_4}
\cite{PhysRevLett.101.076402,Kimetal2009}.

A third direction of research is concerned with fractional topological
insulators.  The fractional quantum Hall state is a famous example of a
topological phase emerging from strong electronic interactions (for a concise
introduction see \cite{RevModPhys.80.1083}). In the latter, strong repulsion
among electrons occupying partially filled, flat Landau levels leads to
intriguing insulating states with quasi-particle excitations that carry
fractional charge and have fractional statistics. Fractional quantum Hall
states have been a major research topic for several decades, with enormous
progress in theory and experiment, and potential applications in quantum
computing \cite{RevModPhys.80.1083}. Remarkably, much of the physics of the
fractional quantum Hall effect can in theory also exist in Chern insulators
\cite{FQHE_sheng2011}, described by models that break time-reversal symmetry
but have zero net magnetic flux, and are hence similar in spirit to Haldane's
model for the integer quantum Hall effect \cite{Haldane98}. Proposals for the
experimental realization of fractional Chern insulators involve materials
with flat bands that allow for an enhancement of electron-electron
interactions \cite{PhysRevLett.108.126405}.  In close analogy to the
noninteracting case, {\it fractional quantum spin Hall insulators} can be
constructed by, in the simplest case, combining two fractional quantum Hall
states \cite{PhysRevLett.103.196803}. The theory of time-reversal invariant
insulators with protected edge states is one of the most active current
research areas (see section~\ref{sec:bulk:fqhe}).

In the context of correlated electrons, the concept of topological order has
yet another, important meaning. Quite generally, a topologically ordered
state is robust with respect to adiabatic transformations of the
Hamiltonian. In the case of the $Z_2$ topological insulators, as realized
for example in the Kane-Mele model \cite{KaMe05a,KaMe05b}, this robustness is
mathematically expressed by a topological invariant. However, even in the
presence of electron-electron interactions, this state---having its origin in
a topological band structure---is adiabatically connected to a noninteracting
band insulator. A classification of such states based on Chern-Simons theory
has recently been given \cite{PhysRevB.86.125119}.  Topological band
insulators are distinct from the class of topological states that emerge from
interactions and are not adiabatically connected to a noninteracting state. A
particularly fascinating example are quantum spin liquids, which are
insulating states---arising from quantum fluctuations---that do not break any
symmetries but have highly nontrivial correlations. For the latter, the term
{\it topological order} implies \cite{Wen_book} (i) the existence of a gap to
excited states, (ii) a topological ground-state degeneracy in a periodic
system (a torus geometry in two dimensions), and (iii) the existence of
fractional quasi-particle excitations with Abelian or non-Abelian
statistics. Whereas (i) is also typical of topological band insulators, (ii)
and (iii) are not. On the other hand, fractional quantum Hall liquids and hence also
fractional topological insulators possess topological order. 
An important recent development is the use of the concept of quantum
entanglement to study and classify topological states \cite{PhysRevB.82.155138}. 
Whereas topological insulators are characterized by short-range quantum
entanglement and fall into the class of {\it symmetry-protected topological
  states} (stable with respect to perturbations that do not break the symmetry
that underlies the topological properties), fractional topological
states and quantum spin liquids have long-range quantum entanglement
\cite{PhysRevB.84.235141}. Whereas any short-range entangled state can be
transformed into a product state by means of {\it local unitary
transformations}, such a transformation is not possible in the presence
of long-range quantum entanglement \cite{PhysRevB.82.155138}. A
classification of topological states, including correlated states, can be
obtained by considering local unitary transformations with different
symmetries. For a more detailed discussion, see
\cite{PhysRevB.82.155138,arXiv:1106.4772,PhysRevB.84.235141,Chen2012,arXiv:1209.4399,arXiv:1212.1726}.

Three-dimensional topological insulators are realized by several different materials
\cite{HaKa10,doi:10.1146/annurev-conmatphys-062910-140432,RevModPhys.83.1057}.
Strong correlation effects haven been argued to play a role in \chem{SmB_6}
(a potential topological Kondo insulator) \cite{arXiv:1211.5104,arXiv:1211.6769} and
in actinide compounds such as \chem{AmN} (identified as a $Z_2$ topological
insulator in theoretical calculations) \cite{Zhang23032012}. A candidate
system for a topological Mott insulator is \chem{Sr_2IrO_4} \cite{PhysRevLett.101.076402,Kimetal2009}.
In contrast, experiments for the two-dimensional case are so far restricted
to HgTe quantum wells \cite{Koenig07,Roth09,Br.Ro.Bu.Ha.Mo.Ma.Qi.Zh.12}, and do not
reveal any substantial electronic correlation effects. Consequently, the
study of correlated quantum spin Hall insulators has so far been a
predominantly theoretical effort. The search for candidate condensed matter
settings is guided by the fact that the spin-orbit coupling increases with
the atomic number. For example, strong and comparable electron-electron and
spin-orbit interactions are expected  for Ir-based transition metal
oxides \cite{Irridates-Nagaosa}, with numerical evidence for a strongly
correlated quantum spin Hall state \cite{Irridates-Nagaosa}, and the
possibility of a topological Mott insulator phase \cite{PeBa10}.  Another
useful experimental setting are heterostructures of transition-metal oxides,
for example a bilayer of \chem{LaNiO_3} sandwiched between \chem{LaAlO_3}
layers \cite{Xi.Zh.Ra.Na.Ok.11}. Such structures provide
substantial tunability \cite{Xi.Zh.Ra.Na.Ok.11,PhysRevB.86.235141}, and have been proposed to
realize topological phases emerging from either intrinsic spin-orbit
coupling \cite{Xi.Zh.Ra.Na.Ok.11}, or from interaction-induced ordering of
complex orbitals \cite{PhysRevB.84.201104,Ru.Fi.11,PhysRevB.85.245131,PhysRevB.84.241103}.
Interaction induced topological phases may also exist in stacked graphene
\cite{PhysRevB.82.115124,PhysRevLett.106.156801}, whereas the intrinsic
spin-orbit coupling may be sufficiently enhanced in decorated graphene
\cite{PhysRevX.1.021001} or in molecular graphene \cite{Go.Ma.Ko.Gu.Ma.12} to
make a topological phase observable.  Although concrete materials for
fractional topological insulators have not yet been identified, several
candidate systems for Chern insulators are known, see
\cite{PhysRevLett.108.126405} and references therein.  Cold atoms in optical
lattices, or on atom chips, provide an alternative route toward correlated
topological states. Using suitable artificial gauge fields
\cite{RevModPhys.83.1523}, $Z_2$ topological insulators with tunable
interactions may be realized \cite{PhysRevLett.107.145301,PhysRevLett.105.255302}.  
The use of Rydberg atoms has also been suggested to realize topological Mott
insulators \cite{PhysRevA.86.053618} and fractional quantum Hall states
\cite{arXiv:1207.3716}. Finally, correlated
topological states may also be engineered with photons in cavity arrays, see
\cite{PhysRevA.82.043811,arXiv:1206.1539}.

Despite the rather short history of topological insulators, the literature is
remarkably rich. Given the background of the authors, this review aims at
providing an overview of work on electronic correlation effects in two
dimensions, with a focus on models with intrinsic spin-orbit coupling and
numerical studies.  It thereby complements earlier reviews by Hasan and Kane
\cite{HaKa10}, Moore \cite{Moore10}, and Qi and Zhang
\cite{RevModPhys.83.1057}, but omits other fascinating topics such as
three-dimensional topological insulators
\cite{HaKa10,RevModPhys.83.1057,doi:10.1146/annurev-conmatphys-062910-140432},
topological Kondo insulators
\cite{DzSuGa10,PhysRevB.85.045130,arXiv:1211.5104}, or topological
superconductors
\cite{PhysRevB.78.195125,Kitaev.09,RevModPhys.83.1057,HaKa10,RevModPhys.83.1057}. For
a self-contained presentation, it includes an introduction to
noninteracting quantum spin Hall insulators using the example of the
Kane-Mele model.  The review is organized as follows. In
section~\ref{sec:models}, the most widely used theoretical models are
introduced. A discussion of noninteracting quantum spin Hall insulators is
provided in section~\ref{sec:noninteracting}. Correlation effects on bulk
properties are considered in section~\ref{sec:bulk}, whereas correlated
helical edge states are the topic of section~\ref{sec:edge}. The calculation
of topological invariants for correlated systems is reviewed in
section~\ref{sec:index}, and section~\ref{sec:conclusions} gives conclusions
and an outlook.

\section{Theoretical models}\label{sec:models}

Since the proposal by Kane and Mele \cite{KaMe05a,KaMe05b}, a number of
theoretical models for the quantum Hall and spin Hall effect have been
discovered. The general structure of such models and possible realizations in
semiconductors were discussed, for example, in \cite{PhysRevB.74.085308}. In
many cases, spinless quantum Hall models can readily be generalized to
describe time-reversal invariant quantum spin Hall states by introducing a
spin-orbit gap with a spin-dependent sign.  Here the emphasis is on some of
the most popular models that support a topological state originating from
spin-orbit coupling in the absence of electronic interactions.  Overviews of
theoretical models can also be found in
\cite{FuKa07,Fi.Ch.Hu.Ka.Lu.Ru.Zy.11}.

A common feature of the quantum spin Hall models considered here is that the
Hamiltonians may be expressed in terms of $4\times 4$ Dirac matrices
\cite{KaMe05a,FuKa07}. Given inversion symmetry, such a representation is
particularly useful to determine the topological invariant in the absence of
interactions \cite{KaMe05a,FuKa07}. The corresponding matrix representations
of the KM and the Bernevig-Hughes-Zhang (BHZ) model can be found in
\cite{FuKa07}. To study local, Hubbard-type interactions, and also for
numerical methods, the tight-binding real-space representations are more
appropriate.

Because of the topological character of the quantum spin Hall state, the
details of the Hamiltonian play a minor role, and different models share
similar low-energy physics. In particular, universal field theory
descriptions can be obtained using Chern-Simons theory for the gapped bulk
\cite{QiHuZh08}, and helical Luttinger liquid theory for the gapless edges
\cite{Wu06,Cenke06}. The above-mentioned requirement of at least four
orbitals per unit cell (here the term orbital may also refer to spin or
pseudospin) or, equivalently, four states at a given momentum $\bm{k}$ in the
noninteracting Bloch picture, can be understood as follows: the two spin
directions are related by time-reversal symmetry and hence degenerate.  A
band insulator, with a gap between filled valence bands and empty conduction
bands and hence not adiabatically connected to the vacuum, then requires two
sets of Kramers degenerate bands, and hence four
bands in total \cite{PhysRevB.85.195116}.

While some features are generic for quantum spin Hall models, there are also
model-specific aspects related to, for example, the crystal lattice. An
important example is the potential quantum spin liquid phase of the KMH
model, which was first reported for the Hubbard model on the honeycomb
lattice \cite{Meng10}. This lattice has the smallest possible coordination
number $z=3$ in two dimensions, rendering quantum fluctuations particularly
important. The honeycomb lattice is also at the centre of attention in
condensed matter physics because of tremendous progress in preparing single
layers or bilayers of graphene \cite{Novoselov22102004,Neto_rev}.

Much of the work reviewed here uses the paradigmatic framework of a local
Hubbard interaction to study electronic correlation effects \cite{Hu63}. This
choice, usually motivated by simplicity, is assumed to capture the dominant
physics.  Models with nonlocal interactions play a key role for topological
Mott insulators with dynamically generated spin-orbit interactions (see
section~\ref{sec:bulk:topmott}), and for spinless models of Chern insulators
(see section~\ref{sec:bulk:fqhe}). Nonlocal interactions may also give rise
to interesting new physics in helical liquids \cite{Cenke06}.

\subsection{Kane-Mele-Hubbard model}

The KM Hamiltonian was originally derived as a model for graphene, and can be
written in the form \cite{KaMe05a,KaMe05b} 
\begin{eqnarray}\label{eq:KM}
  H_{\mbox{\scriptsize{KM}}} 
  &&= 
  -t \sum_{\las i,j \ras} \hat{c}^{\dagger}_{i} \hat{c}^\nag_{j} 
  + 
  \rmi\,\lso \sum_{\llas i,j\rras}
  \hat{c}^{\dagger}_{i}\,
  (\boldsymbol{\nu}_{ij} \cdot \boldsymbol{\sigma})\,
  \hat{c}^\nag_{j} 
  \\\nonumber
  &&\quad+
  \rmi\,\lr \sum_{\las i,j\ras} 
  \hat{c}^{\dagger}_{i}\,
  (\bm{s} \times \hat{\bm{d}}^\nag_{ij})_z \,\hat{c}^\nag_{j}\,. 
\end{eqnarray}
Here ${\hat{c}^{\dagger}_{i} = \big(c^{\dagger}_{i\uparrow},
  c^{\dagger}_{i\downarrow}\big)}$ is a spinor, and $c^{\dagger}_{i\sigma}$
creates an electron in a Wannier state at site $i$ with spin $\sigma$. Pairs
of nearest-neighbour and next-nearest-neighbour lattice sites are indicated
by the symbols $\las i,j\ras$ and $\llas i,j\rras$, which also implicitly
include the Hermitian conjugate terms.

\begin{figure}[t]
  \centering
  \includegraphics[width=0.4\textwidth]{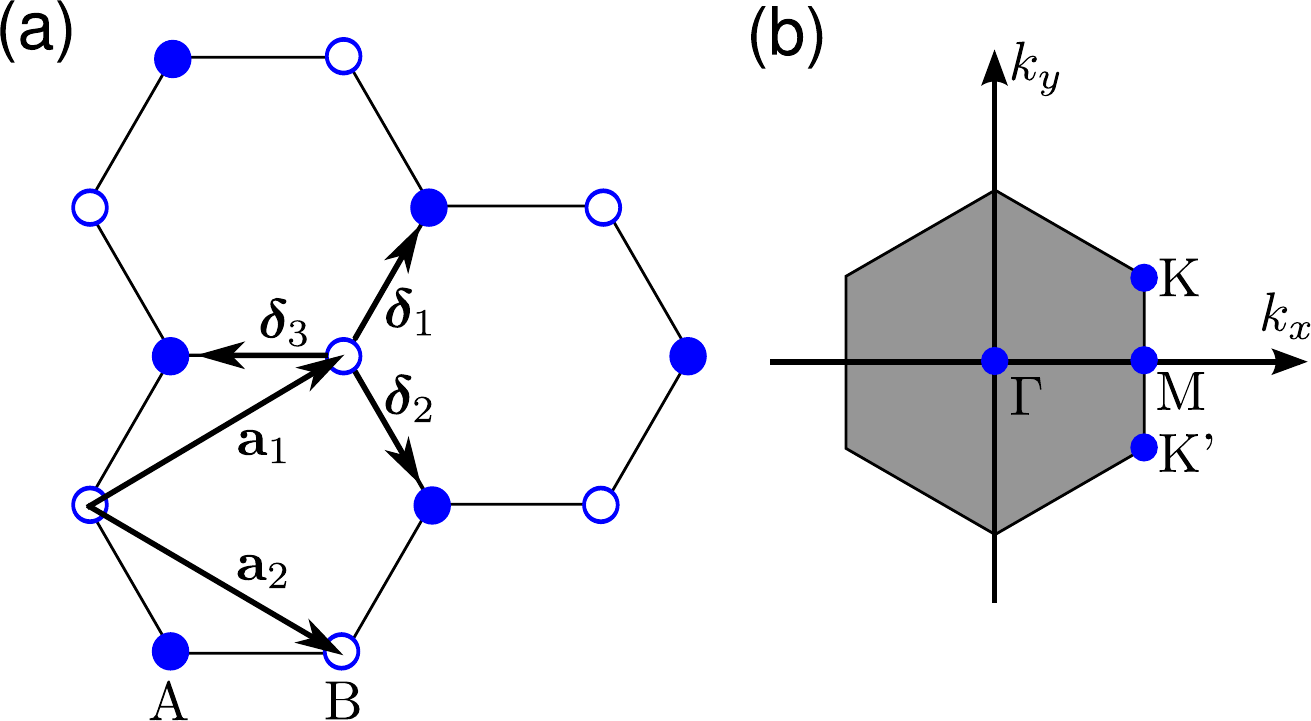}
  \caption{\label{fig:honeycomb}
    (a) The honeycomb lattice with lattice constant $a$ consists 
    of two sublattices A, B and is spanned by the basis vectors $\bm{a}_1=\oh
    a(3,\sqrt{3})$, $\bm{a}_2=\oh a(3,-\sqrt{3})$ (in the conventions of
    \cite{Neto_rev}). Nearest-neighbour lattice sites are connected by the 
    vectors $\bm{\delta}_1=\oh a (1,\sqrt{3})$, $\bm{\delta}_2=\oh a
    (1,-\sqrt{3})$, and $\bm{\delta}_3=a (-1,0)$. The hexagonal first
    Brillouin zone contains the two nonequivalent Dirac points
    $\mathrm{K}=\left(\frac{2\pi}{3a},\frac{2\pi}{3\sqrt{3}a}\right)$ and 
    $\mathrm{K}^\prime=
    \left(\frac{2\pi}{3a},-\frac{2\pi}{3\sqrt{3}a}\right)$.
    }
\end{figure}

The KM model~(\ref{eq:KM}) is defined on the honeycomb lattice spanned by the
lattice vectors $\bm{a}_1$, $\bm{a}_2$, as shown in
figure~\ref{fig:honeycomb}(a) (see caption for details).  Nearest-neighbour
sites are connected by the vectors $\pm \bm{\delta}_i$, $i=1,2,3$. The
honeycomb lattice has two lattice sites per unit cell, and together with the
spin degree of freedom, there are four orbitals per unit cell.  The Brillouin
zone, shown in figure~\ref{fig:honeycomb}(b), contains the Dirac points
$\mathrm{K}$ and $\mathrm{K}'$.

The first term in (\ref{eq:KM}) is the usual tight-binding hopping term
between sites on different sublattices, which leads to the celebrated
graphene band dispersion (here the lattice constant $a=1$) \cite{Wallace47}
\begin{eqnarray}\label{eq:bandstructure}
  \epsilon(\bm{k}) 
  &= \pm |g_{\bm{k}}| \\\nonumber
  &= \pm  t \left[3 + 2\cos(\sqrt{3}k_y) \right.\\\nonumber
  &\qquad+ \left.
    4 \cos(3k_x/2)\cos(\sqrt{3}k_y/2)\right]^{1/2},
\end{eqnarray}
with Dirac cones at  $\mathrm{K}$ and $\mathrm{K'}$. 

\begin{figure}[t]
  \centering
  \includegraphics[width=0.3\textwidth]{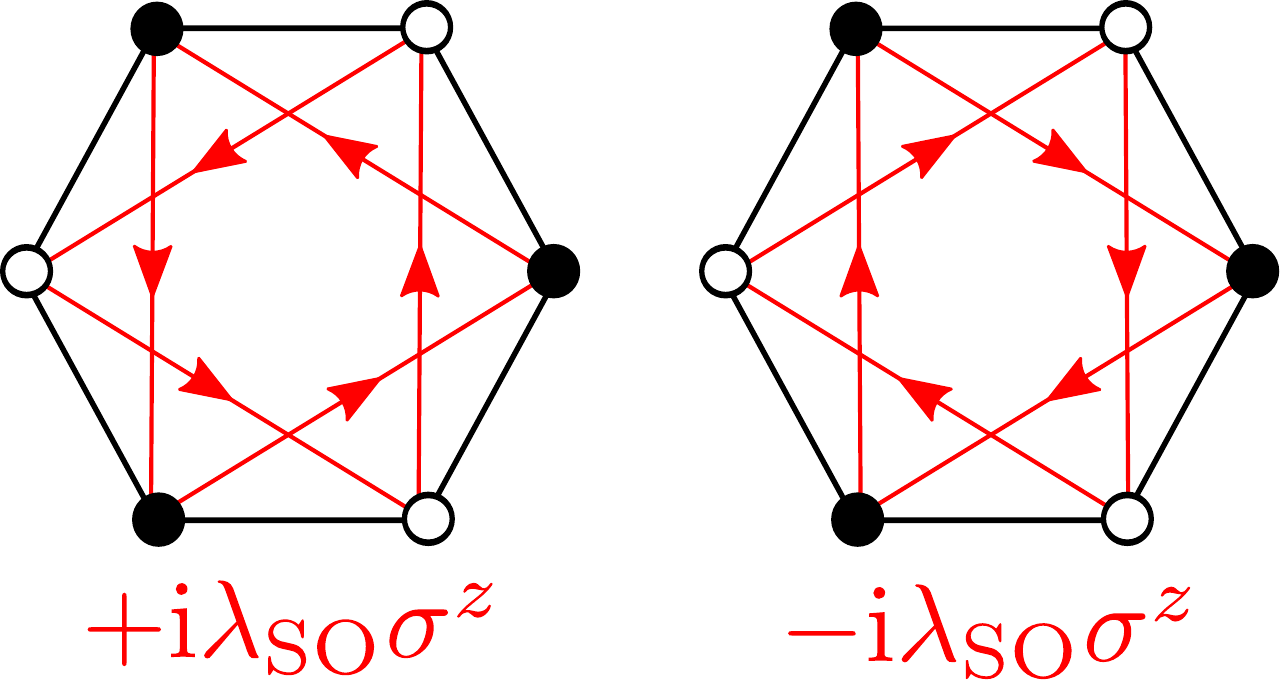}
  \caption{\label{fig:kanemeleso}  
    Sign structure of the spin-orbit term in the Kane-Mele
    model~(\ref{eq:KM}).
  }
\end{figure}

The spin-orbit interaction ($\sim\lso$) is described by a complex-valued
next-nearest-neighbour hopping term with a sign $\pm1$ depending on the
sublattice, the direction of the hop (\ie, left turn or right turn), and the
spin orientation, as illustrated for a single hexagon in
figure~\ref{fig:kanemeleso}. This sign is encoded in $(\boldsymbol{\nu}_{ij}
\cdot \boldsymbol{\sigma})$, where
\begin{equation}
  \boldsymbol{\nu}_{ij} = \frac{\bm{d}_{ik} \times \bm{d}_{kj}}{|\bm{d}_{ik} \times \bm{d}_{kj}|}\,,
\end{equation}
$\bm{d}_{ik}$ is the three-dimensional vector (with vanishing $z$ component)
connecting sites $i$ and $k$, and $k$ is the intermediate lattice site
involved in the hopping process from site $i$ to site $j$. The vector
$\boldsymbol{\sigma}$ is defined by the Pauli matrices as
$\boldsymbol{\sigma} = (\sigma^x,\sigma^y,\sigma^z)$.  Close to the Dirac
points, the spin-orbit term can be written as $H_\text{SO} \sim\lso
\psi^\dagger \sigma^z \tau^z s^z\psi$, where the Pauli matrices refer to the
spin, sublattice, and Dirac point, respectively \cite{KaMe05b}.

The last term in (\ref{eq:KM}) is the Rashba spin-orbit interaction of
strength $\lr$, defined in terms of the spin vector
$\bm{s}=\oh\boldsymbol{\sigma}$, and the unit vector
$\hat{\bm{d}}_{ij}=\bm{d}_{ij}/|\bm{d}_{ij}|$.  The Rashba coupling, which is
purely off-diagonal in spin (hence, spin is no longer conserved), breaks the
$z\mapsto-z$ inversion symmetry, and arises for example, in the presence of a
substrate \cite{KaMe05b}.

For $\lso\neq0$, the KM model with zero chemical potential describes a
quantum spin Hall insulator as long as $\lr<2\sqrt{3}\lso$
\cite{KaMe05a,PhysRevLett.97.036808}, see also
section~\ref{sec:noninteracting}. On a geometry with open edges, the KM
model exhibits helical edge states with a symmetry-protected crossing at
$k=0$ ($k=\pi$) for armchair (zigzag) edges.

The KM-Hubbard (KMH) model corresponds to $H_\text{KM}$ with an additional,
local Hubbard repulsion between electrons, and is written as
\begin{equation}\label{eq:KMH}
  H_{\mbox{\scriptsize{KMH}}} = H_\text{KM} + \oh U \sum_{i} (\hat{c}^{\dagger}_{i} \hat{c}^\nag_{i} -  1 )^2\,.
\end{equation}
It was first considered by Rachel and Le Hur \cite{RaHu10}. Compared to the
Hubbard model ($\lso=0$), the spin-orbit terms reduce the rotational symmetry
from the hexagonal point group $C_6$ to $C_3$, and the spin rotation symmetry from
$SU(2)$ to either $U(1)$ (for $\lr=0$) or $Z_2$ (for $\lr\neq0$).  A detailed
review of previous work on the KMH model is provided in
sections~\ref{sec:bulk} and \ref{sec:edge}, and the noninteracting case is
considered in section~\ref{sec:noninteracting}.

The KMH model has played an important role because it can be investigated
with exact quantum Monte Carlo methods. Moreover, there is a close connection
to the Hubbard model on the honeycomb lattice, which after reports of the
existence of an intermediate spin liquid phase \cite{Meng10} has been at the
focus of intense investigations. The existence of a Mott metal-insulator
transition at a finite critical $U_\text{c}$ on the honeycomb lattice is a
result of the vanishing density of states at the Fermi level, $N(\om)\sim|\om|$
\cite{Neto_rev}, which renders the logarithmic instability toward magnetic
order ineffective. This stability with respect to perturbations is a general
property of Dirac points. In contrast, symmetry breaking and topological
phases can arise for arbitrarily small interactions in models with quadratic
band crossing points \cite{PhysRevLett.103.046811}.

\subsection{Haldane-Hubbard model}

Although it breaks time-reversal symmetry and hence does not describe a
quantum spin Hall state, the Haldane-Hubbard model (the
  name is somewhat misleading, because the interaction is nonlocal) is
included here for two reasons. First, the noninteracting version
\cite{Haldane88} has played a key role for the development of the field of
quantum spin Hall insulators as it underlies the KM model
\cite{KaMe05a,KaMe05b}. Second, the interacting model has been studied in the
context of topological quantum phase transitions
\cite{Wa.Sh.Zh.Wa.Da.Xi.10,VaSuRi10,Va.Su.Ri.Ga.11}.

The Hamiltonian can be written as
\begin{eqnarray}\label{eq:haldane}\nonumber
  H_\text{HH} 
  = &-t_1 \sum_{\las ij \ras} c^\dag_i c^\nag_j -t_2 \sum_{\llas ij \rras}
  \rme^{\rmi \phi_{ij}} c^\dag_i c^\nag_j \\
  &+ V \sum_{\las ij \ras} \on_i \on_j\,.
\end{eqnarray}
The first term describes the nearest-neighbour hopping of {\it spinless}
fermions on the honeycomb lattice. The second term is a spin-orbit coupling
with a phase $\phi_{ij}=\pm\phi$ for clockwise (anticlockwise)
next-nearest-neighbour hopping, equivalent to the spin-orbit term of the KM
model~(\ref{eq:KM}) for a fixed spin direction.  Also included is a repulsion
$V$ between spinless fermions at neighbouring lattice sites. Because of the
restriction to spinless fermions (or, equivalently, to one spin sector), the
Hamiltonian~(\ref{eq:haldane}) breaks time-reversal symmetry, and for $V=0$ realizes the {\it
  integer} quantum anomalous Hall effect at half filling
\cite{Haldane88}. Large enough interactions cause a quantum phase transition
\cite{Wa.Sh.Zh.Wa.Da.Xi.10,VaSuRi10,Va.Su.Ri.Ga.11} to a charge-density-wave
state, see section~\ref{sec:bulk}.

\subsection{Bernevig-Hughes-Zhang-Hubbard model}

The low-energy, continuum model for HgTe quantum wells involves two s and
two p orbitals, each forming a Kramers pair, with a gap at the $\Gamma$ point
\cite{BeHuZh06}. From this model, Bernevig \etal \cite{BeHuZh06} derived a
two-dimensional tight-binding model on the square lattice. A possible
real-space representation, hereafter referred to as the BHZ model, is \cite{FuKa07}
\begin{eqnarray}\label{eq:BHZ}
  H_\text{BHZ} &= \sum_{\bm{i}} \sum_{\alpha} \sum_\sigma \epsilon_\alpha {c}^\dag_{\bm{i},\alpha,\sigma} {c}^\nag_{\bm{i},\alpha,\sigma}\\\nonumber
  &\quad-\sum_{\bm{i},\bm{\delta}} \sum_{\alpha\beta} \sum_\sigma
  {c}^\dag_{\bm{i}+\bm{\delta},\alpha,\sigma}  [t^\sigma_{\bm{\delta}}]^{\alpha\beta} c^\nag_{\bm{i},\beta,\sigma}\,,
\end{eqnarray}
where $a$ labels the four bonds to nearest-neighbour lattice sites,
$\alpha,\beta\in\{\text{s},\text{p}\}$ are orbital indices, and $\sigma$
is a spin index. The $2\times2$ hopping matrix defined in orbital space is
given by (here $\text{sgn}(\sigma)=+1,-1$ for $\sigma=\UP,\DO$)
\begin{eqnarray}
  t_{\pm x}^\sigma
  &=
  \left(
    \begin{array}{cc}
    t_\text{ss} &  \pm t_\text{sp}  \\
    \mp t_\text{sp}  &   -t_\text{pp}
    \end{array}
  \right)\,,\\\nonumber
  t_{\pm y}^\sigma
  &=
  \left(
    \begin{array}{cc}
    t_\text{ss} &   \pm \rmi \,\text{sgn}(\sigma) t_\text{sp} \\
    \pm\rmi \,\text{sgn}(\sigma) t_\text{sp}  &   -t_\text{pp}
    \end{array}
  \right)\,.
\end{eqnarray}
The momentum-space representation of the tight-binding model reduces to the
continuum model for HgTe upon expanding around the $\Gamma$ point
\cite{BeHuZh06}.  A convenient choice of parameters is to set
$\epsilon_\text{s}=-\epsilon_\text{p}\equiv\epsilon/2$, so that $\epsilon$ is
the energy splitting between s and p orbitals, and
$t_\text{ss}=t_\text{pp}\equiv t$. The energy bands are then given by
\begin{eqnarray}
  E(\bm{k}) = &\pm 2 \left\{ 4 t_\text{sp}^2 (\sin^2 k_x + \sin^2 k_y ) \right.\\\nonumber
   &\qquad+\left. [\epsilon /2 - 2t (\cos k_x + \cos k_y)]^2 \right\}^{1/2}\,.
\end{eqnarray}
Writing the BHZ model in terms of Dirac matrices \cite{FuKa07} and
considering the eigenvalues of the parity operator at the time-reversal
invariant points of the Brillouin zone gives a quantum spin Hall phase for
$-8t<\epsilon<8t$, and a trivial insulator else.  This condition, when
translated into the corresponding notations, agrees with previous work
\cite{BeHuZh06,FuKa07,Zaanen12,PhysRevB.85.235449}.  However, as pointed out
in \cite{Zaanen12}, there are in fact two different topological phases,
separated by a metallic point at $\epsilon=0$. These phases have
topological band gaps at the $\mathrm{\Gamma}$ point, $\Delta(\mathrm{\Gamma})
= 4|\epsilon /2 - 4t|$, and at the $\mathrm{M}$ points, $\Delta(\mathrm{M}) =
2|\epsilon|$, respectively. They can be distinguished by their response to
topological defects in the form of dislocations \cite{Zaanen12}. For systems
with edges, the protected crossing point of the helical edge states is
located at $k=0$ ($\mathrm{\Gamma}$ phase) and $k=\pi$ ($\mathrm{M}$ phase),
respectively. Only the $\mathrm{\Gamma}$ phase is observable in HgTe quantum
wells \cite{BeHuZh06}, whereas the $\mathrm{M}$ phase is a feature of the
tight-binding model.

The Hamiltonian~(\ref{eq:BHZ}) neglects terms that couple the two spin
sectors which arise from the breaking of inversion and/or axial
symmetry in HgTe quantum wells, but are expected to be small
\cite{BeHuZh06}. Nevertheless, the spin-conserved case is not
generic. The effects of nonconserving terms have been
discussed, for example, in \cite{Ro.Re.Li.Mo.Zh.Ha.10,PhysRevLett.108.156402}.

Because there are two orbitals per site, the definition of a local Hubbard
interaction term is not unique, and both intra-orbital interactions of the
form $H_U=U\sum_{\bm{i}\alpha} \on_{\bm{i}\alpha\UP}\on_{\bm{i}\alpha\DO}$
\cite{arXiv:1111.6250,Wa.Da.Xi.12,arXiv:1202.3203} as well as intra-- and
interorbital interactions, $H_U=U \sum_{\bm{i}} \on_{\bm{i}}
(\on_{\bm{i}}-1)$ (with
$\on_{\bm{i}}=\sum_{\alpha\sigma}\on_{\bm{i}\alpha\sigma}$)
\cite{PhysRevB.85.235449} have been studied. For realistic calculations, the
full Coulomb interaction (including Hunds rule coupling) should be taken into
account \cite{arXiv:1211.3059}.  Interaction effects on the edge
states of the {\it BHZ-Hubbard (BHZH) model} have been explored in
\cite{arXiv:1202.3203,Wa.Da.Xi.12,PhysRevLett.108.156402}, whereas bulk
properties were investigated in \cite{arXiv:1111.6250,Wa.Da.Xi.12,PhysRevB.85.235449,arXiv:1207.4547v1,arXiv:1211.3059}.

\subsection{Sodium iridate model}

\begin{figure}[t]
  \centering
  \includegraphics[width=0.4\textwidth]{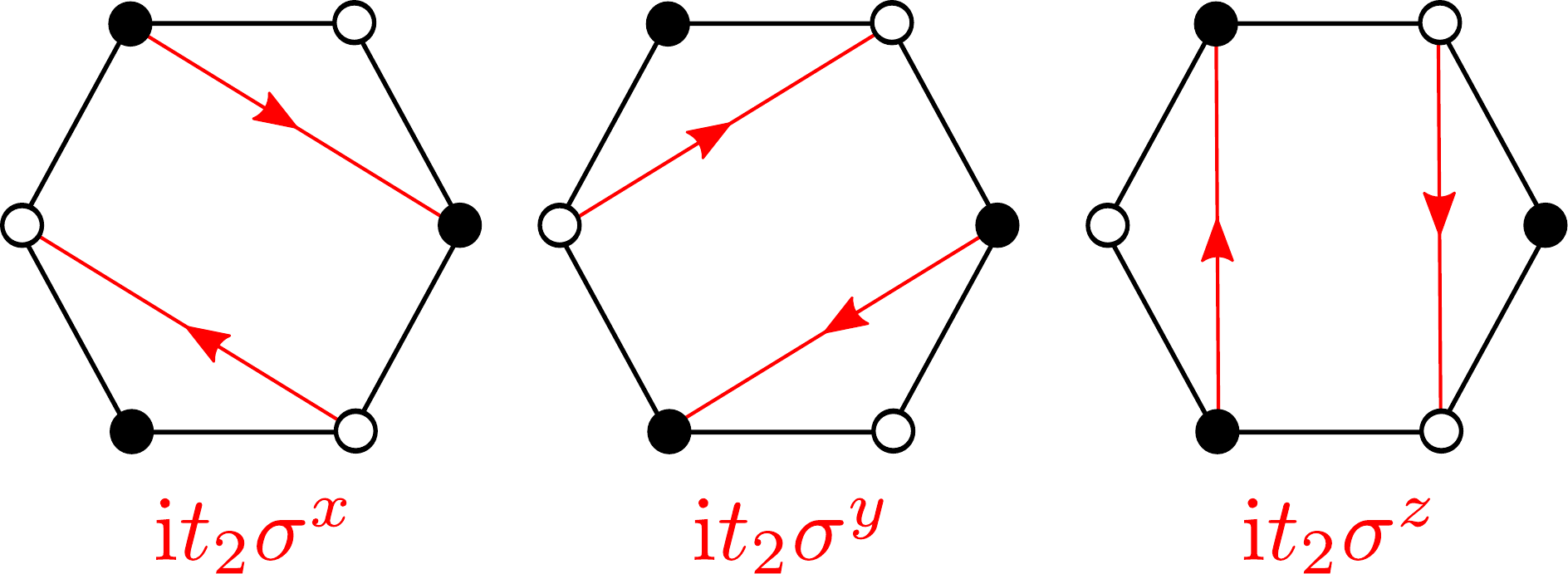}
  \caption{\label{fig:simodel}
    Spin-orbit terms of the SI model~(\ref{eq:SI}), see also \cite{Irridates-Nagaosa}.}
\end{figure}

Shitade \etal \cite{Irridates-Nagaosa} predicted a quantum spin Hall phase
with substantial spin-orbit and electron-electron interactions in the
transition-metal oxide \chem{Na_2IrO_3}. Exploiting the layered structure of
this material, they derived an effective, two-dimensional model on the
honeycomb lattice. The Hamiltonian takes the form \cite{Irridates-Nagaosa,PhysRevLett.108.046401,Re.Th.Ra.12}
\begin{eqnarray}\nonumber\label{eq:SI}
  H_\text{SI} = 
  &-t \sum_{\las ij\ras}\sum_\sigma  c^\dag_{i\sigma}c^\nag_{j\sigma}
  +
  \sum_{\llas ij\rras} \sum_{ss'}
  c^\dag_{is} [t_{ij}]^{ss'} c^\nag_{js'}\\
  &+ U \sum_i \on_{i\UP}\on_{i\DO}\,.
\end{eqnarray}
It describes nearest-neighbour and next-nearest-neighbour electron hopping
between Ir atoms. The indices $s,s'$ refer to pseudospin states
\cite{Irridates-Nagaosa}, but will also be called spin states in the
following. The nearest-neighbour hopping ($t$) is real and diagonal in spin.
The next-nearest-neighbour hopping in general has real and complex
contributions \cite{Irridates-Nagaosa},
\begin{equation}
  [t_{ij}]^{ss'} = -t_1 \delta_{ss'} + \rmi t_2 \sigma^w_{ss'}\,,
\end{equation}
where $w=x,y,z$ selects one of the Pauli matrices depending on the hopping
path, as illustrated in figure~\ref{fig:simodel}. The contribution  $t_1$ 
is sometimes neglected.

In contrast to the KM model (\ref{eq:KM}), there is only one parameter $t_2$
that determines the strength of both spin-conserving and nonconserving
hopping processes. Spin is therefore generically not conserved in the SI
model.  Moreover, both contributions act between next-nearest-neighbour
lattice sites, whereas the spin-orbit and Rashba terms in the KM
model~(\ref{eq:KM}) connect next-nearest and nearest-neighbour sites,
respectively. The helical edge states cross at $k=0$ ($k=\pi$) on
armchair (zigzag) edges.

The model~(\ref{eq:SI}) describes a single honeycomb layer of Ir atoms in
\chem{Na_2IrO_3}, and may be realized at the surface of a three-dimensional
sample \cite{Irridates-Nagaosa}. In contrast to the KMH model for graphene
and the BHZH model for HgTe, strong electron-electron interactions have been
argued to be generically present in \chem{Na_2IrO_3}
\cite{Irridates-Nagaosa}. The SI model has been studied in
\cite{PhysRevLett.108.046401,Re.Th.Ra.12,Ka.La.Fi.12}.  

\subsection{Other models}

Whereas a minimum of four states per unit cell is necessary for a gapped,
time-reversal invariant band structure, there exist several examples of more
complex models that support a quantum spin Hall state. These include the
kagom\'{e} lattice \cite{PhysRevB.80.113102}, the decorated honeycomb lattice
\cite{PhysRevB.81.205115}, the Lieb lattice \cite{PhysRevB.82.085310}, the
square-octagon lattice \cite{PhysRevB.82.085106}, and the ruby lattice
\cite{PhysRevB.84.155116}.  The corresponding multi-band Hamiltonians are of
particular interest for two reasons. First, the decorated honeycomb lattice
has been shown to provide a remarkable connection between a
noninteracting topological state (the quantum spin Hall state) and a strongly
interacting topological state (the chiral spin liquid phase of the Kitaev
model \cite{Kitaev20032,Kitaev20062}), both of which are exactly solvable
\cite{PhysRevB.81.205115,Fi.Ch.Hu.Ka.Lu.Ru.Zy.11}.  Second, such models can
have band structures that include very flat bands, and thereby possess the
potential for fractional topological states arising from electrons in
strongly correlated, partially filled flat bands \cite{PhysRevB.84.155116},
see also section~\ref{sec:bulk:fqhe}. Finally, whereas a quantum spin Hall
state on the honeycomb lattice requires complex next-nearest-neighbour
hopping, {\it nearest-neighbour} hopping is sufficient on more complicated
lattices such as the kagom\'{e} or the decorated honeycomb lattices
\cite{PhysRevB.81.205115}.

Other models that are relevant in the context of correlation effects were
proposed by Ara\`ujo \etal \cite{Ar.Ca.Sa.12} (a bilayer square-lattice model), Goryo
and Maeda \cite{JPSJ.80.044707} (the bilayer KMH model), and Cocks \etal
\cite{Cocks.12} (a time-reversal invariant version of the Hofstadter
problem). Finally, lattice models for topological Mott insulators and
fractional topological insulators will be discussed in
sections~\ref{sec:bulk:topmott} and~\ref{sec:bulk:fqhe}, respectively.

\section{Noninteracting quantum spin Hall insulator}\label{sec:noninteracting}

Although this review is mainly concerned with correlation effects, a brief
summary of the noninteracting case provides the starting point for the
discussion of interacting systems. It also illustrates the key concepts of
time-reversal invariance, the bulk-boundary correspondence, and topological
protection.  For concreteness, the example of the KM model~(\ref{eq:KM}) is
used. More general reviews have been given in
\cite{HaKa10,RevModPhys.83.1057}.

The KM model~(\ref{eq:KM}) can be solved exactly by Fourier
transformation. In the absence of a Rashba term, it can be written in the
form $H_\text{KM}=\sum_{\bm{k}} \Psi^\dag_{\bm{k}} H(\bm{k})
\Psi^\nag_{\bm{k}}$, with
\begin{eqnarray}\label{eq:kspace}\nonumber
H(\bm{k}) &= 
\left(
  \begin{array}{cccc}
    \gamma_{\bm{k}} & -g_{\bm{k}} & 0 & 0 \\
    -g^*_{\bm{k}} & -\gamma_{\bm{k}} & 0 & 0\\
    0 & 0 & -\gamma_{\bm{k}} & -g_{\bm{k}}\\
    0 & 0 & -g^*_{\bm{k}}  & \gamma_{\bm{k}}
  \end{array}
\right)\\
&= 
\left(
  \begin{array}{cc}
     H^\UP(\bm{k}) & 0_{2\times2} \\
     0_{2\times2} & H^\DO(\bm{k}) \\
  \end{array}
\right)
\end{eqnarray}
and
$\Psi^\dag_{\bm{k}}=(a^\dag_{\bm{k}\UP},b^\dag_{\bm{k}\UP},a^\dag_{\bm{k}\DO},b^\dag_{\bm{k}\DO})$.
Here $a$ and $b$ operators refer to the two sublattices of the honeycomb
lattice, see figure~\ref{fig:honeycomb}; $g_{\bm{k}}=t\sum_i
e^{\rmi\bm{k}\cdot \delta_i}$ is related to the nearest-neighbour hopping
(see (\ref{eq:bandstructure})), whereas the spin-orbit coupling enters via
$\gamma_{\bm{k}} =2\lso [ 2\cos (3k_x/2) \sin (\sqrt{3}k_y/2) -
\sin(\sqrt{3}k_y)]$. From (\ref{eq:kspace}) it is quite obvious that the KM
model can be regarded as two decoupled models for the $\UP$ and $\DO$ spins,
each equivalent to the spinless Haldane model~(\ref{eq:haldane}), and
described by the $2\times2$ matrices $H^\sigma(\bm{k})$. The two spin sectors
become coupled in the presence of Rashba terms, see~(\ref{eq:KM}).

Because time-reversal symmetry is inherently linked to the quantum spin Hall
insulator, it is worth considering the corresponding transformation of
(\ref{eq:kspace}). For spin-$1 / 2$ fermions, the time-reversal operator can
be written as $\hat{\Theta}=\exp(\rmi\pi \sigma^{y}) \hat{K}$
\cite{FuKa07,HaKa10}, where $\sigma^{y}$ is a Pauli matrix and $\hat{K}$
denotes complex conjugation. Application of $\hat{\Theta}$ to a
single-particle Bloch state leads to $\bm{k}\mapsto-\bm{k}$ and flipping of
the spin. Interchanging the $\UP$ and $\DO$ sectors of (\ref{eq:kspace}),
taking the complex conjugate, and using $\gamma_{-\bm{k}}=-\gamma_{\bm{k}}$
and $g_{-\bm{k}}=g^*_{\bm{k}}$, one can verify that the KM Hamiltonian indeed
has the symmetry $H(-\bm{k}) = \hat{\Theta} H(\bm{k}) \hat{\Theta}^{-1}$.  In
contrast, this is not the case for the spinless Haldane model
\cite{Haldane98} (equation~(\ref{eq:haldane})).

The eigenvalues of (\ref{eq:kspace}) are
\begin{equation}\label{eq:kmbands}
E^\pm(\bm{k})
=
\pm
\sqrt{|g_{\bm{k}}|^2+\gamma_{\bm{k}}^2}\,,
\end{equation}
so that the spectrum has two bands, each of which has a Kramers degeneracy
between ${\sigma=\UP}$ and ${\sigma=\DO}$. For $\lso=0$, the band structure
reduces to (\ref{eq:bandstructure}), and the familiar gapless Dirac spectrum
is recovered at the points $\mathrm{K}$, $\mathrm{K}'$. Equation~(\ref{eq:kmbands}) shows that
any nonzero spin-orbit coupling $\lso$ opens a gap
$\Delta_\text{SO}={3}\sqrt{3}\lso$ at the Dirac points \cite{KaMe05b,RaHu10}.
For $\lso/t>1/(3\sqrt{3})\sim0.2$, the minimal gap of size $\Delta=2t$ is
instead found at the $\mathrm{M}$ points \cite{RaHu10}.

In the Haldane model, the spin-orbit term gives rise to an integer Chern
invariant, a chiral edge mode, and a Hall conductivity $\sigma_{xy}=\pm
e^2/h$ \cite{Haldane98}. The key observation of Kane and Mele
\cite{KaMe05a,KaMe05b} was that combining two copies of the
Haldane model, with Hall conductivities $\sigma^{\UP}_{xy}=-\sigma^\DO_{xy}$
(the opposite sign is apparent from the diagonal matrix elements in
(\ref{eq:kspace})), leads to a model of spinful fermions that preserves
time-reversal symmetry. Remarkably, this construction---dubbed the {\it
  Haldane-Kane-Mele correspondence} in
\cite{Fi.Ch.Hu.Ka.Lu.Ru.Zy.11}---extends to interacting systems with and
without intrinsic spin-orbit coupling, and even to fractional states
\cite{PhysRevLett.103.196803}. Whereas the Hall conductivity
$\sigma_{xy}=\sigma^{\UP}_{xy}+\sigma^{\DO}_{xy}=0$, as required by the fact
that $\sigma_{xy}$ is odd under time reversal, the spin Hall conductivity
$\sigma^\text{s}_{xy}= (\hbar/2e) (\sigma^{\UP}_{xy}-\sigma^{\DO}_{xy}$),
related to the spin current
$\bm{J}_\text{s}=(\hbar/2e)(\bm{J}_\UP-\bm{J}_\DO)$, takes on nonzero values
$\sigma^\text{s}_{xy}=\pm e/2\pi$ \cite{KaMe05b}. The sign of
$\sigma^\text{s}_{xy}$ depends on the
sign of the spin-orbit coupling $\lso$ \cite{PhysRevLett.97.036808}.  The
quantization of $\sigma^\text{s}_{xy}$ holds as long as spin is conserved,
but a nonzero spin Hall conductivity exists under more general conditions
\cite{KaMe05b}.

Remarkably, the quantum spin Hall phase of the KM model, which does not break
any symmetries, represents a new state of matter that is topologically
distinct from a simple band insulator \cite{KaMe05a,KaMe05b}. This
distinction and the topological character of the quantum spin Hall state are
a consequence of time-reversal symmetry, which gives rise to a
symmetry-protected topological state \cite{PhysRevB.84.235141}, see
section~\ref{sec:intro}.

In close analogy to the integer quantum Hall effect
\cite{PhysRevLett.49.405,PhysRevB.31.3372}, a topological invariant can be
defined. An explicit mathematical definition is most easily achieved when
spin is conserved and the quantum spin Hall state decouples into
two integer quantum Hall states, as described by (\ref{eq:kspace}).  For a
noninteracting, integer quantum Hall system, the Chern invariant for the
$m$-th band is given by
\cite{PhysRevLett.49.405,PhysRevLett.51.51,Kohmoto1985343,RevModPhys.82.1959}
\begin{equation}\label{eq:chern:1}
  C_m = \frac{1}{2\pi} \int \rmd^2 \bm{k}\, \mathcal{F}_m(\bm{k})\,,
\end{equation}
where $\mathcal{F}_m(\bm{k})=\nabla_{\bm{k}} \times \mathcal{A}_m(\bm{k})$ is
the so-called {\it Berry flux} or {\it Berry curvature} (a gauge-invariant,
local quantity), and $\mathcal{A}_m=\rmi\bra{u_m}\nabla_{\bm{k}}\ket{u_m}$ is
the {\it Berry phase} or {\it Berry connection} (whose line-integral around a
closed loop in momentum space gives the phase picked up by the Bloch state
$\ket{u_m}$) \cite{RevModPhys.82.1959}. A particularly simple representation
can be obtained for a $2\times 2$ Hamiltonian, such as $H^\sigma(\bm{k})$ in
(\ref{eq:kspace}). Writing the Hamiltonian as
$H^\sigma(\bm{k})=\bm{h}^\sigma(\bm{k})\cdot\bm{\sigma}$, and defining the
unit vector field $\hat{\bm{h}}^\sigma(\bm{k}) =
\bm{h}^\sigma(\bm{k})/|\bm{h}^\sigma(\bm{k})|$, the Chern number reads
\cite{PhysRevA.64.052101,Volovik}
\begin{equation}\label{eq:chern:2}
  C^\sigma = \frac{1}{4\pi} \int \rmd^2 \bm{k} \,
  \left[ \partial_{k_x} \hat{\bm{h}}^\sigma(\bm{k}) \times \partial_{k_y}
    \hat{\bm{h}}^\sigma(\bm{k}) \right] \cdot \hat{\bm{h}}^\sigma(\bm{k})\,.
\end{equation}
In this case, the topological invariant corresponds to the winding number of
the mapping $\hat{\bm{h}}^\sigma(\bm{k})$ between the Brillouin zone and the
unit sphere; mathematically, nonzero values of $C^\sigma$ require three
nonzero components of the vector $\hat{\bm{h}}^\sigma(\bm{k})$, which
physically corresponds to broken time-reversal symmetry \cite{Volovik}.

For a quantum spin Hall insulator with time-reversal symmetry, as described
by (\ref{eq:kspace}) for $\lso\neq0$, the Chern indices have opposite sign,
so that $C^\UP+C^\DO=0$. However, the {\it spin Chern number}
\cite{PhysRevLett.91.116802,PhysRevLett.97.036808}
$C^\text{s}=(C^\UP-C^\DO)/2=\pm1$ is nonzero, and the $Z_2$ topological
invariant defined by \cite{HaKa10}
\begin{equation}\label{eq:Z2}
  \nu = C^\text{s}\, \mathrm{mod}\, 2
\end{equation}
takes on the values $0,1$. If spin is conserved ($\lr=0$), $\nu$ is directly
related to the quantized spin Hall conductivity, $\sigma^\text{s}_{xy}=\pm
\nu e/2\pi$. Hence $\nu=1$ corresponds to the topologically nontrivial state
that exhibits the quantum spin Hall effect. When spin is not conserved, it is
no longer possible to calculate $\nu$ from the Chern indices $C^\UP$ and
$C^\DO$, and $\sigma^\text{s}_{xy}$ is no longer quantized to $\pm e /2\pi$.
Nevertheless, a quantum spin Hall phase with quantized $\nu=1$ persists
\cite{KaMe05a,PhysRevLett.97.036808}.  The calculation of the topological
invariant of systems without spin conservation, and with interactions and/or
disorder will be the topic of section~\ref{sec:index}.

The existence of a topological state in the KM model relies on the special
form of the spin-orbit term. The topological band structure arises from the
opposite sign of the spin-orbit gap at $\mathrm{K}$ and $\mathrm{K}'$ in the
presence of inversion symmetry. Explicitly,
$\gamma_{\mathrm{K}'}=-\gamma_{\mathrm{K}}$, as can be verified using the
definition of $\gamma$ given after (\ref{eq:kspace}) and the definition of
the Dirac points in figure~\ref{fig:honeycomb}. As a counter example, a gap
in the spectrum can also be opened by adding a staggered sublattice
potential, $\lambda_v\sum_i \xi_i \hat{c}^\dag_i \hat{c}^\nag_i$, with
$\xi_i=\pm1$ depending on the sublattice. Such a term, which preserves
time-reversal symmetry but breaks inversion symmetry
\cite{PhysRevLett.53.2449,Haldane98}, has the same sign at $\mathrm{K}$ and
$\mathrm{K}'$, and leads to a topologically trivial insulator
with $\nu=0$ \cite{Haldane98,KaMe05a,KaMe05b}. It is expected to be
generically present in silicene as a result of buckling
\cite{PhysRevLett.107.076802}. The topological phase can be destroyed when
the gap is closed by competing terms such as Rashba spin-orbit
coupling \cite{KaMe05a,PhysRevLett.97.036808}.

Similar to the quantum Hall effect, the fact that the quantum spin Hall phase
is topologically distinct from a simple band insulator has important
consequences. As implied by the term topological, such as state cannot be
connected to a nontopological state via adiabatic transformations of the
Hamiltonian. Instead, a transition from a quantum spin Hall state to a
topologically trivial state can occur either via a closing of the bulk band
gap, or via the breaking of time-reversal symmetry \cite{PhysRevB.84.235141}.

Because a closing of the gap is required in a system with time-reversal
symmetry, an interface between a quantum spin
Hall insulator and a trivial insulator such as the vacuum necessarily gives
rise to metallic states \cite{HaKa10}. The existence of these states may also
be inferred from a flux insertion argument \cite{PhysRevB.74.195312}. Whereas
all electrons at a given edge of a quantum Hall sample move in the same
direction ({\it chiral edge state}) \cite{PhysRevB.41.12838}, quantum spin
Hall insulators have {\it helical edge states} \cite{Wu06}: at each edge,
electrons with spin $\UP$ move in one direction and electrons with spin $\DO$
move in the opposite direction. Hence, there is a one-to-one correspondence
between spin and direction of motion in helical edge states
\cite{KaMe05b,Wu06}. The existence of pairs of edge states with opposite
chirality follows directly from equation~(\ref{eq:kspace}), which combines
two independent integer quantum Hall states that possess chiral edge modes of
opposite velocity.

For noninteracting systems, the number of pairs of edge states (modulo 2) is
directly linked to the value of the $Z_2$ topological invariant $\nu$
\cite{HaKa10,KaMe05a}. Moreover, the so-called {\it bulk-boundary
  correspondence} expresses the fact that the existence of edge states is
guaranteed by the topological nature of the bulk system.  This correspondence
goes back to the work of Halperin \cite{PhysRevB.25.2185}. It also follows
from topological field theory when imposing gauge invariance \cite{Wen92},
see below.  Intuitively, it can be understood as a consequence of the
protection of the topological state with respect to adiabatic deformations,
implying that the only way to achieve a change of the topological invariant
is by closing the bulk band gap. Such a closing, which can either occur as a
function of a parameter (phase transition), or at the interface with another
material (including the vacuum), is tantamount to the existence of gapless
excitations.  Hence, edge properties can in principle be studied to infer the
properties of the bulk, but as discussed in sections~\ref{sec:edge}
and~\ref{sec:index}, this equivalence can break down in interacting systems.
For the latter, important progress has been made using the concept of
symmetry-protected topological states
\cite{PhysRevB.82.155138,arXiv:1106.4772,PhysRevB.84.235141,Chen2012,arXiv:1209.4399,arXiv:1212.1726}.

\begin{figure}[t]
  \centering
  \includegraphics[width=0.45\textwidth]{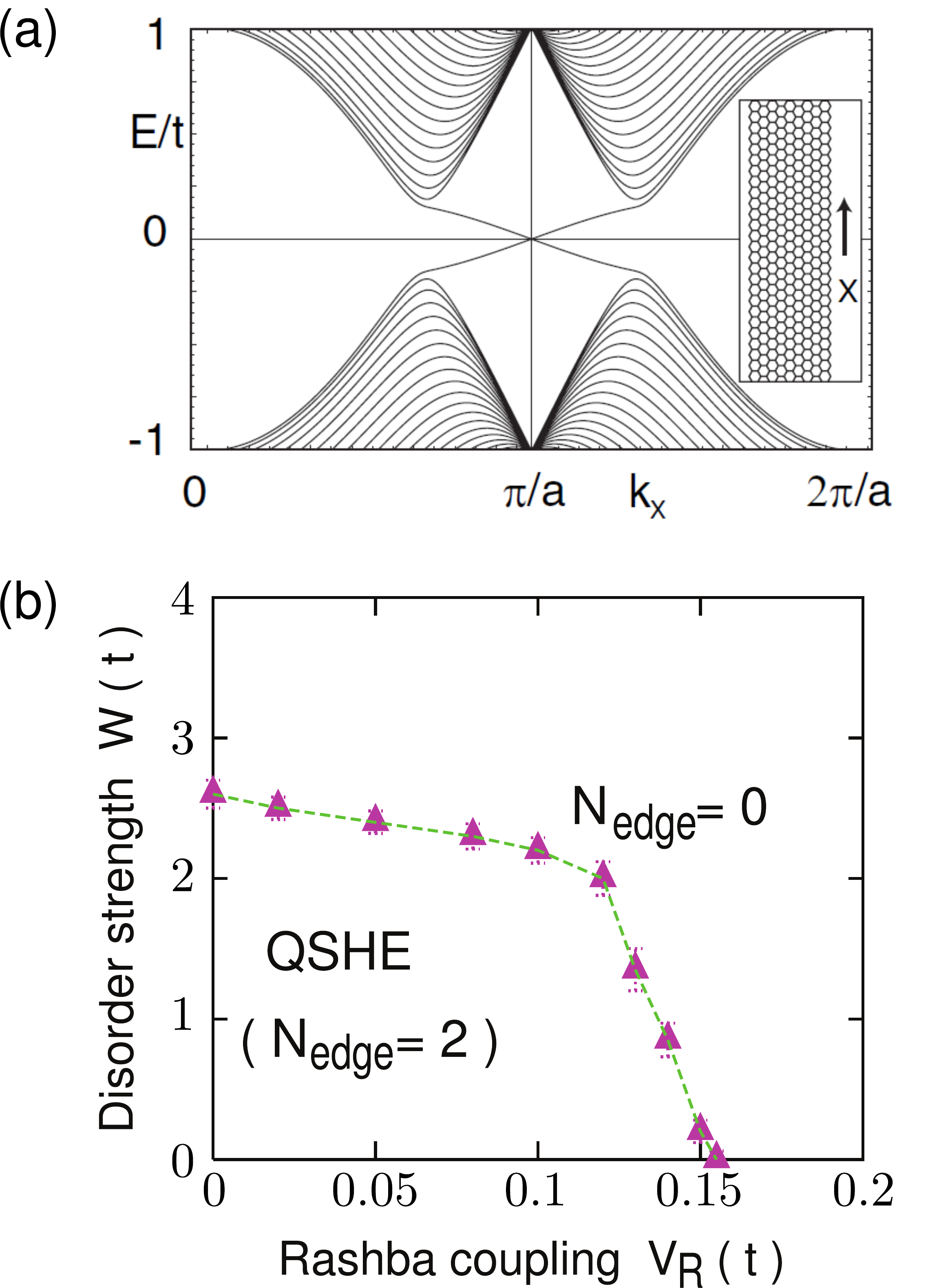}
  \caption{\label{fig:kmedgestates}
    (a) Edge states of the KM model on a zigzag ribbon
    with open boundary conditions in the $y$ direction (see inset) and
    $\lso/t=0.03$ \cite{KaMe05b}. 
    (Reprinted with permission from \cite{KaMe05b}. Copyright 2005
    by the American Physical Society).
    (b) Phase diagram of the KM model as a function of  Rashba coupling
    $V_\text{R}$ and disorder strength $W$, at fixed spin-orbit
    coupling $V_\text{SO}/t=0.05$ \cite{PhysRevLett.97.036808}. 
    (Adapted with permission from \cite{PhysRevLett.97.036808}. Copyright 2006
    by the American Physical Society).
}
\end{figure}

As long as time-reversal symmetry is preserved, the quantum spin Hall state
cannot be destroyed by small perturbations such as disorder or
electron-electron interactions \cite{KaMe05a,KaMe05b,Wu06,Cenke06}. Exploiting the bulk-boundary
correspondence, this stability can be understood by considering the helical
edge states. Denoting the band dispersion of the edge states by
$\epsilon_\sigma(k)$ (where $k$ is the conserved component of the
two-dimensional crystal momentum along the edge) time-reversal symmetry leads
to the condition
\begin{equation}
  \epsilon_\DO(-k)=\epsilon_\UP(k)\,.
\end{equation}
At the special, time-reversal invariant momenta $k=0,\pi$ (these
points are mapped onto themselves under the action of $\hat{\Theta}$), this
condition implies 
\begin{equation}\label{eq:helical-tri-pi}
  \epsilon_\UP(k)=\epsilon_\DO(k)\,,
  \quad k=0,\pi\,,
\end{equation}
and hence a symmetry-protected, gapless crossing point of edge states.  It
turns out that in the topological phase of the KM model, there exists a
single pair of edge states that cross either at $k=0$ for an armchair edge,
or at $k=\pi$ for a zigzag edge \cite{KaMe05a,KaMe05b}.  The latter case is
illustrated in figure~\ref{fig:kmedgestates}(a) for a ribbon with open
boundaries in the $y$ direction, whereas the armchair case can be seen in
figure~\ref{fig:edgestates-armchair}.  The existence of only one crossing
illustrates the fact that these states are edge states, and hence do not cover
the entire Brillouin zone. The edge states in
figure~\ref{fig:kmedgestates}(a) connect the Dirac points $\mathrm{K}$ and
$\mathrm{K}'$, and their helicity becomes apparent by plotting the
spin-resolved single-particle spectrum.

Because time-reversal symmetry implies degenerate edge states at either
at $k=0$ or $k=\pi$ \cite{KaMe05b}, perturbations that preserve
time-reversal symmetry cannot open a gap. Most importantly, single-particle
backscattering, which is the most relevant perturbation in ordinary
one-dimensional metals \cite{PhysRevLett.68.1220}, is not allowed in a
helical liquid because the single-particle states $\ket{k,\UP}$ and
$\ket{-k,\DO}$ are Kramers partners, and therefore orthogonal. Consequently,
helical edge states are symmetry-protected from elastic, single-particle
backscattering \cite{KaMe05b,Wu06,Cenke06,RevModPhys.83.1057}, but the
protection is weaker than that of chiral edge states in a quantum Hall
insulator where all electrons at a given edge propagate in the same
direction, and there is no phase space for backscattering. The presence of
counter-propagating electrons at a given edge has important implications for
strongly interacting systems, see section~\ref{sec:edge}.  Finally,
single-particle  backscattering is not forbidden by symmetry if an even
number of edge states exists for each spin direction. This case corresponds to a
trivial phase ($\nu=0$) \cite{Wu06,Cenke06}.

The stability of the edge states with respect to disorder and Rashba
spin-orbit coupling has been numerically verified
\cite{KaMe05a,PhysRevLett.97.036808,PhysRevLett.98.076802}, and the resulting
phase diagram is reproduced in figure~\ref{fig:kmedgestates}(b).  The
topological phase exists as long as the Rashba coupling does not exceed
the value $\lr=2\sqrt{3}\lso$ \cite{KaMe05b} (note that the coupling
constant $\lso$ used here and $V_\text{SO}$ used in
\cite{PhysRevLett.97.036808} differ by a prefactor), and up to a critical
value of the disorder strength. This result is crucial for the experimental
realization of the quantum spin Hall effect \cite{Koenig07,Roth09} because
neither Rashba coupling nor disorder can be completely excluded. In the
presence of interactions, the stability of the edge states is no longer
guaranteed, as discussed in section~\ref{sec:edge}.

As shown in this section, many aspects of integer quantum Hall and spin Hall
insulators can be understood in the framework of topological band theory
\cite{RevModPhys.83.1057}. However, the latter is inherently restricted to
noninteracting systems. An alternative approach to topological states of
matter is based on Chern-Simons field theory, a low-energy theory of gauge
fields that is applicable to interacting and even fractional states.  A
comprehensive introduction to Chern-Simons theory can be found, for example,
in \cite{RevModPhys.80.1083}. For the extension to topological insulators
with time-reversal symmetry see \cite{QiHuZh08}.

To illustrate how Abelian Chern-Simons theory explains the quantization of
$\sigma_{xy}$ as well as the presence of edge states in a quantum Hall
insulator, it is sufficient to consider the noninteracting Haldane model in
the form $H=\sum_{ij} t_{ij} c^\dag_i c^\nag_j$, cf.~(\ref{eq:haldane}), 
minimally coupled to an electromagnetic field,
\begin{eqnarray}
  H(A^\mu)  
  = 
  &\sum_{ij}   t_{ij} c^{\dagger}_{i} c^{}_{j} 
  \exp\left[  \frac{2 \pi \rmi}{\Phi_0}  \int_{i}^{j} \mathbf{A}(\mathbf{l},t) \cdot \rmd \mathbf{l} \right]
  \\\nonumber
  &+   
  e c \sum_{i}  A_{0}(i,t)  c^{\dagger}_{i} c^{}_{i} \,.
\end{eqnarray}
The scalar and vector potentials may be combined in the form $A^{\mu} \equiv
( A_0, -\mathbf{A})$ with $A_0=\Phi/c$ and $\mathbf{A}=(A_1,A_2)$. Gauge
invariance implies $\partial_{\mu} j^{\mu} = 0 $ with $j^{\mu} = - \delta H /
\delta A_{\mu}$.  Linear response theory then directly gives the quantized
Hall conductivity.

The starting point for the derivation of Chern-Simons theory is the Grassmann
path integral \cite{Wen_book}
\begin{equation}
  Z (A_{\mu})   =     \int  {   \prod_{i} \rmd c^{\dagger}_{i} \rmd c^\nag_{i} } \rme^{ \rmi S (A_{\mu})}
\end{equation}
with the action
\begin{eqnarray}\label{eq:action1}
  S(A_{\mu}) 
  &= 
  \int \rmd t \left\{ \sum_{i}    c^{\dagger}_{i} (t) 
  \left[ \delta_{ij} \rmi \partial_t - t_{ij}\right] c_{j}(t) \right.
  \\\nonumber
  &\qquad\quad+
  \left.
  \sum_{i} j^{\mu}(i,t) A_{\mu}(i,t) \right\}\,. 
\end{eqnarray}
Equation~(\ref{eq:action1}) follows from a gradient expansion in $A^\mu$.
The fermions can be integrated out, at the expense of a Gaussian integral, to
obtain $Z (A_{\mu})  =  \rme^{\rmi S_\text{eff}(A_\mu) }$  with
\begin{eqnarray}
  S_\text{eff}(A_\mu) 
  = \int\int \rmd t \rmd t'  \sum_{ij}   &A_{\mu}(i,t) \\\nonumber
  &\times P^{\mu\nu}(i,t,j,t') A_{\nu}(j,t')\,. 
\end{eqnarray}
Here $P^{\mu\nu}$ is the current-current correlation function, which is
short-ranged in an insulating state, and a summation over repeated Greek
indices is implied. Hence, to a first approximation, retardation effects can
be omitted by setting $t'=t$. Taking the continuum limit, all possible gauge
invariant terms can be derived. The relevant contribution for topological
states is the Chern-Simons term
\begin{equation}\label{eq:CS}
S^\text{CS}_\text{eff}(A_\mu) = C\frac{e^2 }{4 \pi}  \epsilon^{\mu \nu \rho} \int
\rmd^2 x \int \rmd t   A_{\mu} \partial_\nu A_\rho 
\end{equation}
from which all the low-energy properties of the integer quantum Hall state
follow. The invariance of the theory with respect to twists of the boundary
conditions by a multiple of the flux quantum $\Phi_0$ pins the first Chern
number $C$ to an integer \cite{PhysRevLett.49.405,PhysRevLett.51.51}.  Using
$j_{\mu} = \delta S^\text{CS}_\text{eff} / \delta A_{\mu}$, and assuming
$A_0=0$, gives the familiar quantized Hall response
\begin{equation}
j_{x}   = \sigma_{xy} E_y\,,\quad \sigma_{xy}=C\frac{e^{2}}{ h }\,.
\end{equation} 

On a torus geometry, the Chern-Simons action~(\ref{eq:CS}) is invariant under
the gauge transformation $A_{\mu} \rightarrow A_{\mu} + \partial_{\mu}
\chi$. However, in the presence of a boundary (\eg, on a cylinder), the gauge
transformation leads to an additional boundary term that is directly related
to the chiral edge state of a quantum Hall insulator
\cite{RevModPhys.80.1083}.

Finally, a Chern-Simons theory of a noninteracting, spin-conserving quantum
spin Hall insulator (as appropriate for the KM model) takes the form
\begin{equation}\label{eq:CSQSH}
S^\text{CS}_\text{QSH} = 
\frac{e^2}{4 \pi}  \epsilon^{\mu \nu \rho} \int
\rmd^2 x \int \rmd t   \left( A^\UP_{\mu} \partial_\nu A^\UP_\rho - A^\DO_{\mu} \partial_\nu A^\DO_\rho\right)\,.
\end{equation}
For more general cases, including fractional topological states, see
\cite{PhysRevB.84.165107} and references therein.

\section{Bulk correlation effects}\label{sec:bulk}

The integer quantum spin Hall effect can be understood purely in terms of 
the band structure. Much of the recent theoretical work has focused on the
impact of electronic correlations. Given the scope of this review, the
discussion of bulk correlation effects begins with results for
models with intrinsic spin-orbit coupling that support an integer quantum
spin Hall state in the noninteracting limit. The investigation of such
systems is particularly rewarding as the corresponding models are amenable to
 numerical methods, including exact quantum Monte Carlo simulations that
permit to investigate, for example, the critical behaviour at quantum phase
transitions. After an overview of phenomena using the example of the
intensely studied KMH model, a detailed discussion of correlated quantum spin
Hall phases and interaction-driven phase transitions with or without symmetry
breaking is given, followed by two shorter sections on interaction-driven
topological insulators, and fractional topological insulators.

\subsection{Phases of the Kane-Mele-Hubbard model}\label{sec:phasesKMH}

The KMH model provides a framework to study the effects of electronic
correlations on a quantum spin Hall insulator. It is of particular interest
because of connections with graphene \cite{Neto_rev} and the Hubbard model on the
honeycomb lattice \cite{Meng10}. Given its historical importance and the
possibility of applying exact quantum Monte Carlo methods, the KMH model is
arguably the best understood model among those introduced in
section~\ref{sec:models}.

\begin{figure}
  \centering
  \includegraphics[width=0.45\textwidth]{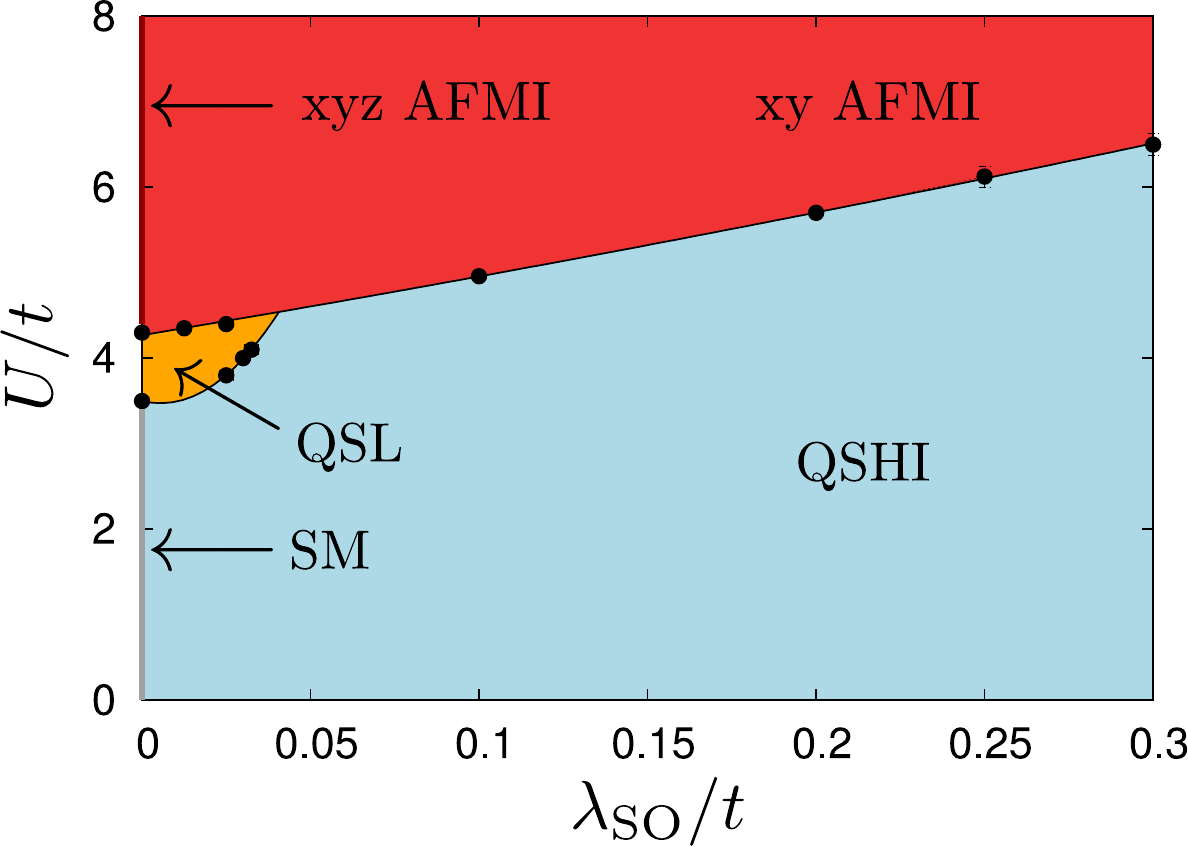}
  \caption{\label{fig:phasediagram}
  Phase diagram of the half-filled KMH model
  from quantum Monte Carlo simulations
  \cite{Hohenadler10,Ho.Me.La.We.Mu.As.12,As.Be.Ho.2012}. The phases 
  correspond to a semimetal (SM, for $\lso=0$), a quantum spin Hall insulator (QSHI), a
  quantum spin liquid (QSL), and an antiferromagnetic Mott insulator (AFMI) with
  either Heisenberg (for $\lso=0$) or easy plane (for $\lso\neq0$)
  order. Here $\lr=0$. Data taken from \cite{Hohenadler10,Ho.Me.La.We.Mu.As.12,As.Be.Ho.2012}.
  }
\end{figure}

Figure~\ref{fig:phasediagram} shows the numerically obtained
\cite{Hohenadler10,Ho.Me.La.We.Mu.As.12,As.Be.Ho.2012} phase diagram of the
KMH model at half filling, and with conserved spin ({$\lr=0$}).  Most
importantly, the numerical results confirm the stability of the topological
state of the KM model ({$U=0$}) up to Hubbard interactions $U/t\gtrsim4$, as
first predicted based on mean-field theory \cite{RaHu10}. As discussed in
more detail below, the numerical results suggest that the topological phase
at $U>0$ is adiabatically connected to the noninteracting case. Even stronger
electron-electron repulsion drives a quantum phase transition to a magnetic
state discussed in detail in section~\ref{sec:trans}. The other phases of the
KMH model shown in figure~\ref{fig:phasediagram} are a semimetal (SM) that
exists for vanishing spin-orbit coupling (with a massless Dirac spectrum
similar to (\ref{eq:bandstructure})), and possibly a gapped quantum spin
liquid (QSL) phase at intermediate $U/t$ \cite{Meng10} to be discussed in
section~\ref{sec:trans}.  Contrary to analytical predictions \cite{Gr.Xu.11},
numerical studies so far indicate that all phase transitions in
figure~\ref{fig:phasediagram} are continuous \cite{Ho.Me.La.We.Mu.As.12}.

The phase diagram of the KMH model has been investigated by a variety of
theoretical methods. Mean-field theory captures the magnetic transition, and
the increase of the critical $U$ with increasing $\lso$ \cite{RaHu10}.
Quantum Monte Carlo simulations provide the exact phase boundary for the
magnetic phase, demonstrate the absence of magnetic order in the $z$
direction, and suggest the existence of a quantum spin liquid phase
\cite{Hohenadler10,Zh.Wu.Zh.11,Ho.Me.La.We.Mu.As.12}. (Figure~\ref{fig:edgephasediagram}
shows the numerical phase diagram obtained by Zheng \etal
\cite{Zh.Wu.Zh.11}.) Approximate variational Monte Carlo results are also
available at $\lso/t=0.1$ \cite{PhysRevB.83.205122}.  Quantum cluster methods
such as the cluster dynamical mean-field theory \cite{Wu.Ra.Li.LH.11} and the
variational cluster approach \cite{Yu.Xie.Li.11,arXiv:1203.2928} are able to
capture the overall structure of the phase diagram. However, care has to be
taken regarding the choice of the cluster shape and size (a detailed
discussion of the impact of nonlocal correlations in the Hubbard model on
the honeycomb lattice can be found in \cite{PhysRevB.83.035113}). Analytical
approaches include slave-boson \cite{We.Ka.Va.Fi.11}, slave-spin
\cite{Ma.Va.Va.11} and slave-rotor methods \cite{PhysRevB.85.235449}, as well
as mappings to spin models to explore the regime of large $U/t$
\cite{PhysRevB.85.195126,Re.Th.Ra.12,Ka.La.Fi.12}.  Using field theory and
symmetry arguments, predictions about the phases and phase transitions of the
KMH model were made by Griset and Xu \cite{Gr.Xu.11} as well as by Lee
\cite{PhysRevLett.107.166806}.

With the exception of the quantum spin liquid phase, whose existence is still
under debate \cite{Meng10,Hohenadler10,Ho.Me.La.We.Mu.As.12,So.Ot.Yu.} and
expected to be specific to the honeycomb lattice, the phase diagram in
figure~\ref{fig:phasediagram} contains many features that are generic for
models of correlated quantum spin Hall insulators. In addition to the $Z_2$
topological phase, a semimetal also exists in
the SI model (\ref{eq:SI}) with $t_1=t_2=0$ (where it reduces to
the Hubbard model on the honeycomb lattice), and in the BHZH model at the
metallic point $\epsilon=0$. A quantum phase
transition to a magnetic state with easy-plane order (as a result of
spin-orbit coupling) at large $U/t$ has also been observed in the BHZH model
\cite{PhysRevB.85.235449,arXiv:1111.6250}, the SI model
\cite{PhysRevLett.108.046401}, and a spinful version of the Hofstadter
butterfly problem with a Hubbard interaction \cite{Cocks.12}.  The existence of a
magnetic phase is a direct consequence of the nonfrustrated exchange
interactions that emerge on bipartite lattices. Similarly, a
correlation-driven transition from a topological to a magnetic state is also
expected in three dimensions \cite{PeBa10}.

\subsection{Correlated topological band insulators}\label{sec:corrTBI}

\begin{figure}
  \centering
  \includegraphics[width=0.45\textwidth]{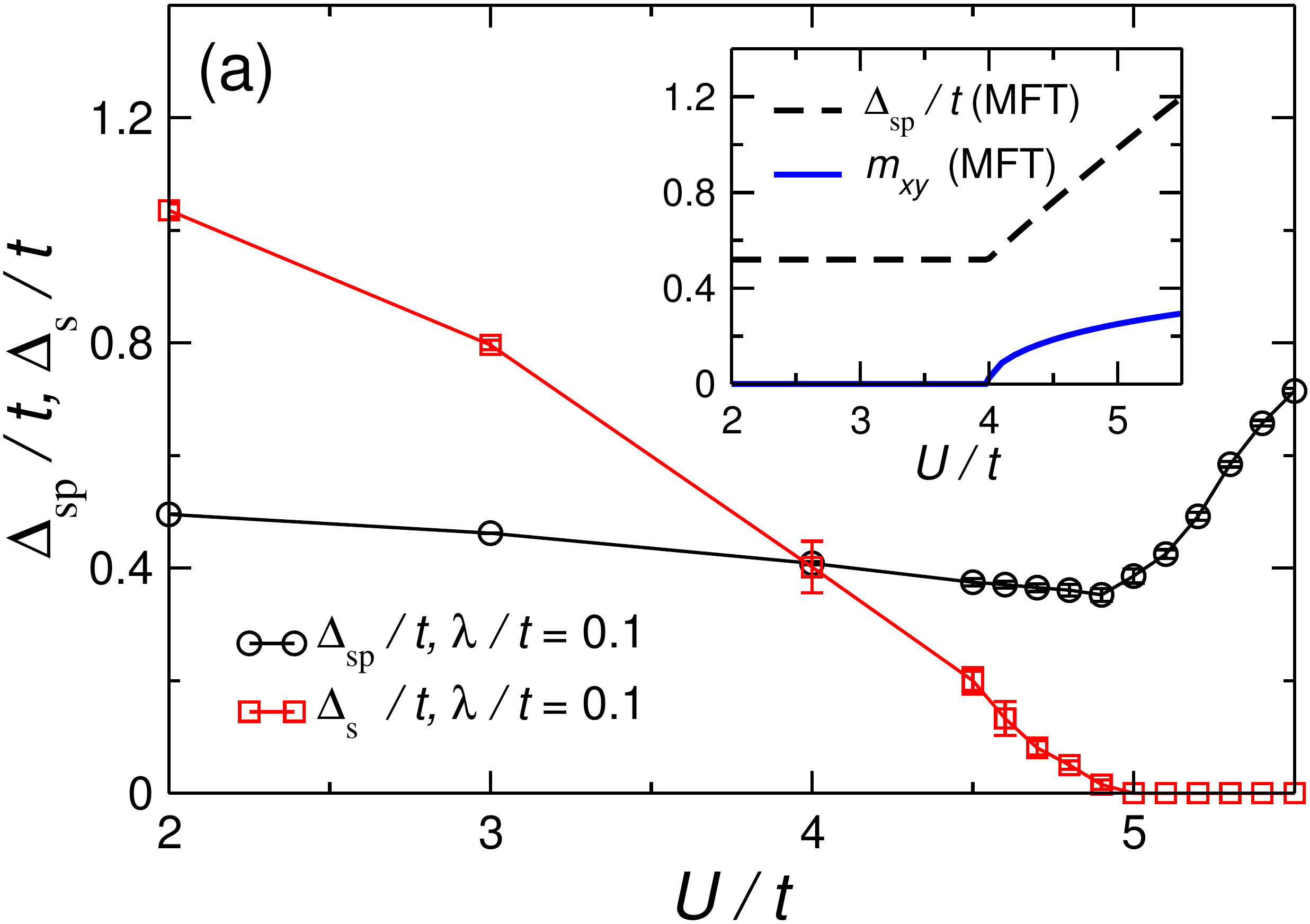}
  \caption{\label{fig:tbi-afmi-gaps}
    Single-particle gap $\Delta_\text{sp}$ and spin gap $\Delta_\text{s}$ 
    of the KMH model as a function of $U$ at $\lambda\equiv\lso=0.1t$ \cite{Ho.Me.La.We.Mu.As.12}. 
    At the transition ($\Uc/t=4.95(5)$), $\Delta_\text{sp}$ shows a dip,
    whereas $\Delta_\text{s}$ closes. Inset: $\Delta_\text{sp}$ and
    magnetization from mean-field theory.
    (Reprinted with permission from \cite{Ho.Me.La.We.Mu.As.12}. Copyright 2012
      by the American Physical Society). 
}
\end{figure}

In section~\ref{sec:noninteracting}, the bulk-boundary correspondence was
invoked to argue that the quantum spin Hall phase should be stable as long as
time-reversal symmetry is not broken. This idea is borne out by the phase
diagram in figure~\ref{fig:phasediagram}, which shows that (with the possible
exception of the region around the spin liquid phase) the topological state
is eventually destroyed when time-reversal symmetry is spontaneously broken. 
However, throughout the quantum spin Hall phase, the bulk gap remains open
upon switching on the Hubbard interaction, see Fig.~\ref{fig:tbi-afmi-gaps}.
The topological character of this phase at $U>0$ also manifests itself
through its response to $\pi$ fluxes \cite{As.Be.Ho.2012}, see section~\ref{sec:index}.
These results firmly establish an adiabatic connection between the quantum
spin Hall states with $U=0$ and $U>0$. Such behaviour seems to be typical of
models that conserve spin \cite{Re.Th.Ra.12}. On the other hand, an
interaction-driven transition to a fractional topological state that
preserves time-reversal symmetry has been reported for the SI model
\cite{PhysRevLett.108.046401} (see section~\ref{sec:trans}).

\begin{figure}[t]
  \centering
  \includegraphics[width=0.45\textwidth]{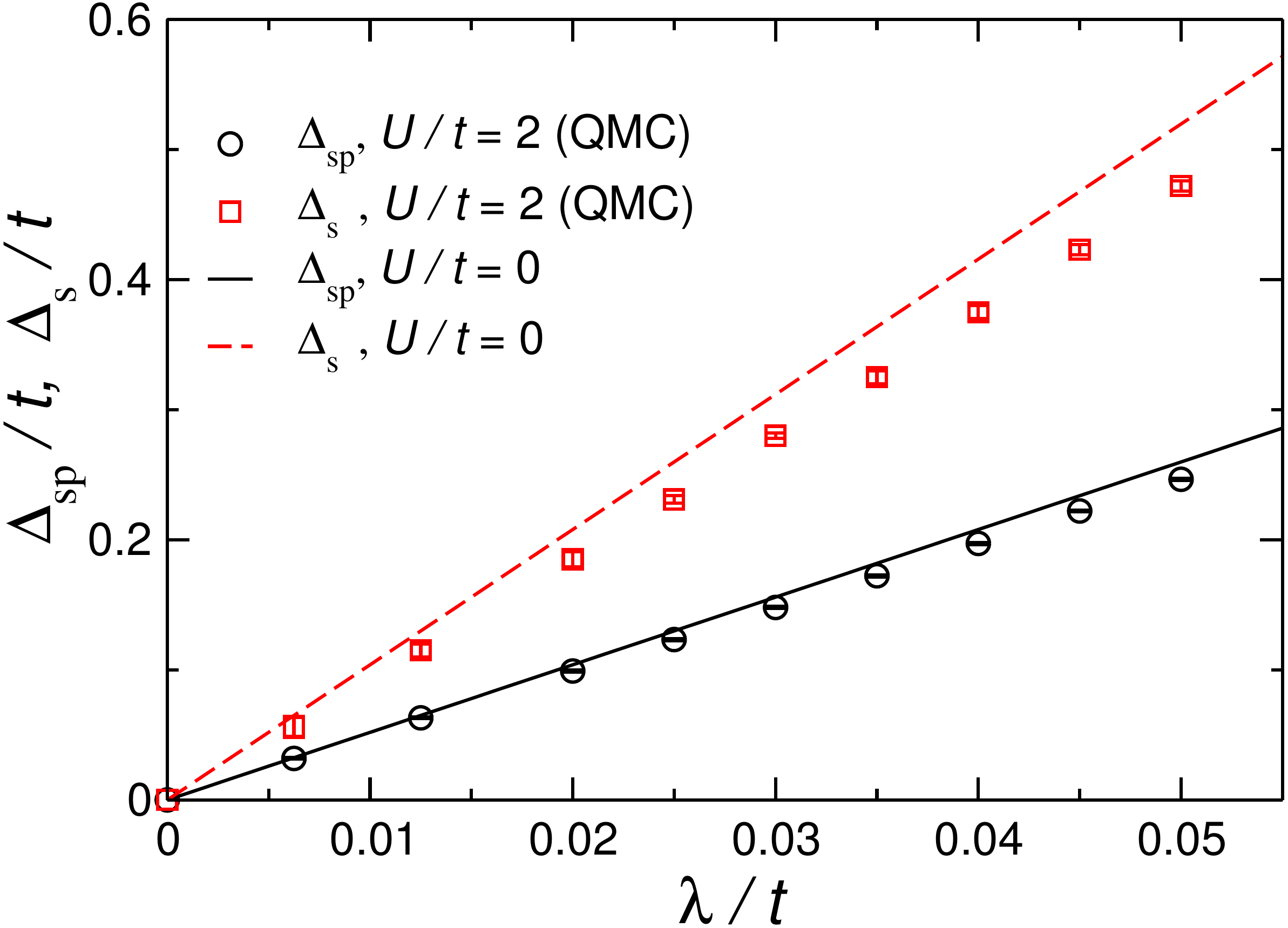}
  \caption{\label{fig:sm-tbi}
    Single-particle gap $\Delta_\text{sp}$ and spin gap
    $\Delta_\text{s}$ as a function of spin-orbit coupling $\lambda\equiv\lso$
    from quantum Monte Carlo simulations of the KMH model with $U/t=2$
    \cite{Ho.Me.La.We.Mu.As.12}. Also shown are the gaps
    of the noninteracting KM model (see text). 
    (Reprinted with permission from \cite{Ho.Me.La.We.Mu.As.12}. Copyright 2012
      by the American Physical Society).
  }
\end{figure}

Given a correlated quantum spin Hall insulator that exists up to rather large
values of $U/t$, an important question is to what extent this phase resembles
that of the noninteracting KM model (even if the phase is
adiabatically connected to the noninteracting state, there could be
substantial quantitative differences). A partial answer is provided by the
quantum Monte Carlo results for the single-particle gap $\Delta_\text{sp}$
and the spin gap $\Delta_\text{s}$ shown in figure~\ref{fig:sm-tbi}.
Starting in the correlated semimetal phase at $U/t=2$, any nonzero spin-orbit
coupling $\lso$ opens both gaps simultaneously. Importantly, for small
$\lso$, both gaps follow very closely the corresponding results for the
{$U=0$} KM model (indicated by the lines in figure~\ref{fig:sm-tbi}), namely
$\Delta_\text{sp}=\Delta_\text{SO}=3\sqrt{3}\lambda$ and
$\Delta_\text{s}=2\Delta_\text{sp}$. This observation suggests that the
quantum spin Hall states with $U/t=2$ and $U/t=0$ are very similar.  Minor
correlation effects are apparent from the fact that the gaps for $U/t=2$ are
systematically smaller than those for $U/t=0$, and that
$\Delta_\text{s}<2\Delta_\text{sp}$.

\begin{figure}[t]
  \centering
  \includegraphics[width=0.4\textwidth]{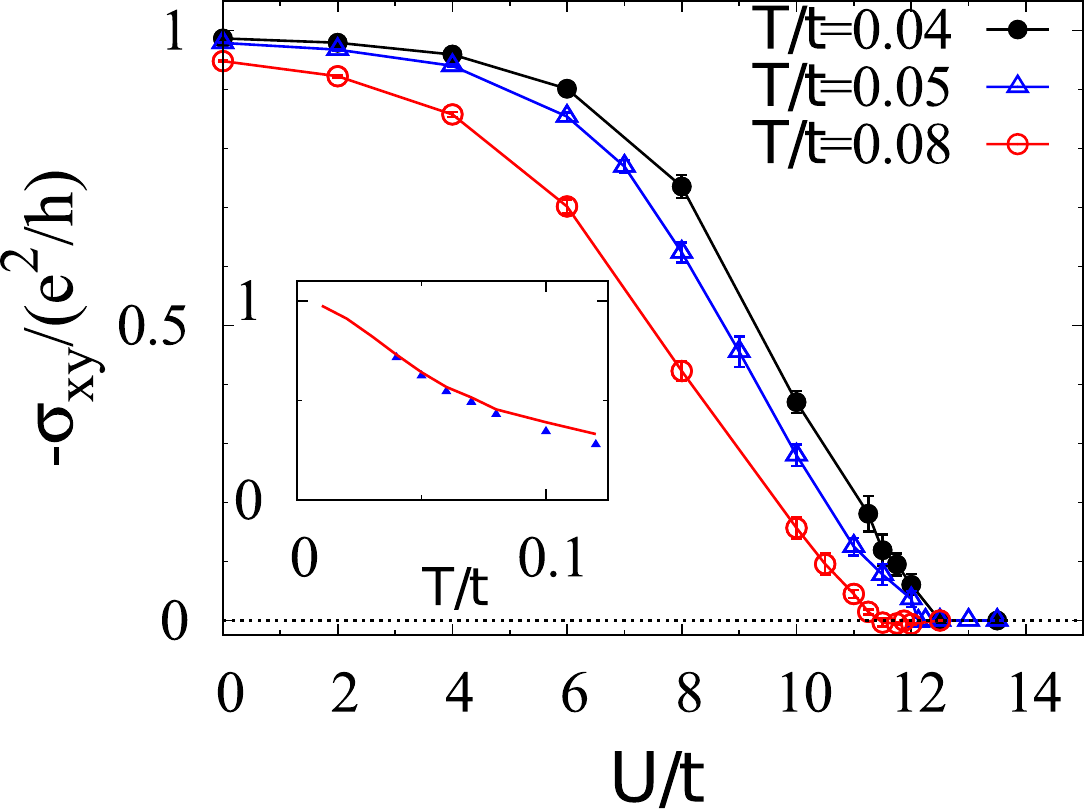}
  \caption{\label{fig:sigma_bhz}
    Spin Hall conductivity as a function of interaction strength and
    for different temperatures, as obtained from dynamical mean-field
    calculations for the BHZH model with intra-orbital repulsion $U$ \cite{arXiv:1111.6250}.
    (Reprinted with permission from \cite{arXiv:1111.6250}. Copyright 2012
    by the American Physical Society).}
\end{figure}

Yoshida \etal \cite{arXiv:1111.6250} have used dynamical mean-field theory to
calculate the spin Hall conductivity of the BHZH model as a function of
temperature and the intra-orbital repulsion $U$. The results, shown in
figure~\ref{fig:sigma_bhz}, reveal a clear suppression of $\sigma_{xy}^s$ due
to thermal fluctuations and electronic correlations, and that $\sigma_{xy}^s$
eventually vanishes at $U/t\sim12$.  However, as demonstrated in the inset,
for small enough $U/t$, the conductivity extrapolates to the quantized value
$e^2/h$ for $T\to0$. Consequently, although interaction effects are clearly
visible at finite temperatures, the correlated quantum spin Hall phase of the
BHZH model resembles the noninteracting state at $T=0$. As pointed out in
section~\ref{sec:noninteracting}, a nonquantized Hall conductivity is also
observed in the noninteracting case when spin is not conserved; however, the
topological invariant retains its integer value $\nu=1$
\cite{PhysRevLett.97.036808}. It would be interesting to confirm this
explicitly in an interacting model, see also
section~\ref{sec:index}. Temperature effects on the quantum spin Hall phase
of the KMH model have been discussed in \cite{Wu.Ra.Li.LH.11}. The
topological nature of the quantum spin Hall phase at $U>0$ has also been
demonstrated for the KMH model both at $T=0$ and $T>0$ with the help of $\pi$
fluxes \cite{As.Be.Ho.2012}. Finally, the topological invariant of the KMH
model has been calculated using the variational cluster approach
\cite{arXiv:1203.2928}.

The properties of the correlated quantum spin Hall phase have further been
explored by numerically calculating the single-particle spectrum in systems
with edges, see section~\ref{sec:edge}. Both for the KMH model
\cite{Yu.Xie.Li.11,Wu.Ra.Li.LH.11} and the BHZH model \cite{Wa.Da.Xi.12},
gapless helical edge states are observed even for substantial values of
$U/t$, although interaction effects modify the spectral weight, edge
transport and magnetic correlations (see section~\ref{sec:edge}).  In
summary, in the spin-conserving KMH model, the quantum spin Hall phase
remains qualitatively unaffected by correlations up to the point where it
breaks down, in accordance with the expected adiabatic connection to
the noninteracting spin Hall state. This finding has been exploited to
construct an effective model of the edge states in which a Hubbard
interaction is only taken into account at the edge, whereas the
noninteracting bulk establishes the topological phase \cite{Hohenadler10,Ho.As.11}.

\subsection{Interaction-driven phase transitions}\label{sec:trans}

Although the quantum spin Hall phase of, for example, the KMH model persists
even in the presence of rather strong interactions, it eventually breaks down
at the onset of long-range magnetic order that spontaneously breaks both the
$U(1)$ spin symmetry and time-reversal symmetry
\cite{RaHu10,Hohenadler10,Zh.Wu.Zh.11}. Such a magnetic transition is a
generic feature of many quantum spin Hall models. In the KMH
model, it can be studied in detail by quantum Monte Carlo simulations.
Whereas the magnetic transition involves a well-defined order parameter and
can be captured already in mean-field theory, interactions can also lead to
more complex phenomena in which quantum fluctuations play a key role. These
include topological Mott insulators \cite{RaQiHo08}, quantum spin Hall states
with Mott insulating edges \cite{PeBa10,RaHu10}, or quantum spin liquids
\cite{Hohenadler10}. It is even possible to have a coexistence of a
topological insulator with protected edge states and long-range magnetic
order \cite{PhysRevB.81.245209}.

\subsubsection{Symmetry-breaking transitions}
The best studied example of an interaction-driven quantum phase transition
that involves the breaking of time-reversal symmetry is the magnetic
transition observed in the KMH model. As visible from the phase diagram in
figure~\ref{fig:phasediagram}, the Hubbard repulsion induces a transition
from the quantum spin Hall phase to a Mott insulator with long-range magnetic
order at a critical value $U_\text{c}/t$. The existence of this magnetic
transition has first been demonstrated using mean-field theory \cite{RaHu10}.

Already for $U<U_\text{c}$, inside the correlated quantum spin Hall phase,
the Hubbard interaction gives rise to the formation of local moments and
substantial magnetic correlations.  These correlations can be measured by the
transverse spin structure factor
\cite{Hohenadler10,Ho.Me.La.We.Mu.As.12,Zh.Wu.Zh.11}
\begin{eqnarray}\label{eq:SAF}
 \Sxy
 &\equiv \sum_{\alpha} [\Sxy]^{\alpha\alpha}\,,
 \\\nonumber
 [\Sxy]^{\alpha\beta}
 &= \frac{1}{L^2}\sum_{\bm{r}\bm{r}'} (-1)^{\alpha} (-1)^{\beta} \\\nonumber
 &\qquad\times \langle \Psi_0
 | S^{+}_{\bm{r}\alpha} S^{-}_{\bm{r}'\beta} + S^{-}_{\bm{r}\alpha} S^{+}_{\bm{r'}\beta} | \Psi_0 \rangle\,,
\end{eqnarray}
and, similarly, by the longitudinal spin structure factor $\Szz \equiv \sum_\alpha [\Szz]^{\alpha\alpha}$ with
\begin{equation}\label{eq:SAFzz}
 [\Szz]^{\alpha\beta}
 = \frac{1}{L^2}\sum_{\bm{r}\bm{r}'} (-1)^{\alpha} (-1)^{\beta} 
 \langle \Psi_0
 | S^{z}_{\bm{r}\alpha} S^{z}_{\bm{r}'\beta} | \Psi_0 \rangle\,.
\end{equation}
Here $\bm{r},\bm{r}'$ are vectors that indicate unit cells of the honeycomb
lattice, $\alpha,\beta\in\{A,B\}$ are sublattice indices,  $(-1)^\alpha=1$ ($-1$) for
$\alpha=A$ ($B$), and the trace of the $2\times2$ matrices is taken. 

\begin{figure}[t]
  \centering
  \includegraphics[width=0.45\textwidth]{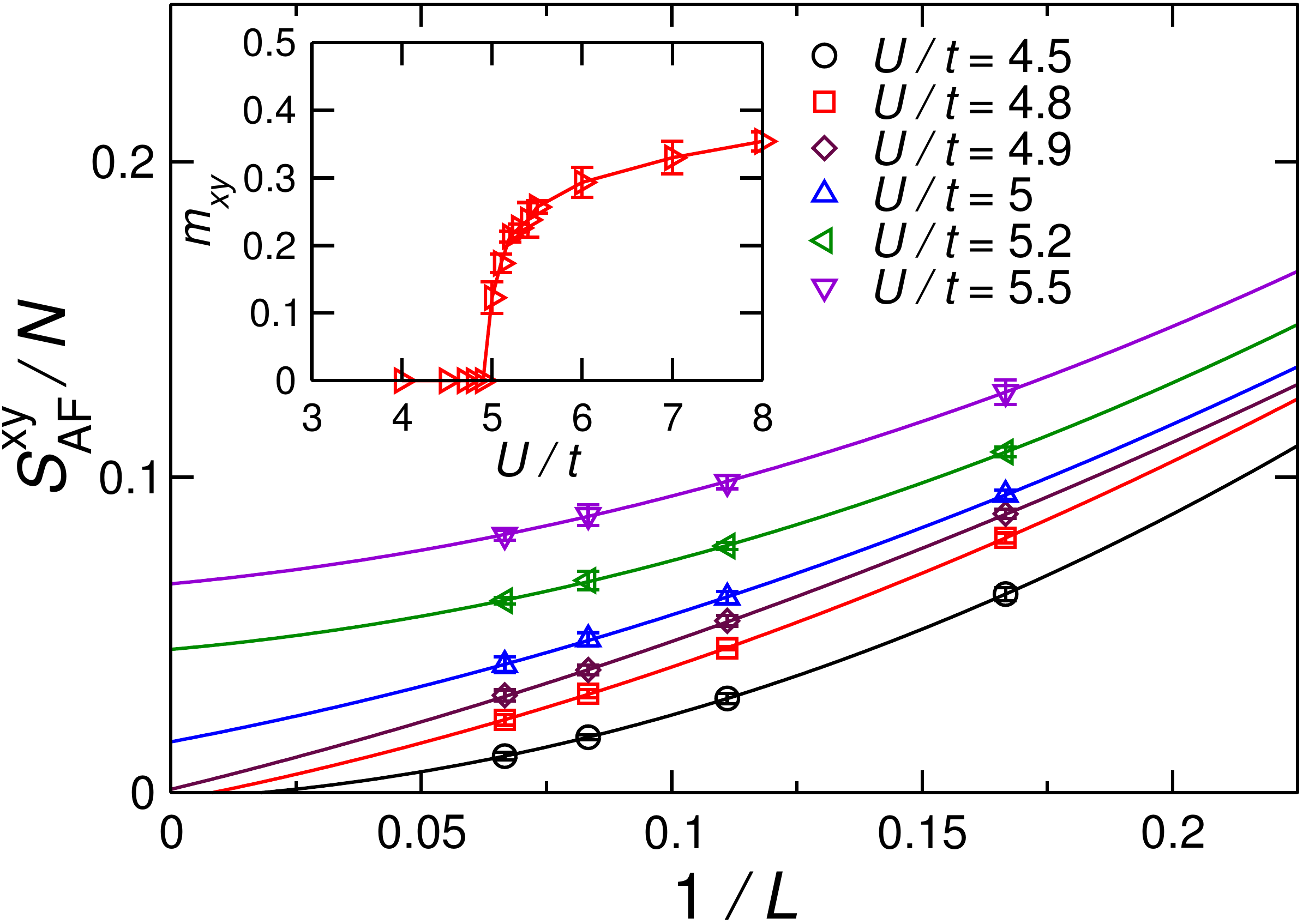}
  \caption{\label{fig:magneticorder}
    Finite-size scaling of the rescaled magnetic structure factor
    $\Sxy/N$ (\ref{eq:SAF}) in the KMH model
    with  $\lso/t=0.1$ and for different values of $U/t$ \cite{Ho.Me.La.We.Mu.As.12}. The quantity $\Sxy/N$
    extrapolates to zero in the quantum spin Hall phase and to a finite value in the magnetic phase; the critical value is  $\Uc/t=4.95(5)$.
    Inset: order parameter $m^{xy}$ as obtained from the size-extrapolated values.
    (Reprinted with permission from \cite{Ho.Me.La.We.Mu.As.12}. Copyright 2012
      by the American Physical Society).
   }
\end{figure}

Quantum Monte Carlo results for the quantity $\Sxy /N$ (with $N=2L^2$ being
the number of lattice sites), which is related to the transverse
magnetization by $m_{xy}^2=\Sxy /N$, are shown in
figure~\ref{fig:magneticorder} for different values of $U/t$ and for
$\lso/t=0.1$ \cite{Ho.Me.La.We.Mu.As.12}. Up to $U/t=4.9$, magnetic
correlations are substantial on small length scales but the structure factor
extrapolates to zero in the thermodynamic limit $L\to\infty$. For $U/t\geq
5$, long-range magnetic order is found. The critical value can be estimated
as $U_\text{c}/t=4.96(4)$ \cite{Ho.Me.La.We.Mu.As.12}, and the transition is
well visible from the magnetization results displayed in the inset of
figure~\ref{fig:magneticorder}.  A similar analysis for $\Szz /N$ reveals
that no long-range order exists up to very large values of $U/t$
\cite{Ho.Me.La.We.Mu.As.12}, in contrast to predictions of longitudinal order
based on cluster calculations \cite{Yu.Xie.Li.11}. A very similar picture
arises for other values of $\lso/t$ \cite{Ho.Me.La.We.Mu.As.12}.  At first
sight, the increase of $\Uc$ with increasing $\lso$ visible in
figure~\ref{fig:phasediagram}, and observed in several other works
\cite{RaHu10,Hohenadler10,Zh.Wu.Zh.11,Yu.Xie.Li.11,Wu.Ra.Li.LH.11,PhysRevB.85.195126,PhysRevB.85.235449,arXiv:1203.2928},
may be attributed to the increase of the spin-orbit gap. However, since $\Uc$
also increases from $\Uc/t=5.70(3)$ at $\lso/t=0.2$ \cite{As.Be.Ho.2012} to
$\Uc/t=6.5(1)$ at $\lso/t=0.3$ \cite{Hohenadler10} even though the minimal
gap at the M points remains constant \cite{RaHu10}, the origin of the
increase of $\Uc$ is instead expected to be related to the competition
between kinetic energy and magnetic order.

The fact that the magnetic ordering occurs in the $xy$ direction can be
understood from an effective spin model derived in the limit $U/t\to\infty$
where charge degrees of freedom are frozen out \cite{RaHu10}. The spin
Hamiltonian takes the form \cite{RaHu10,PhysRevB.85.195126,Re.Th.Ra.12}
\begin{equation}\label{eq:KMspin}
  H^\infty_\text{KMH} = J \sum_{\las ij\ras} \bm{S}_i \cdot \bm{S}_j 
  + J'\sum_{\llas ij \rras} (S^z_i S^z_j - S_i^x S_j^x - S_i^y S_j^y)\,.
\end{equation}
The first term is an antiferromagnetic coupling between neighbouring lattice
sites emerging from nearest-neighbour hopping ($J=4t^2/U$), whereas the second
term originates from the spin-orbit interaction
($J'=4\lambda^{2}_\text{SO}/U$) and couples next-nearest-neighbour sites
ferromagnetically in the $xy$ direction and antiferromagnetically in the $z$
direction.  Hence, at least for large $U/t$, the KMH model corresponds to a
frustrated spin model with respect to the $z$ component of spin, but allows
for nonfrustrated magnetic order in the $xy$ direction. Exact numerical
results \cite{Hohenadler10,Ho.Me.La.We.Mu.As.12,Zh.Wu.Zh.11} confirm this
strong-coupling picture. For the parameter range of
figure~\ref{fig:phasediagram}, no order in the $z$ direction was found down
to $\lso/t=0.002$ \cite{Ho.Me.La.We.Mu.As.12}.

\begin{figure}[t]
  \centering
  \includegraphics[width=0.45\textwidth]{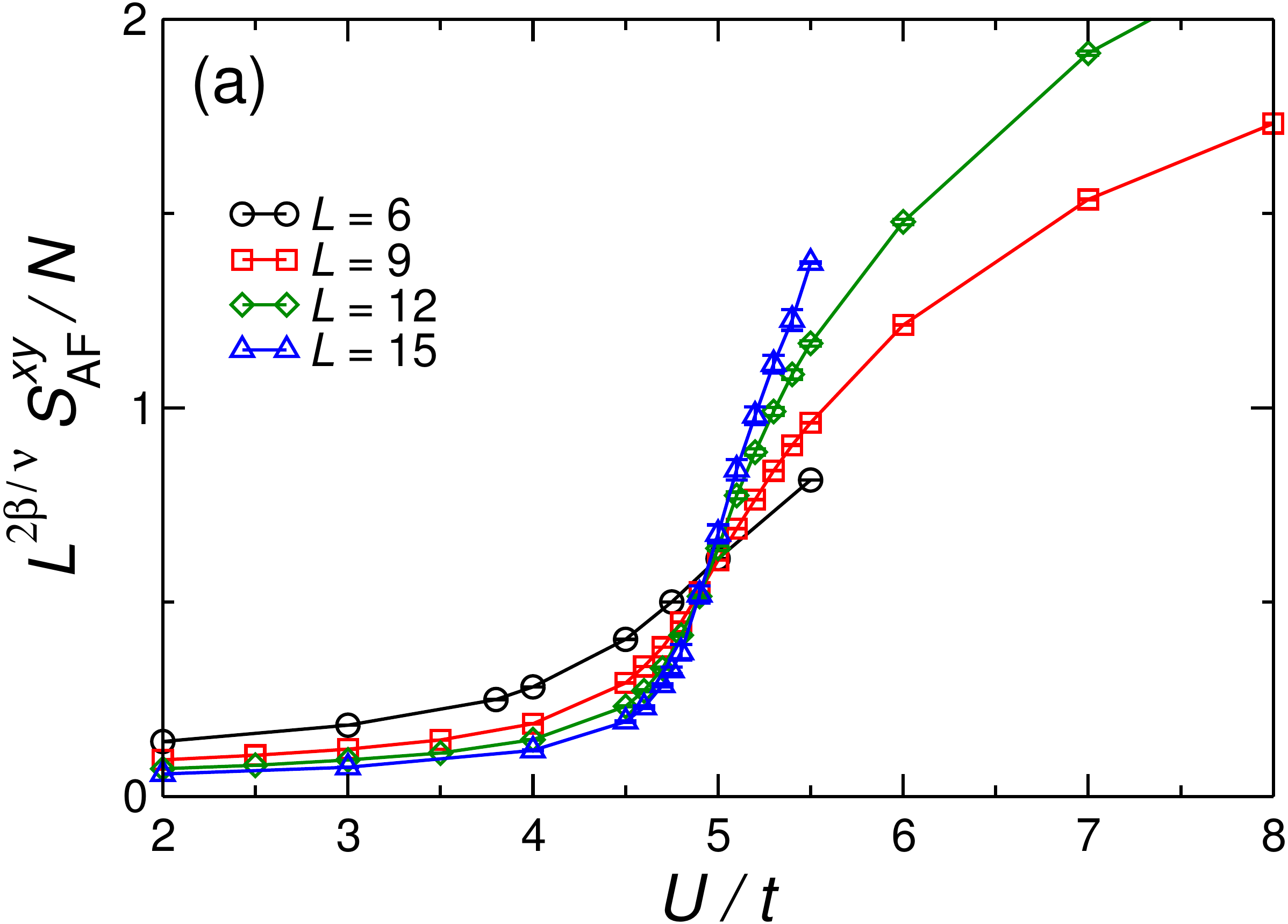}\\
  \includegraphics[width=0.45\textwidth]{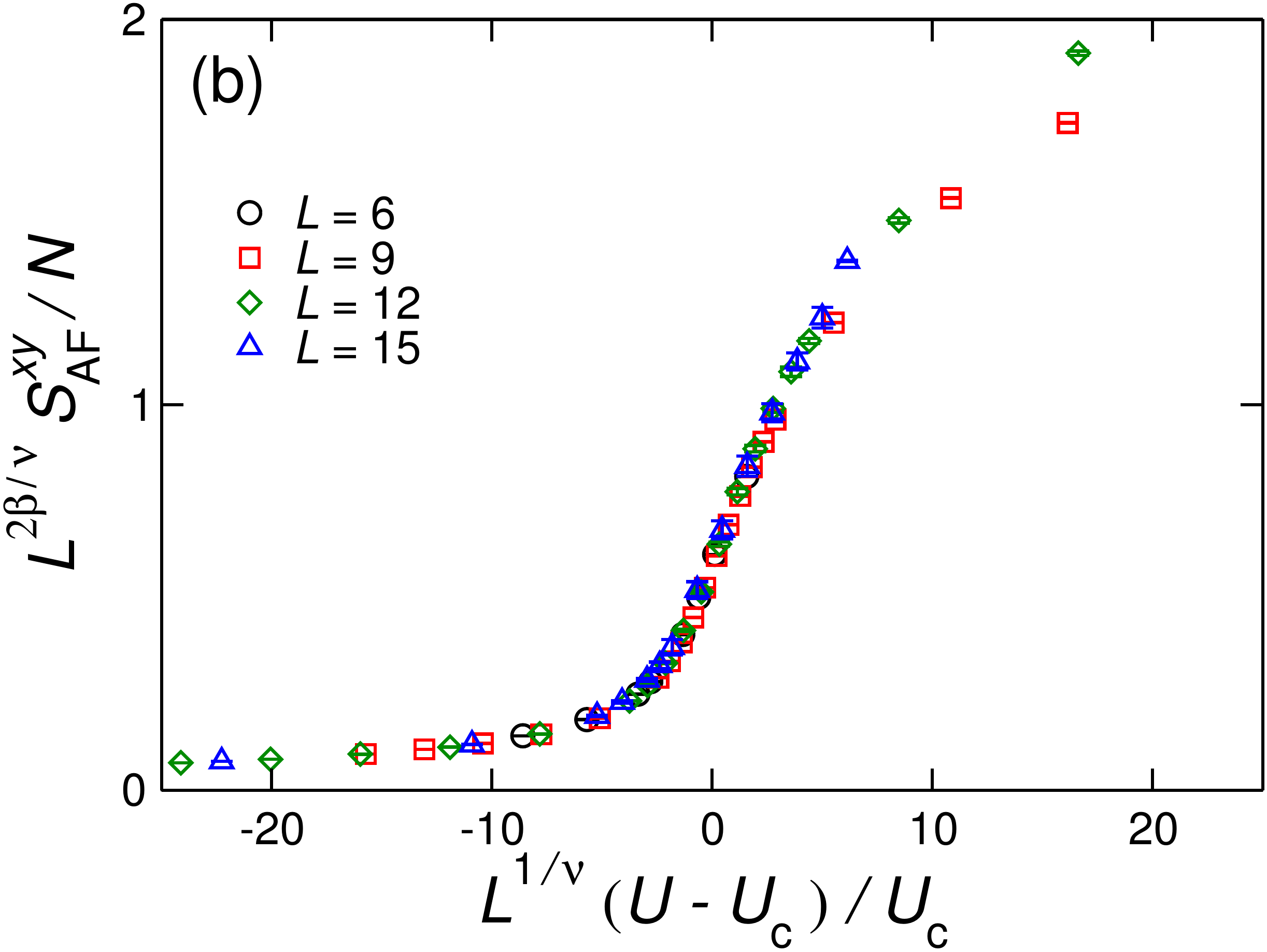}
  \caption{\label{fig:3dxy}
    (a)
    Rescaled transverse magnetic structure factor $\Sxy/N$, see (\ref{eq:SAF}),
    as a function of $U$ in the KMH model with $\lso/t=0.1$. 
    Assuming 3D XY behaviour and using the critical exponents of the 3D XY
    model \cite{PhysRevB.74.144506}, a clean intersection of curves for
    different system sizes $L$ at the critical point $\Uc/t=4.96(4)$ is obtained. 
    (b)       
    Scaling collapse of $L^{2\beta/\nu} \Sxy/N$ as a function of $L^{1/\nu} (U-\Uc)/\Uc$
    using $\Uc/t=4.96(4)$ and 3D XY exponents.
    (Reprinted with permission from \cite{Ho.Me.La.We.Mu.As.12}. Copyright 2012
      by the American Physical Society). 
    }
\end{figure}

Considering the spin-conserved case ($\lr=0$), the spin-orbit coupling in the
KMH model reduces the spin symmetry from $SU(2)$ to $U(1)$. Therefore,
the magnetic transition is expected to be in the universality class of the 3D XY
model \cite{Hohenadler10,Gr.Xu.11,PhysRevLett.107.166806}. 
The quantum Monte Carlo method permits one to test this conjecture. Assuming that
the transition is of the 3D XY type, the order parameter $\Sxy/N$ is expected
to fulfil the scaling equation \cite{Ho.Me.La.We.Mu.As.12}
\begin{equation}\label{eq:scaling:S}
  \Sxy/N  = L^{-2\beta/\nu} f_1[(U-\Uc) L ^{1/\nu}]
\end{equation}
where $z=1$, $\nu=0.6717(1)$ and $\beta=0.3486(1)$ are the critical exponents
of the 3D XY model \cite{PhysRevB.74.144506}. In particular,
(\ref{eq:scaling:S}) implies that at the critical point, $U=U_\text{c}$,
$\Sxy/N$ should become independent of the system size $L$.  As illustrated in
figure~\ref{fig:3dxy}(a) for $\lso/t=0.1$, the numerical results
\cite{PhysRevB.74.144506} indeed reveal the expected behaviour, showing a
well-defined intersection of curves for different $L$ at the critical
point. Moreover, a clean scaling collapse is obtained upon
rescaling the $U$ axis according to (\ref{eq:scaling:S}), see figure~\ref{fig:3dxy}(b). 
The 3D XY scaling of $\Sxy/N$ has also
been verified for other values of $\lso/t$
\cite{Ho.Me.La.We.Mu.As.12,As.Be.Ho.2012}, as well as for the spin gap
$\Delta_\text{s}$ \cite{Ho.Me.La.We.Mu.As.12}. The 3D XY criticality can be
expected to be generic for quantum spin Hall models with conserved spin. If
spin is not conserved, the symmetry is reduced to the $Z_2$ Ising
universality class, allowing magnetic order at finite temperatures.

At the magnetic transition, time-reversal symmetry---underlying the
topological protection of the quantum spin Hall state---is broken. Therefore,
a change of the topological invariant from nontrivial to trivial does not
require the closing of any gaps. Indeed, quantum Monte Carlo simulations
reveal only a slight dip of the single-particle gap near the critical point,
see figure~\ref{fig:tbi-afmi-gaps}. Very similar behaviour is also found in
mean-field theory, see inset of figure~\ref{fig:tbi-afmi-gaps}. On the other
hand, the spin gap $\Delta_\text{s}$ does close at the magnetic transition,
corresponding to the condensation of magnetic excitons
\cite{PhysRevLett.107.166806,Ho.Me.La.We.Mu.As.12,As.Be.Ho.2012}.  For
$U<U_\text{c}$, excitons have a finite energy, and can be used in combination
with $\pi$ fluxes (see section~\ref{sec:index}) to construct quantum spin
models with tunable interactions \cite{As.Be.Ho.2012}. The coupling between
the helical edges states and the magnetic excitons at the critical point was
recently discussed by Grover and Vishwanath \cite{GrVi12}.

An interaction-driven magnetic transition has also been reported for the BHZH
model \cite{arXiv:1111.6250,Wa.Da.Xi.12,PhysRevB.85.235449}. Whereas existing
work has focused on equal bandwidths for hopping via the s and p
orbitals ($t_\text{ss}=t_\text{pp}$), a much richer picture can be expected
for more general parameters. In the absence of the complex spin-orbit
coupling ($t_\text{sp}=0$), the BHZH model reduces to the two-band Hubbard
model.  The latter has been studied intensely as it may describe consecutive
Mott transitions within the heavier and lighter band
\cite{e20020021,PhysRevLett.91.226401,PhysRevLett.92.216402,PhysRevB.72.201102}.
The interplay of orbital-selective Mott transitions and spin-orbit coupling
represents a fascinating topic for future research. The SI model
(\ref{eq:SI}) is expected to undergo a magnetic transition in the Ising
universality class if the spin-orbit coupling is not too large
\cite{PhysRevLett.108.046401}, see also figure~\ref{fig:qshstar}.

Time-reversal symmetry is a key concept for the understanding of topological
insulators, and underlies many of their properties. Remarkably, topological
insulators can exist even in systems where time-reversal symmetry is broken
by long-range magnetic order. As first discussed by Mong \etal
\cite{PhysRevB.81.245209} for three dimensions, a topological state
with protected edge states can exist in the absence of time-reversal symmetry
$\Theta$ if $S=\Theta T_{1/2}$ remains conserved, where $T_{1/2}$ is a
translation symmetry defined by the unit cell of magnetic order. A simple
example is given by an antiferromagnet in which the magnetic order leads to a
doubling of the unit cell. Three-dimensional antiferromagnetic topological
insulators may be realized in \chem{GdBuOt} \cite{PhysRevB.81.245209} as well
as with ultra-cold atoms \cite{PhysRevB.85.195116}.  In two dimensions, the
existence of edge states protected against disorder has been shown to
additionally require spin conservation \cite{PhysRevB.83.045114}.  According
to Guo \etal \cite{PhysRevB.83.045114}, nonmagnetic or magnetic staggered
potentials can drive the noninteracting BHZ model into a topological
phase. Surprisingly, for the same model, a correlated topological phase with
coexisting magnetic order has recently been reported by Yoshida \etal
\cite{arXiv:1207.4547v1}.  If this phase indeed exists, it is interesting to
understand why it is absent in the KMH model.  However, another possibility
is that the coexistence is due to the fact that the dynamical mean-field
theory applied in \cite{arXiv:1207.4547v1} does not allow for magnetic order
in the $xy$ direction (as preferred for systems with spin-orbit coupling).
Indeed, at the level of static mean-field theory, two different critical
points can be observed when decoupling the interaction in terms of the
magnetization in the $z$ and the $xy$ direction, respectively.  Phases in
which topological order coexists with magnetic order, for example {\it
  topological spin density waves}, have also been found in the interacting,
spinful Haldane model (also known as the topological Hubbard model)
\cite{PhysRevB.84.035127,He.Wa.Ko.12,PhysRevLett.108.046806}.  In contrast to
the BHZH model of \cite{arXiv:1207.4547v1}, the topological Hubbard model
breaks time-reversal symmetry from the outset.
 
A related theoretical problem, yet with important differences, is that of the
Haldane-Hubbard model defined in (\ref{eq:haldane}). Without spin-orbit
coupling ($t_2=0)$, the nearest-neighbour repulsion drives a second-order
transition to a charge-density-wave insulator that spontaneously breaks
inversion symmetry
\cite{PhysRevLett.97.146401,RaQiHo08,PhysRevLett.100.146404}.  The two
degenerate ground states have all electrons on either the A or the B
sublattice.  Because inversion symmetry is in principle not relevant for the
stability of the topological state of the Haldane model, in contrast to the
role of time-reversal symmetry for the quantum spin Hall state, the question
how the transition will occur for nonzero $t_2$ is very interesting. This
problem was studied using mean-field theory, exact diagonalization, and
constrained-path Monte Carlo simulations
\cite{Wa.Sh.Zh.Wa.Da.Xi.10,VaSuRi10,Va.Su.Ri.Ga.11}. Although the results do
not fully agree, there is evidence that the Chern number changes from
nontrivial to trivial, and that this change and the onset of long-range
charge order can either coincide, or happen in two distinct transitions. It
is not yet clear if a gap closes at the critical point. However, because the
topological phase does not rely on inversion symmetry, and the topological
protection is therefore not destroyed by the charge order, one may expect to
see the closing of an excitation gap at the point where the Chern number
changes. This appears to be confirmed by recent results
\cite{Va.Su.Ri.Ga.11}, which relate the closing of the first excitation gap
($E_1-E_0$, the difference between the two lowest eigenvalues) to a
topologically protected level crossing that exists even in finite systems.

\subsubsection{Symmetry-conserving transitions}

The magnetic and charge-density-wave transitions discussed above involve a
well-defined, local order parameter, and in general may or may not coincide
with the destruction of topological order (in the sense of a nonzero spin
Chern or Chern number, respectively). Here the possibility of interaction-driven
transitions without symmetry breaking is discussed, including transitions
across which topological order either remains or emerges.

In the context of the KMH model, the possible quantum spin liquid phase shown
in figure~\ref{fig:phasediagram} has received a lot of attention. Quantum
Monte Carlo results suggest that this phase has nonzero single-particle and
spin gaps, but does not break any symmetries \cite{Meng10}. Furthermore,
numerical results are compatible with a closing of the single-particle and
spin gaps at the transition to the quantum spin Hall state, which suggests
that the two phases are not adiabatically connected
\cite{Hohenadler10,Ho.Me.La.We.Mu.As.12}. However, despite the remarkable
system sizes accessible in numerical simulations, the length scales for spin
correlations are still not under control
\cite{Ho.Me.La.We.Mu.As.12,So.Ot.Yu.}, making extrapolations to the
thermodynamic limit very delicate. The existence of the quantum spin liquid
phase in the Hubbard model ($\lso=0$) has recently been challenged based on
quantum Monte Carlo results on lattices with up to $36\times36$ unit cells
\cite{So.Ot.Yu.}. Signatures of the spin liquid phase in the KMH model have
also been reported from quantum cluster simulations
\cite{Wu.Ra.Li.LH.11,Yu.Xie.Li.11}, although the latter cannot describe a
true spin liquid phase because of the inherent breaking of translation
symmetry; instead, the observed paramagnetic insulator may be regarded as a
valence bond solid.  Furthermore, the quantum spin liquid phase of the KMH
model has been related to a topological Mott insulator
\cite{We.Ka.Va.Fi.11}. Because the KMH and SI models coincide for
$\lso=t_1=t_2=0$, the same phase may also be expected to exist in the SI
model. At finite spin-orbit coupling, the frustrated spin exchange
interactions of the SI model (see below) may even provide a more favourable
setting for such a state. A review of the connections between spin liquids
and topological insulators has been given by Fiete \etal
\cite{Fi.Ch.Hu.Ka.Lu.Ru.Zy.11}.

For two-dimensional models with explicit spin-orbit coupling, the possibility
of a topological Mott insulator has first been discussed by Rachel and Le Hur
\cite{RaHu10}, following an earlier suggestion by Pesin and Balents
\cite{PeBa10}.  Within a slave-rotor approach to the KMH model, the charge
degrees of freedom can be gapped out, whereas the spin degrees of freedom
inherit the topological band structure, and hence preserve time-reversal
symmetry \cite{RaHu10}.  Such an interaction-generated state is expected to
have topological order.  For the KMH model, the corresponding phase was
suggested to lie in between the quantum spin Hall and antiferromagnetic Mott
insulator regions of figure~\ref{fig:phasediagram}, but is not visible in
numerical simulations \cite{Ho.Me.La.We.Mu.As.12}.  Its absence can be
attributed to the role of gauge fluctuations in two dimensions
\cite{RaHu10}. A topological Mott insulator with gapped charge {\it and} spin
excitations in the KMH model \cite{We.Ka.Va.Fi.11}---presumably stable with
respect to gauge fluctuations---was argued to be related to the quantum spin
liquid phase shown in figure~\ref{fig:phasediagram}.  Transitions from
topologically trivial to nontrivial states upon increasing the Hubbard
repulsion $U$ have also been reported for the BHZH model \cite{Wa.Da.Xi.12,arXiv:1211.3059}
and for a bilayer square-lattice model \cite{Ar.Ca.Sa.12}, in both cases
starting with band parameters outside the respective quantum spin Hall phase.
Of particular interest is the fact that such a transition can be
either of the mean-field type (via a renormalization of single-particle
parameters), or driven by quantum fluctuations \cite{arXiv:1211.3059}.

\begin{figure}
  \centering
  \includegraphics[width=0.4\textwidth]{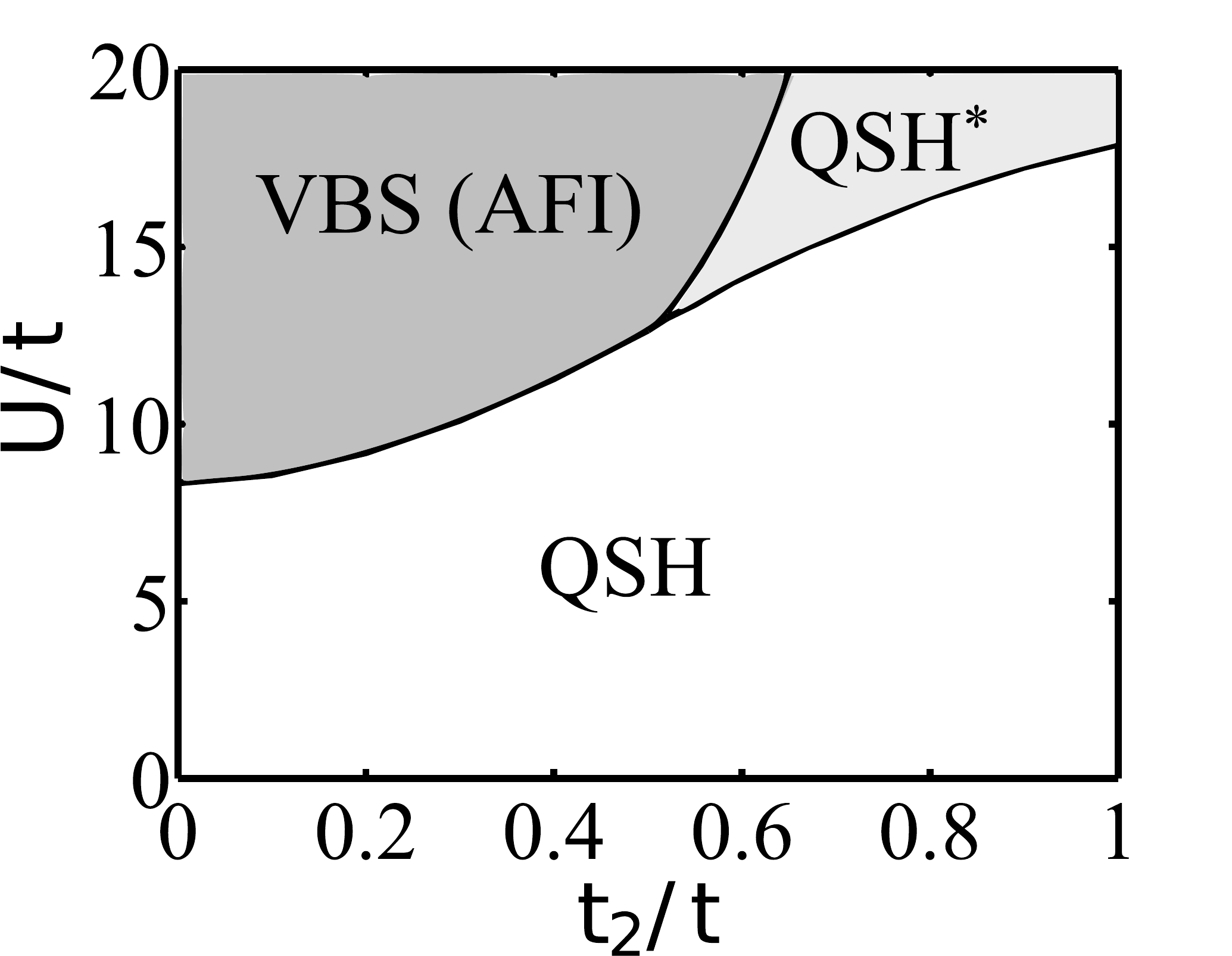}
  \caption{\label{fig:qshstar}
    Mean-field phase diagram of the SI model, showing a quantum spin
    Hall phase (QSH), a valence bond solid (VBS), and a fractional
    QSH* phase \cite{PhysRevLett.108.046401}. (Adapted with permission from
      \cite{PhysRevLett.108.046401}. Copyright 2012
      by the American Physical Society). 
    }
\end{figure}

A particularly interesting, strongly correlated topological phase has been
argued to exist in the SI model (\ref{eq:SI}).  Following the suggestion
of a strongly correlated quantum spin Hall phase resulting from the coupling
to a dynamical $Z_2$ gauge field by Ran \etal \cite{PhysRevLett.101.086801},
R\"uegg and Fiete \cite{PhysRevLett.108.046401} used a self-consistent $Z_2$
slave-spin mean-field theory to demonstrate the existence of the so-called
QSH* phase in the SI model. Their phase diagram, reproduced in
figure~\ref{fig:qshstar}, exhibits an extended regular QSH phase, and a
valence bond solid at large $U/t$. The latter is expected to become an
antiferromagnetic Mott insulator (AFI) phase similar to
figure~\ref{fig:phasediagram} when the possibility of time-reversal
symmetry breaking is added to the theory \cite{PhysRevLett.108.046401}. For large
spin-orbit coupling $t_2$ and large Hubbard $U$, the QSH* phase emerges. It
has been pointed out that because the QSH--QSH* transition corresponds to an
order-disorder transition of the slave spins \cite{PhysRevLett.108.046401},
the QSH* phase should be considered an {\it orthogonal metal} rather than a
Mott insulator \cite{PhysRevB.86.045128}.

According to \cite{PhysRevLett.108.046401}, the QSH* phase is characterized
by a bulk band gap, and protected helical edge states for charge and spin
excitations. It supports fractional, bosonic quasi-particle excitations with
semionic mutual braiding statistics \cite{PhysRevLett.108.046401}. The phase
does not break any symmetries, but has a four-fold topological
degeneracy on a torus \cite{PhysRevLett.108.046401}. It is thus topologically
ordered in the sense of section~\ref{sec:intro}, and hence distinguished from
the QSH phase by the presence of long-range quantum entanglement \cite{PhysRevB.84.235141}.
Fractional time-reversal invariant liquids will be further discussed in
section~\ref{sec:bulk:fqhe}. Because of the strong interactions, the edge
states are expected to be strongly correlated \cite{PhysRevLett.108.046401},
leading to a suppression of the Drude weight in agreement with results for
other models (section~\ref{sec:edge:correl}).

The reliability of the mean-field results in \cite{PhysRevLett.108.046401} is
not yet clear. Moreover, because spin is not conserved, the SI model can
presently not be investigated by quantum Monte Carlo methods. This situation
has motivated two recent studies of the limit $U/t=\infty$, in which an
effective spin model can be derived \cite{Re.Th.Ra.12,Ka.La.Fi.12}. It has
the form \cite{Re.Th.Ra.12,Ka.La.Fi.12}
\begin{eqnarray}\nonumber\label{eq:SIspin}
  H^\infty_\text{SI} 
  &=
  J_1
  \sum_{\las ij\ras} 
  \bm{S}_i \cdot \bm{S}_j 
  +
  J_2
  \sum_{\las\las ij\ras\ras}
  \bm{S}_i \cdot \bm{S}_j\\ 
  &\quad-
  J_3
  \sum_{\las\las ij\ras\ras}
  (
  \bm{S}_i \cdot \bm{S}_j - 2 S_i^w S_j^w
  )\,.
\end{eqnarray}
Here the exchange constants are $J_1=4t^2/U$, $J_2=4{t_1}^2/U$, and
$J_3=4t_2^2/U$. The coupling $J_2$ was neglected in \cite{Re.Th.Ra.12}.
Importantly, the term $S_i^w S_j^w$ depends on the link that connects sites
$i$ and $j$, leading to frustration and the absence of any preferred
direction for magnetic order \cite{Re.Th.Ra.12}. In contrast, in the
$U/t=\infty$ limit of the spin-conserving KMH model, (\ref{eq:KMspin}), order
in the spin-$z$ direction is frustrated, but easy-plane order is not. The
Hamiltonian~(\ref{eq:SIspin}) has close connections to the Kitaev-Heisenberg
model \cite{PhysRevB.84.100406,Chaloupka10}, an exactly solvable model with a
spin liquid ground state.

For small values of $J_2$, the model~(\ref{eq:SIspin}) undergoes transitions
from stripe order to N\'eel order to spiral order upon increasing $J_3$ from
negative to positive values \cite{Re.Th.Ra.12,Ka.La.Fi.12}. For the repulsive
SI model, the transition from N\'eel to spiral order is of particular
interest, and has been interpreted as a strong-coupling signature of the VBS
(or AFI) to QSH* transition \cite{Re.Th.Ra.12,Ka.La.Fi.12}. A magnetic
transition from the QSH* to the spiral state is expected to occur with
increasing $U/t$ \cite{Re.Th.Ra.12}.  The spin-nonconserving SI model hence
appears to describe phenomena beyond the numerically studied KMH model with
$\lr=0$, although incommensurate order has also been predicted for the latter
analytically \cite{PhysRevB.85.195126}. An interesting question for future
research is if fractional topological states and exotic magnetic states can
also occur in the KMH model with $\lr\neq 0$ \cite{Re.Th.Ra.12}. The
corresponding spin Hamiltonian has been derived in \cite{Re.Th.Ra.12}.

\subsection{Interaction-driven topological insulators}\label{sec:bulk:topmott}

Spin-orbit coupling plays a crucial role for topological insulators
\cite{KaMe05a,KaMe05b,BeHuZh06}.  In general, materials with heavy atoms are
required for the effects of spin-orbit coupling to be observable.  On the
theoretical side, the existence of topological phases in models with
intrinsic spin-orbit coupling can be understood in the framework of
topological band theory \cite{RevModPhys.83.1057} without considering
electronic correlations. A completely different approach to create quantum
spin Hall phases is based on the theoretical concept of a {\it topological
  Mott insulator}, a state with an interaction-generated band gap and
protected edge states \cite{RaQiHo08}.  In addition to providing a much
richer theoretical setting to study correlated topological phases, this
concept promises the existence of topological phases in a wider range of
materials by abandoning the need for strong intrinsic spin-orbit coupling.

\begin{figure}
  \centering
  \includegraphics[width=0.45\textwidth]{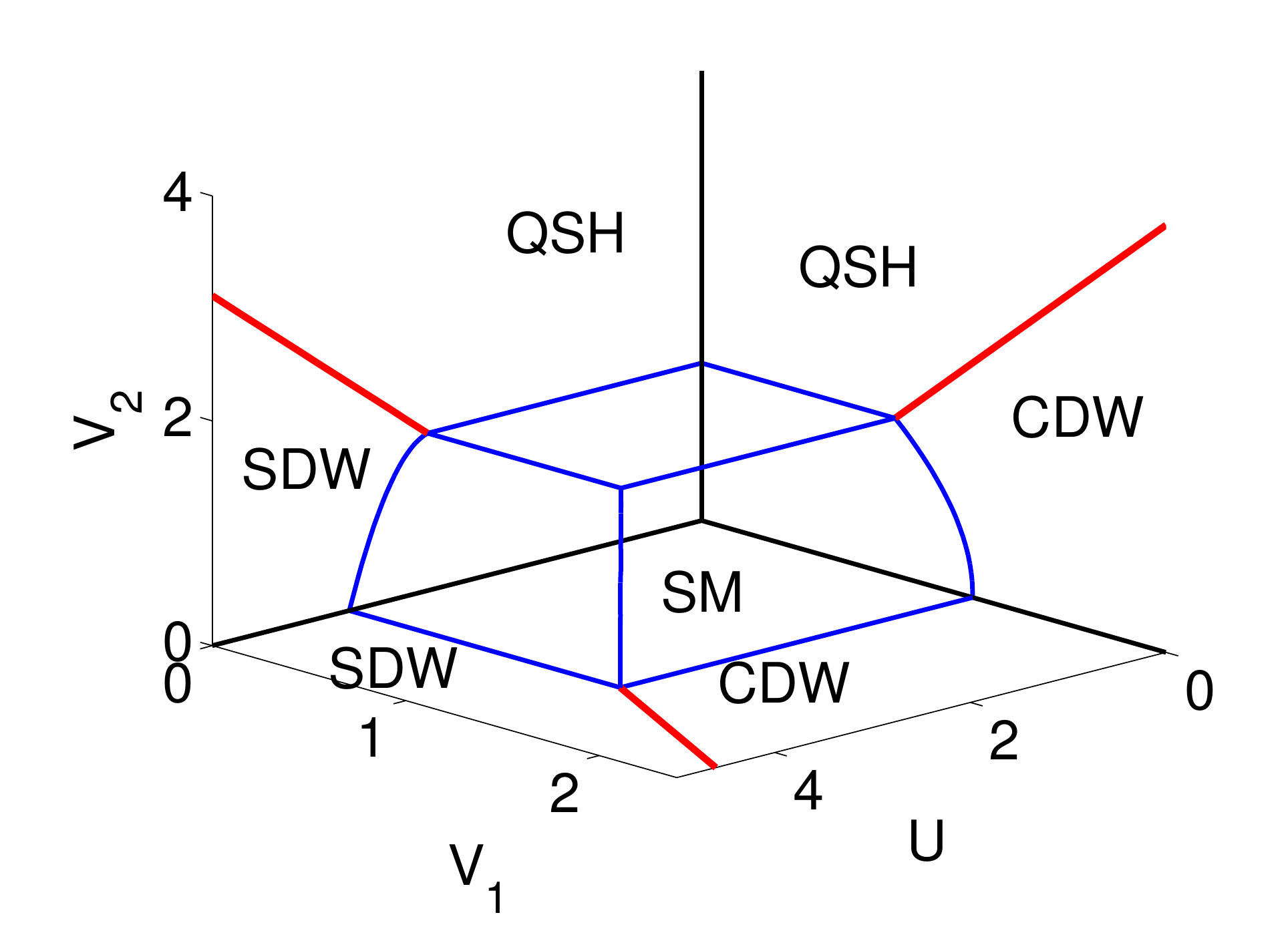}
  \caption{\label{fig:raghu} Mean-field phase diagram of the extended Hubbard
    model (with repulsions $U$, $V_1$, $V_2$) on the honeycomb lattice
    \cite{RaQiHo08}. The phases correspond to a semimetal (SM),
    charge-density-wave order (CDW), spin-density-wave order (SDW), and a
    quantum spin Hall state (QSH) emerging from dynamically generated
    spin-orbit coupling.  (Reprinted with permission from
    \cite{RaQiHo08}. Copyright 2008 by the American Physical Society).  }
\end{figure}

Raghu \etal \cite{RaQiHo08} considered an extended Hubbard model on the
honeycomb lattice with the usual nearest-neighbour hopping term (see
(\ref{eq:KM})) as well as onsite ($U$), nearest-neighbour ($V_1$) and
next-nearest-neighbour ($V_2$) repulsion in the form of density-density
interaction terms. In the spinless case, the repulsion $V_2$ frustrates the
charge-density-wave order driven by $V_1$, and a mean-field treatment
suggests that large values of $V_2$ lead to a state with spontaneously broken
time-reversal symmetry characterized by a nonzero, imaginary value of the
bond order parameter $\chi_{ij}=\las \hat{c}^\dag_i \hat{c}^\nag_j \ras$
\cite{RaQiHo08}.  This state is fully equivalent to the integer quantum Hall
state of the Haldane model \cite{Haldane98}. In the spinful case, the nonzero
bond order parameter can either have the same or opposite signs for the two
spin directions. The former case leads to a quantum anomalous Hall state,
whereas the latter case corresponds to a quantum spin Hall state resulting
from a dynamically generated spin-orbit coupling that breaks the $SU(2)$ spin
symmetry \cite{PhysRevLett.93.036403,RaQiHo08}. At the mean-field level, the quantum
Hall and quantum spin Hall states are degenerate.  Taking into account
quantum fluctuations within the random-phase approximation reveals that the
quantum spin Hall state is favoured because of the existence of a Goldstone
mode, and leads to the phase diagram shown in figure~\ref{fig:raghu}
\cite{RaQiHo08}. In addition to the quantum spin Hall phase, the latter
contains semimetal, spin-density-wave and charge-density-wave states. Similar
results and the possibility of a spin-dependent Kekul\'{e} order parameter
were discussed by Weeks and Franz \cite{PhysRevB.81.085105}. In contrast to
models with intrinsic spin-orbit coupling, the quantum spin Hall state from
dynamical spin-orbit interactions is characterized by a local order
parameter.  Nevertheless, it will in general be necessary to calculate the
topological invariant to distinguish such a state from one that has a nonzero
order parameter but is topologically trivial. Whether the quantum spin Hall
state arising from a dynamically generated spin-orbit coupling can be
adiabatically connected to a noninteracting QSH state (or instead supports
fractional excitations) is an
open question. Finally, for spinless fermions on a honeycomb lattice with
extended interactions, mean-field theory predicts the existence of a quantum
anomalous Hall phase \cite{RaQiHo08,PhysRevB.81.085105,PhysRevA.86.053618};
the resulting order parameter is equivalent to the single-particle hopping
terms in the Haldane model \cite{Haldane98}.
A possible realization with Rydberg atoms in an optical lattice is discussed
in \cite{PhysRevA.86.053618}.

Interaction-driven topological states in models without intrinsic spin-orbit
coupling have also been studied on the checkerboard
\cite{PhysRevLett.103.046811}, kagome
\cite{PhysRevB.82.045102,PhysRevB.82.075125}, and decorated honeycomb
lattices \cite{PhysRevB.82.075125}. Most importantly, these more complicated
geometries support both Dirac band crossing points and quadratic band
crossing points depending on the band filling in the noninteracting
limit. For example, the low-energy physics on the kagome lattice is
determined by a Dirac point for $n=1/3$ and by a quadratic crossing point for
$n=2/3$. Both filling fractions support a quantum spin Hall phase in the
presence of intrinsic spin-orbit coupling \cite{PhysRevB.80.113102}. However,
whereas topological order due to interaction-generated spin-orbit terms is
the leading instability for $n=2/3$, it competes with other ordering
phenomena and requires strong interactions and fine-tuning in the case
$n=1/3$ due to the vanishing density of states at the Fermi level
\cite{PhysRevB.82.075125}.  In principle, because these lattices support
partially filled flat bands, the interacting models are good candidates for
fractional topological phases.  A topological phase arising from spontaneous
ordering of complex orbitals has been found theoretically in a model of
transition-metal heterostructures \cite{Ru.Fi.11}. Importantly, in such
multi-orbital models, topological order can emerge from purely local
interactions, whereas nonlocal interactions are typically required for
single-orbital models \cite{Ru.Fi.11}. Another possible setting to observe
interaction-generated topological states is stacked graphene
\cite{PhysRevB.82.115124,PhysRevLett.106.156801}.  Finally, three-dimensional
topological insulators can also be generated from interactions
\cite{PhysRevB.79.245331}.

Topological states arising from electronic interactions have also been
discussed in the context of systems with intrinsic spin-orbit coupling that
undergo a Mott transition from a regular to an exotic quantum spin Hall phase
of spinons, in which interaction-generated low-energy spin excitations
inherit the topological band structure
\cite{PhysRevB.78.125316,PeBa10,RaHu10,We.Ka.Va.Fi.11}, see also
section~\ref{sec:trans}. Alternatively, interactions can give rise to a
transition from a nontopological state to an integer quantum spin Hall state,
as in \cite{Wa.Da.Xi.12,Ar.Ca.Sa.12}, essentially via a renormalization of
the effective spin-orbit coupling.  In relation to fractional states to be
discussed next, it is important to distinguish {\it fractionalized} and {\it
  fractional } topological insulators. The former are states with spin-charge
separation but typically integer charge excitations, whereas the latter are
states that exhibit charge fractionalization as in the fractional quantum
Hall effect.

\subsection{Fractional topological insulators}\label{sec:bulk:fqhe}

Given the close relation between integer quantum Hall states and quantum spin
Hall insulators, Bernevig and Zhang \cite{PhysRevLett.96.106802} envisaged
the possibility of fractional time-reversal invariant insulators arising from
strong electronic interactions \cite{PhysRevLett.96.106802}.  Levin and Stern
\cite{PhysRevLett.103.196803} considered the case of independent fractional
quantum Hall states for $\UP$ and $\DO$ spins, and established that the
criterion for the existence of edge states protected by time-reversal
symmetry is that $\sigma^s_{xy}/e^*$ is odd, where $e^*$ is the elementary
charge of the fractional Hall state.  However, a combination of two
fractional quantum Hall states is not necessarily a topological insulator
with protected edge modes, but can instead also lead to {\it fractional
  trivial insulators} with $\sigma^s_{xy}/e^*$ even
\cite{PhysRevLett.103.196803}.  In the latter, edge modes can be gapped out
by time-reversal invariant perturbations.

A lattice model that supports the integer quantum Hall effect without a
magnetic field was introduced by Haldane \cite{Haldane98}, see
(\ref{eq:haldane}). Analogous models for fractional quantum Hall states were
discovered only recently, for example, in
\cite{PhysRevLett.106.236802,PhysRevLett.106.236803,PhysRevLett.106.236804,FQHE_sheng2011,Le.Bu.KJ.St.11,PhysRevB.84.155116,La.Li.Be.Mo.12}.
A common feature of such models is the existence of flat bands with a nonzero
Chern number (so-called {\it Chern bands}) that mimic the Landau levels of
more traditional fractional quantum Hall states observed in strong magnetic
fields \cite{RevModPhys.80.1083}. The flatness allows electronic interactions
to be large compared to the kinetic energy but small compared to the band gap
(thereby avoiding level mixing effects and allowing for a projection onto a
single-band problem). A number of Haldane-like models with these properties
where suggested
\cite{PhysRevLett.106.236802,PhysRevLett.106.236803,PhysRevLett.106.236804,FQHE_sheng2011,He.Wa.Ko.12},
and the existence of a fractional quantum Hall phase was demonstrated in
\cite{PhysRevLett.106.236804,FQHE_sheng2011,PhysRevX.1.021014}. The prospect of fractional
quantum Hall states without magnetic fields has also inspired the search for
materials with partially filled Chern bands arising from spontaneous orbital
order, see \cite{PhysRevLett.108.126405} and references therein. Similar to
the case of topological Mott insulators (section~\ref{sec:bulk:topmott}), the
necessary spin-orbit coupling (or spin-dependent magnetic flux) can arise
from interaction-induced spontaneous symmetry breaking
\cite{PhysRevLett.108.126405}.  An adiabatic connection between Chern
insulators and the Hofstadter problem is discussed in \cite{Wu.Ja.Su.12},
and a proof for an adiabatic connection between fractional Chern insulators
and fractional quantum Hall states has been given in \cite{PhysRevLett.109.246805}.

In the integer quantum Hall effect, the chemical potential lies in the gap
between the highest filled and the lowest unfilled energy levels. As long as
the energy scale for electronic interactions is smaller than the gap to
excited states, interaction effects can be considered as a small
perturbation. In contrast, electronic correlations are at the heart of
fractional quantum Hall liquids and hence also fractional topological
insulators; for a review see \cite{RevModPhys.80.1083}. At the special
filling fractions for which $\sigma_{xy}$ exhibits a plateau, a family of
ground states with well-defined topological degeneracy is spontaneously
selected from the highly degenerate set of many-body states. This process is
inherently nonperturbative, and the fractional state has quasiparticle
excitations with fractional charge and fractional statistics. The extremely
rich physics known from the fractional quantum Hall effect stimulates
research on fractional topological insulators. A previously encountered
example of such a fractional spin Hall liquid is the QSH* phase of the SI
model \cite{PhysRevLett.108.046401}, see section~\ref{sec:trans}.

Neupert \etal \cite{PhysRevB.84.165107} considered the general case of a
fractional liquid with time-reversal symmetry but no spin conservation.  Such
states (including the noninteracting, integer quantum spin Hall states) are
described at low energies by Abelian Chern-Simons theories, whereas the
special case with spin conservation is described by a $U(1)\times U(1)$
Chern-Simons or BF theory \cite{QiHuZh08,Cho20111515}. If spin is not
conserved, the existence of protected edge states can be related to a $Z_2$
invariant independent of $\sigma^s_{xy}$ that reduces to the usual $Z_2$
index in the integer, noninteracting limit \cite{PhysRevB.84.165107}.  A
general Chern-Simons theory of two-dimensional, time-reversal invariant
insulators with Abelian quasiparticles and a classification of fractional
topological insulators was given by Levin and Stern \cite{Le.St.12}.  Recent
work along these lines includes parton \cite{PhysRevB.85.165134} and
composite fermion \cite{Fe.Vi.11} descriptions, the relation between
fractional topological insulators and quantum Hall states in Landau levels
\cite{Roy.12}, and an effective field theory of general fractional,
two-dimensional topological insulators with spin-orbit coupling
\cite{Ni.11,Ni.12}.  Because the fractional states arise from strong
interactions, topological order will typically compete with other many-body
instabilities such as charge-density-wave formation. For Chern insulators,
the stability of the topological state has been found to be enhanced by a
certain amount of band dispersion or for bands with higher Chern numbers
\cite{Gr.Ne.Ch.Mu.12}. Finally, the experimental detection of fractional
topological insulators, including the determination of the filling fraction
from transport measurements, is discussed in \cite{PhysRevLett.108.206804}.

An important open problem is the question under what circumstances two
coupled fractional Hall liquids form a fractional topological insulator with
protected edge modes. The interactions between the two fractional states may
completely destroy the insulating state, give rise to spontaneous breaking of
time-reversal symmetry, or lead to novel, non-Abelian states
\cite{PhysRevB.84.165107}. It will also be interesting to see if the physics
beyond the Laughlin states recently reported for Chern insulators
\cite{arXiv:1206.2626,La.Li.Be.Mo.12} carries over to fractional topological insulators.
Most of the numerical work on fractional topological insulators so far relies
on exact diagonalization, with an exponential scaling of computational effort
with system size. A possible alternative is the use of tensor network states
\cite{PhysRevLett.106.156401}.

\section{Edge correlation effects}\label{sec:edge}

The existence of metallic, protected edge states is one of the most obvious
manifestations of the topological properties of quantum spin Hall systems
\cite{KaMe05a,KaMe05b}, and has played a crucial role for the experimental
verification of the existence of topological insulators
\cite{Koenig07,Roth09}.  The fundamental properties of these edge states can
be understood in the noninteracting limit, as reviewed in
section~\ref{sec:noninteracting}.  Because of time-reversal symmetry, the edge
states always exist as Kramers pairs, and electrons of opposite spin
propagate in opposite directions along a given sample edge. For an odd number
of such helical edge state pairs, time-reversal symmetry provides a
protected, gapless crossing point \cite{KaMe05b}. In contrast,
single-particle backscattering is possible for an even number of pairs, and
the edge states are not protected \cite{KaMe05a,KaMe05b}. The parity of the
number of edge state pairs can therefore be related to the $Z_2$ topological
invariant \cite{KaMe05a}.

Compared to the gapped bulk, electron-electron interactions can be expected
to strongly affect the gapless edge states \cite{Cenke06}. Most importantly,
in the presence of interactions, inelastic scattering becomes possible, and
is not forbidden by Kramers theorem, as it connects states of different
energy \cite{Cenke06}.  Moreover, whereas time-reversal symmetry forbids
elastic single-particle processes, two-particle backscattering is allowed
even for a single Kramers pair. The possibility of such processes, which can
spontaneously break time-reversal symmetry at the edge \cite{Wu06,Cenke06},
is a crucial difference to the quantum Hall effect, where counter-propagating
states (and hence backscattering) are completely absent
\cite{PhysRevB.41.12838}. As pointed out in \cite{Cenke06,Wu06}, the
qualitative, topological distinction between odd and even numbers of edge
state pairs in the noninteracting case becomes a quantitative distinction
(depending on the interaction strength) in the presence of correlations.
However, as discussed below, for the class of models with conserved spin and
without impurities, the topological protection and the bulk-boundary
correspondence remain intact.

An odd number of Kramers edge state pairs, and hence a topologically
protected quantum spin Hall state, can be realized only at the boundary of a
two-dimensional system \cite{Wu06}, giving rise to the notion of a {\it
  holographic liquid} \cite{Wu06}. Hence, edge states of a quantum spin Hall
insulator may be regarded and theoretically described (to a good
approximation) as one-dimensional liquids that inherit their topological
properties from the bulk \cite{Wen92,Cenke06}.  Similar to other
quasi-one-dimensional systems such as quantum wires, quantum fluctuations
play a key role in determining the physics of edge states. Consequently,
whereas mean-field theory provided valuable insight into the effect of bulk
correlation effects, it is insufficient for edge physics and in fact gives
misleading results as a result of instabilities
\cite{PhysRevLett.107.166806}. Similarly, given the non-Fermi liquid
character of the metallic edge states, perturbative approaches fail.
Instead, Luttinger liquid theory \cite{Voit94} provides a rather accurate
description of many aspects \cite{Wu06,Cenke06}. Although a strictly
one-dimensional description seems justified by the exponential suppression of
scattering to bulk states, signatures of the holographic nature of the
helical edge states are visible as quantitative deviations from bosonization
predictions \cite{Ho.As.11}.

In this section, an introduction to the helical Luttinger model is given
\cite{Wu06,Cenke06} (for a previous review see \cite{RevModPhys.83.1057}),
followed by a discussion of analytical and numerical results with an emphasis
on theoretical aspects of the intensely studied integer quantum spin Hall
insulators. A more experimentally oriented review of transport properties was
given by Tkachov and Hankiewicz \cite{Tk.Ha.12}. Other interesting
developments not addressed in the following include Majorana edge states of
topological superconductors \cite{GrVi12}, and edge states of fractional
topological insulators \cite{PhysRevLett.103.196803,PhysRevLett.108.206804}.
Finally, the stability of edge states in magnetically ordered topological
insulators is discussed in
\cite{PhysRevB.81.245209,PhysRevB.84.035127,PhysRevB.83.045114,PhysRevB.85.195116,He.Wa.Ko.12}.

\subsection{Luttinger liquid description}\label{sec:edge:lutt}

The low-energy description of helical edge states in terms of Luttinger
liquid theory was established in two early papers by Wu \etal \cite{Wu06} and
Xu and Moore \cite{Cenke06}. The corresponding theory and hence the physics
is independent of the particular model for the bulk as long there is an
adiabatic connection to the noninteracting case. (For edge state theories of
fractional topological insulators see
\cite{PhysRevB.84.165107,PhysRevLett.103.196803,PhysRevLett.108.206804}.)
However, details such as the number of edge states, the allowed interactions
and the location of the crossing in momentum space are determined by the
$Z_2$ topological invariant, the spin symmetry, the bulk band structure and
the edge topology (for example, zigzag or armchair for honeycomb lattice
models). For a generic topological insulator, the low-energy model for the
edge states can be derived from Chern-Simons theory \cite{Wen92,Cenke06}, see
also section~\ref{sec:noninteracting}.

In the following, a single pair of edge states is considered, assuming a
geometry that is wide enough for the two edges to be independent.
Correlation effects on even numbers of edge state pairs, corresponding to a
topologically trivial state in the noninteracting case, have been discussed
by Xu and Moore \cite{Cenke06}. Surprisingly, in this situation, the edge
states can be stabilized against localization by interactions
\cite{Cenke06}. Effects of inter-edge tunnelling
\cite{PhysRevLett.103.166403,arXiv:1202.3203,PhysRevLett.107.166806} will be
addressed in section~\ref{sec:edge:correl}.

The following discussion follows Wu \etal \cite{Wu06} (see also
\cite{RevModPhys.83.1057}). Assuming a system with periodic boundaries in the
$x$ direction and open boundaries in the $y$ direction, and linearizing the
edge dispersion as $\epsilon_\sigma(k)=\pm \vF (k-\kF)$, the free continuum
theory takes the form
\begin{equation}
  H_0 = \vF \int \rmd x \, (\psi^\dag_{\mathrm{R}\UP} \rmi \partial_x
  \psi^\nag_{\mathrm{R}\UP}
  - \psi^\dag_{\mathrm{L}\DO} \rmi \partial_x
  \psi^\nag_{\mathrm{L}\DO})\,.
\end{equation}
The helicity of the edge states is apparent from the existence of right
moving electrons (R) with velocity $\vF$ and only  spin $\UP$, and left moving
electrons (L) with velocity $-\vF$ and only spin $\DO$. The spin directions can be
defined with respect to an arbitrary quantization axis even if spin is not
conserved. The Fermi velocity $\vF$ is related to the bulk band gap and the
edge topology, but analytical expressions for specific models are in general
not known (see, for example, \cite{1210.4818}). Because of the correspondence
between the direction of motion and spin, a spin index is not required, and
drops out of the effective low-energy Luttinger liquid theory (see below). An
interesting consequence of the reduced number of degrees of freedom compared
to ordinary Luttinger liquids is the predicted possibility to observe
fractional charges \cite{QiHuZh.08}.

To derive the low-energy model, it is necessary to consider the
possible interactions. Elastic perturbations which are odd under time
reversal, most importantly single-particle backscattering, are not
allowed. Inelastic single-particle scattering is allowed, but requires a spin
flip. It can arise, for example, from magnetic impurities, Rashha coupling in
combination with phonons \cite{PhysRevLett.108.086602}, or a coupling to
spin-$1$ bosons (for example, spin fluctuations \cite{Ho.As.11}). Forward
scattering as well as elastic or inelastic two-particle processes are
generally allowed \cite{Cenke06,Wu06}.  Although a discussion of scattering
processes based on the single-particle picture is physically intuitive, a
many-body formulation \cite{Cenke06} is more appropriate for interacting
systems, and leads to the same low-energy Luttinger model.  A continuum
theory of helical edge states was presented in
\cite{0953-8984-24-35-355001}. Bosonization results for non-helical
one-dimensional systems with spin-orbit coupling can be found in
\cite{PhysRevLett.84.4164,PhysRevB.68.075107}.

Forward scattering is described by  \cite{Wu06}
\begin{equation}\label{eq:fw}
  H_\text{fw} = g \int \rmd x \, \psi^\dag_{\mathrm{R}\UP}
  \psi^\nag_{\mathrm{R}\UP} \psi^\dag_{\mathrm{L}\DO} \psi^\nag_{\mathrm{L}\DO}\,.
\end{equation}
In clean systems, the most important two-particle term is umklapp scattering
(\ie, backscattering from a periodic lattice potential), given by  \cite{Wu06}
\begin{eqnarray}\label{eq:um}\nonumber
  H_\text{um} &= g_u \int \rmd x \, \rme^{-\rmi 4\kF x} 
  \psi^\dag_{\mathrm{R}\UP}(x)
  \psi^\dag_{\mathrm{R}\UP} (x+a) \\
  &\qquad\qquad\times
  \psi^\nag_{\mathrm{L}\DO}(x+a)
  \psi^\nag_{\mathrm{L}\DO}(x)
  + \text{H.c.}
\end{eqnarray}
Whereas $H_\text{fw}$ is generically present in helical liquids, the umklapp
term only plays a role at half filling ($4\kF=2\pi l$, $l$
integer) and in the presence of spin-nonconserving terms in
the Hamiltonian, such as the Rashba spin-orbit interaction \cite{Rasha}. It
is therefore generically absent in models with $U(1)$ spin symmetry, even at
half filling. Two-particle backscattering can also arise from magnetic
impurities \cite{Wu06,Maciejko09,PhysRevB.85.245108,PhysRevB.86.121106} or
quenched disorder \cite{Wu06,KaMe05b}, see
section~\ref{sec:edge:correl}. Finally, in systems without spin conservation,
combined single-particle forward- and backscattering can become important
\cite{PhysRevLett.108.156402}.

Under the assumption that the edge states remain quasi-one-dimensional,
interaction effects can be studied in the powerful framework of bosonization.
If only forward scattering is allowed, electron-electron interactions can be
accounted for by the Luttinger liquid parameters $K=\sqrt{(\vF-g)/(\vF+g)}$
and $v=\sqrt{\vF^2-g^2}$ (these expressions are valid for $g\ll \vF$). The
corresponding helical Luttinger model reads
\begin{equation}\label{eq:HLL}
H_\text{HL}= \frac{v}{2} \int \rmd x 
\left[
K^{-1} (\partial_x \phi)^2 + K (\partial_x \theta)^2\right]\,,
\end{equation}
where $\phi=\phi_\text{R}+\phi_\text{L}$ and
$\theta=\phi_\text{R}-\phi_\text{L}$. For a given spin direction,
only one chirality is allowed, in contrast to a regular Luttinger liquid
where for each spin direction there are left and right moving
electrons.

Because umklapp-- and backscattering are excluded, the quadratic and hence
exactly solvable Luttinger model~(\ref{eq:HLL}) describes correlated but
metallic helical edge states, as realized in a number of spin-conserving
theoretical models without disorder. Electron-electron interactions lead to a
change of the parameters $K$ and $v$. Generally, $K=1$ and $v=\vF$ for noninteracting
electrons, and $K<1$ ($K>1$) for repulsive (attractive) interactions.
However, no quantum phase transitions occur, and all correlation functions
decay with power laws with exponents determined by $K$. Explicitly, the
equal-time correlation functions can be derived as
\begin{eqnarray}\label{Luttinger_liquid_corr.eq}
  S^{xx}(r)&= \las S^{x}(r) S^{x}(0) \ras 
          \sim \frac{1}{r^{2K}} \cos(2\kF r )\,,\\\nonumber
  S^{zz}(r)&= \las S^{z}(r)  S^{z}(0) \ras 
          \sim  \frac{1}{r^2}\,, \\\nonumber
  N(r)    &= \las n(r) n(0) \ras 
          \sim \frac{1}{r^2}\,,\\ \nonumber
  P(r)    &= \las \psi^\nag_{\text{R}\UP}(r) \psi^\nag_{\text{L}\DO}(r)
             \psi^\dag_{\text{R}\UP}(0) \psi^\dag_{\text{L}\DO}(0)
             \ras \sim \frac{1}{r^{2/K}}\,,
\end{eqnarray}
corresponding to transverse spin, longitudinal spin, charge and pairing
correlations, respectively.  The results in~(\ref{Luttinger_liquid_corr.eq})
imply that transverse spin correlations (described by $S^{xx}(r)$) are the
slowest decaying correlations for repulsive interactions, whereas pairing
correlations $P(r)$ dominate for attractive interactions. The longitudinal
spin correlations $S^{xx}(r)$ (here $S^{zz}\neq S^{xx}$ reflects the broken
$SU(2)$ spin symmetry in the presence of spin-orbit coupling) and charge
correlations $N(r)$ decay exactly as in a Fermi liquid. Finally, only
$S^{xx}(r)$ depends on the Fermi momentum via the term $\cos(2\kF r)$, which
can be related to spin-flip processes \cite{Ho.As.11}. Whereas a
nonlocal electron-electron interaction of finite range is qualitatively
equivalent to the local interaction considered here, correlation functions
deviate from the power-law Luttinger liquid form in the case of a long-range
Coulomb potential \cite{PhysRevB.86.165121,Schulz93}.

\begin{figure}[t]
  \centering
  \includegraphics[width=0.45\textwidth]{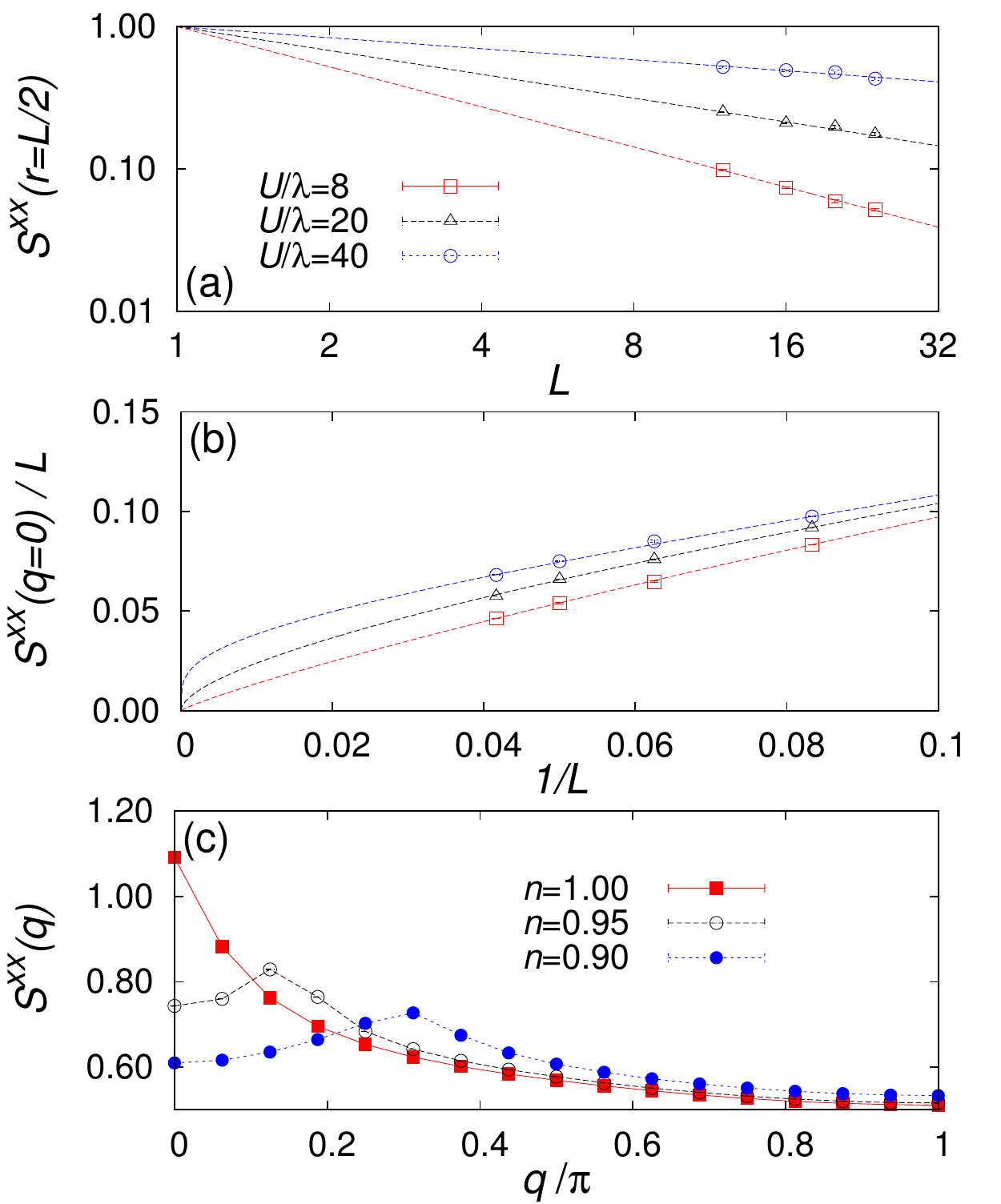}
  \caption{\label{fig:magneticorder_edge}
    (a) Transverse spin correlator for different interaction strengths
    $U /\lambda$ ($\lambda$: spin-orbit coupling), showing a power-law 
    decay with an interaction-dependent exponent in accordance with
    (\ref{Luttinger_liquid_corr.eq}). Lines are fits to the form
    $A/r^\eta$. Data are normalized to 1 at $r=1$ for comparison.
    (b) Renormalized transverse spin structure factor at $q=0$, showing the
    absence of long-range magnetic order in the thermodynamic limit. Lines
    are fits to the form $S^{xx}(q=0)/L=b/L + c/L^{\eta}$, with $\eta$ taken
    from (a).
    (c)
    Transverse spin structure factor at $U/\lambda=8$ for different
    band fillings $n$. 
    All results were obtained from zero-temperature quantum Monte Carlo simulations
    of an effective model with Hubbard interactions only at the edge of a
    zigzag ribbon.
    Data taken from \cite{Ho.As.11}.
  }
\end{figure}

\begin{figure}[t]
  \centering
  \includegraphics[width=0.45\textwidth]{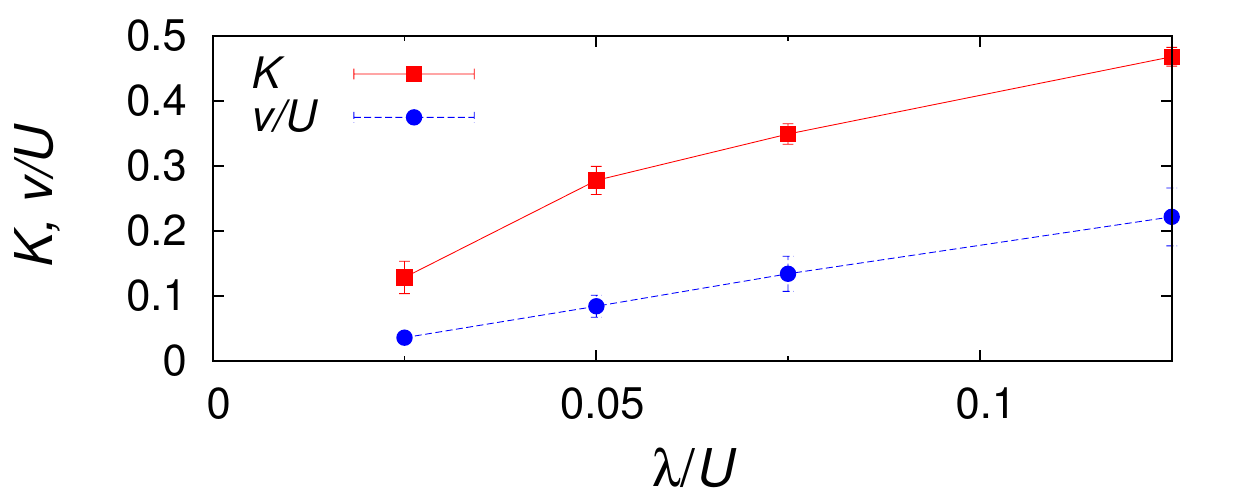}
  \caption{\label{fig:luttinger} Luttinger liquid parameters $K$ and $v$ for
    an effective model of helical edge states, as obtained from quantum Monte
    Carlo simulations and finite-size extrapolation.  Data taken from
    \cite{Ho.As.11}.}
\end{figure}

\begin{figure}[t]
  \centering
  \includegraphics[width=0.425\textwidth]{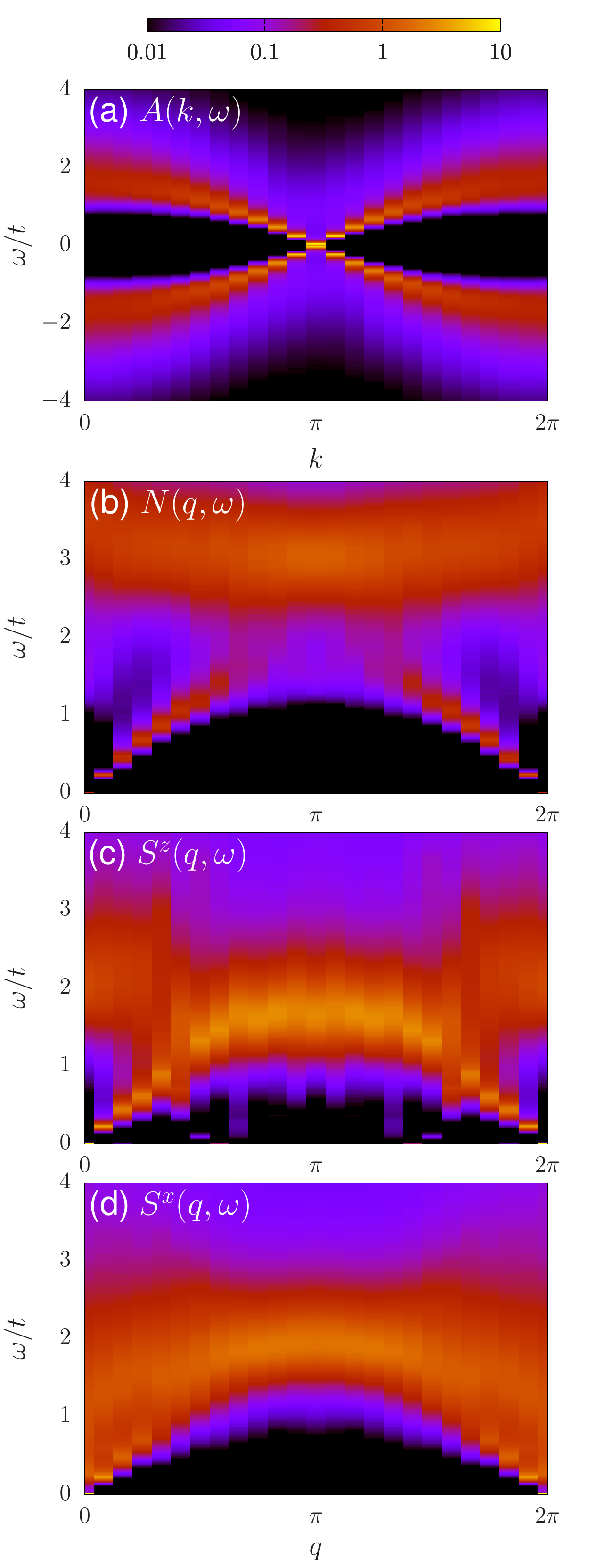}
  \caption{\label{fig:dynamicsU2} 
    Dynamical correlation functions for (a) single-particle, (b) charge, (c)
    longitudinal spin, and (d)  transverse spin excitations.  Results
    obtained from zero-temperature quantum Monte Carlo simulations of the KMH
    model on a $24\times64$ zigzag ribbon with a Hubbard $U$ only at the edge sites
    \cite{Hohenadler10,Ho.As.11}. Here $U/t=2$ and $\lso/t=0.2$.
  }
\end{figure}

The bosonization predictions~(\ref{Luttinger_liquid_corr.eq}) have been
tested numerically by considering the quantum spin Hall phase of microscopic,
interacting models. To verify the characteristic interaction-dependent
power-law decay of transverse spin and pairing correlation functions, a
hallmark feature of one-dimensional liquids predicted by
(\ref{Luttinger_liquid_corr.eq}), rather large system sizes are
required. Existing work along these lines is based on quantum Monte Carlo
simulations of the KMH model \cite{Zh.Wu.Zh.11}, and of an effective model
that takes into account electron correlations only at the edge
\cite{Hohenadler10,Ho.As.11}. This approach is justified by the minor role of
bulk interactions in the topological band insulator phase, as discussed in
section~\ref{sec:bulk}. The effective model allows one to access large system
sizes without any further approximations. (The results shown in
\cite{Hohenadler10,Ho.As.11} were obtained with an incorrectly implemented
bulk Green function, and are therefore not for the KM model but for a
different model of a quantum spin Hall insulator. However, using the
KM Green function leads to very similar
results, and none of the conclusions are affected, see errata of
\cite{Hohenadler10,Ho.As.11}. In the following, figure~\ref{fig:dynamicsU2}
shows results for the KM model with edge interactions, whereas
figures~\ref{fig:magneticorder_edge},~\ref{fig:luttinger},~\ref{fig:drude}
correspond to previously obtained results for the alternative model.)

Results for $S^{xx}(r)$ are reproduced in
figure~\ref{fig:magneticorder_edge}(a) \cite{Ho.As.11}. For a local Hubbard
interaction, the forward scattering matrix element $g$ depends on $U$,
whereas the Fermi velocity is determined by the spin-orbit gap and hence by
$\lso$. The effective interaction strength for the edge states is therefore
set by the ratio $U/\lso$ \cite{Ho.As.11}.
Figure~\ref{fig:magneticorder_edge}(a) reveals that transverse spin
correlations $S^{xx}(r)$ decay increasingly slower with increasing $U/\lso$,
corresponding to a reduction of $K$ in (\ref{Luttinger_liquid_corr.eq}). All
other correlation functions decay much faster \cite{Ho.As.11}. Similar
results have been obtained in a full bulk calculation of the KMH model
\cite{PhysRevB.83.205122,Zh.Wu.Zh.11}.  Despite the strongly enhanced
transverse spin correlations, the Luttinger model predicts the absence of
long-range order. This prediction is consistent with the finite-size scaling
of the spin structure factor shown in
figure~\ref{fig:magneticorder_edge}(b). Figure~\ref{fig:magneticorder_edge}(c)
confirms the doping dependence of $S^{xx}(r)$ by showing its Fourier
transform (the spin structure factor) at different particle densities. Away
from half filling ($n=1$, $\kF=\pi$), the $\cos(2\kF r)$ factor in
(\ref{Luttinger_liquid_corr.eq}) becomes visible.  From the numerical
results, and given the absence of logarithmic corrections to
(\ref{Luttinger_liquid_corr.eq}), the Luttinger parameters $K$ and $v$ can be
determined using finite-size scaling. The results in figure~\ref{fig:luttinger} demonstrate a
strong suppression of $K$ from its noninteracting value $K=1$, and
a slight renormalization of the velocity in agreement with
\cite{Yu.Xie.Li.11,Wu.Ra.Li.LH.11}.  The parameter $K$ has also been
determined from quantum Monte Carlo simulations of the KMH model by fitting
power laws to correlation functions at a fixed system size
\cite{Zh.Wu.Zh.11}, see figure~\ref{fig:edgephasediagram}. The
renormalization of the velocity $v$ was also studied in \cite{Yu.Xie.Li.11}.

While the numerical value of $K$ for given microscopic parameters $U$ and
$\lso$ will depend on the details of the model, repulsive interactions
generically lead to dominant $2\kF$ transverse spin fluctuations on helical
edges
\cite{Wu06,Cenke06,Hohenadler10,Ho.As.11,Zh.Wu.Zh.11,PhysRevLett.107.166806}. In
the light of the bulk results discussed in section~\ref{sec:bulk}, transverse
spin correlations are therefore a hallmark feature of correlated topological
insulators. This finding is expected to hold even in the presence of Rashba
coupling, as long as interactions are not too strong (see below). Although
mean-field theory is not appropriate to describe correlated edge states, it
does provide some insight into the differences between bulk and edge
correlations. For a system with periodic boundaries, mean-field theory gives
a finite critical value $U_\text{c}$ at which the topological band insulator
undergoes a quantum phase transition to a magnetic Mott insulator
\cite{RaHu10}, see inset of figure~\ref{fig:tbi-afmi-gaps}.  In contrast, a
system with edges is unstable toward the opening of a gap due to a
nesting-related, logarithmically diverging susceptibility, and shows a gapped
magnetic state with long-range order, $\las S^{xx} \ras\neq 0$, for any
finite $U$ \cite{PhysRevLett.107.166806}.  This so-called edge magnetism has
been studied in the context of graphene
\cite{PhysRevB.82.161302,PhysRevLett.106.226401,PhysRevB.83.195432,Schmidt2012}. Lee
\cite{PhysRevLett.107.166806} has shown that for weak enough interactions,
taking into account fluctuations around the mean-field saddle point restores
the expected power-law decay at the edges. Although long-range order is
absent, exact numerical results for the KMH model expose that magnetic
correlations are still strongest at the edge, and decay quickly with the
distance from the edge \cite{Zh.Wu.Zh.11}.

The effective model with interactions only at the edge sites provides a
faithful description of the low-energy physics inside the quantum spin Hall phase
\cite{Hohenadler10,Ho.As.11}. In addition to capturing the power-law decay of
correlations expected from Luttinger liquid theory, this approach also
permits one to calculate dynamical correlation functions on large
systems. Figure~\ref{fig:dynamicsU2}(a) shows the single-particle spectral
function of the KMH model with edge Hubbard interactions on a zigzag ribbon
at $\lso/t=0.2$ and $U/t=2$.  As in the noninteracting case
(figure~\ref{fig:kmedgestates}(a)), there is a gapless crossing point at
$k=\pi$.  The absence of any features related to spin-charge separation is
a generic feature of one-dimensional systems with spin-orbit coupling
\cite{PhysRevLett.84.4164}.  The single-particle spectrum of a correlated
helical liquid has also been obtained with cluster dynamical mean-field
theory on zigzag and armchair edges \cite{Wu.Ra.Li.LH.11}, see
figure~\ref{fig:edgestates-armchair},
for the KMH model
using the variational cluster approach \cite{Yu.Xie.Li.11}, and for the BHZH
model \cite{arXiv:1202.3203,Wa.Da.Xi.12,PhysRevLett.108.156402}. The
real-space Green function of the helical Luttinger model has been calculated
analytically \cite{PhysRevB.85.035136}.

Whereas dynamical two-particle correlation functions are not reliably
accessible with cluster methods \cite{Yu.Xie.Li.11,Wu.Ra.Li.LH.11} and
difficult to obtain within bosonization, they can be calculated with high
momentum resolution using quantum Monte Carlo simulations of the effective
model \cite{Hohenadler10,Ho.As.11}. Figure~\ref{fig:dynamicsU2} shows results
for the KMH model with edge Hubbard interactions. The charge structure factor
in figure~\ref{fig:dynamicsU2}(b) and the longitudinal spin structure factor
in figure~\ref{fig:dynamicsU2}(c) are characterized by a linear mode at long
wavelengths that has the same velocity as the edge state visible in the
single-particle spectrum shown in figure~\ref{fig:dynamicsU2}(a).  The
corresponding spectral weight is much smaller in the charge channel than in
the spin channel. The linear mode is related to spin-conserving excitations
within the helical states, whereas excitations at higher energies involve
bulk states. The transverse spin structure factor shown in
figure~\ref{fig:dynamicsU2}(d) involves spin flips---connecting the two
helical edge modes---and therefore exhibits a continuum of low-energy
excitations near $q=0$ \cite{Ho.As.11}.

As confirmed by the explicit numerical calculations discussed in this
section, for weak enough interactions, the fixed point described by the forward
scattering model~(\ref{eq:HLL}) qualitatively describes the physics of
helical edges. As long as $K$ does not fall below the respective critical
values to be discussed below, the model remains valid even for systems with
nonconserved spin \cite{PhysRevLett.108.156402}, disorder
\cite{KaMe05b,Wu06}, or magnetic impurities \cite{Wu06,Maciejko09,Maciejko09}.

\subsection{Strongly correlated regime}\label{sec:edge:correl}

\begin{figure}[t]
  \centering
  \includegraphics[width=0.45\textwidth]{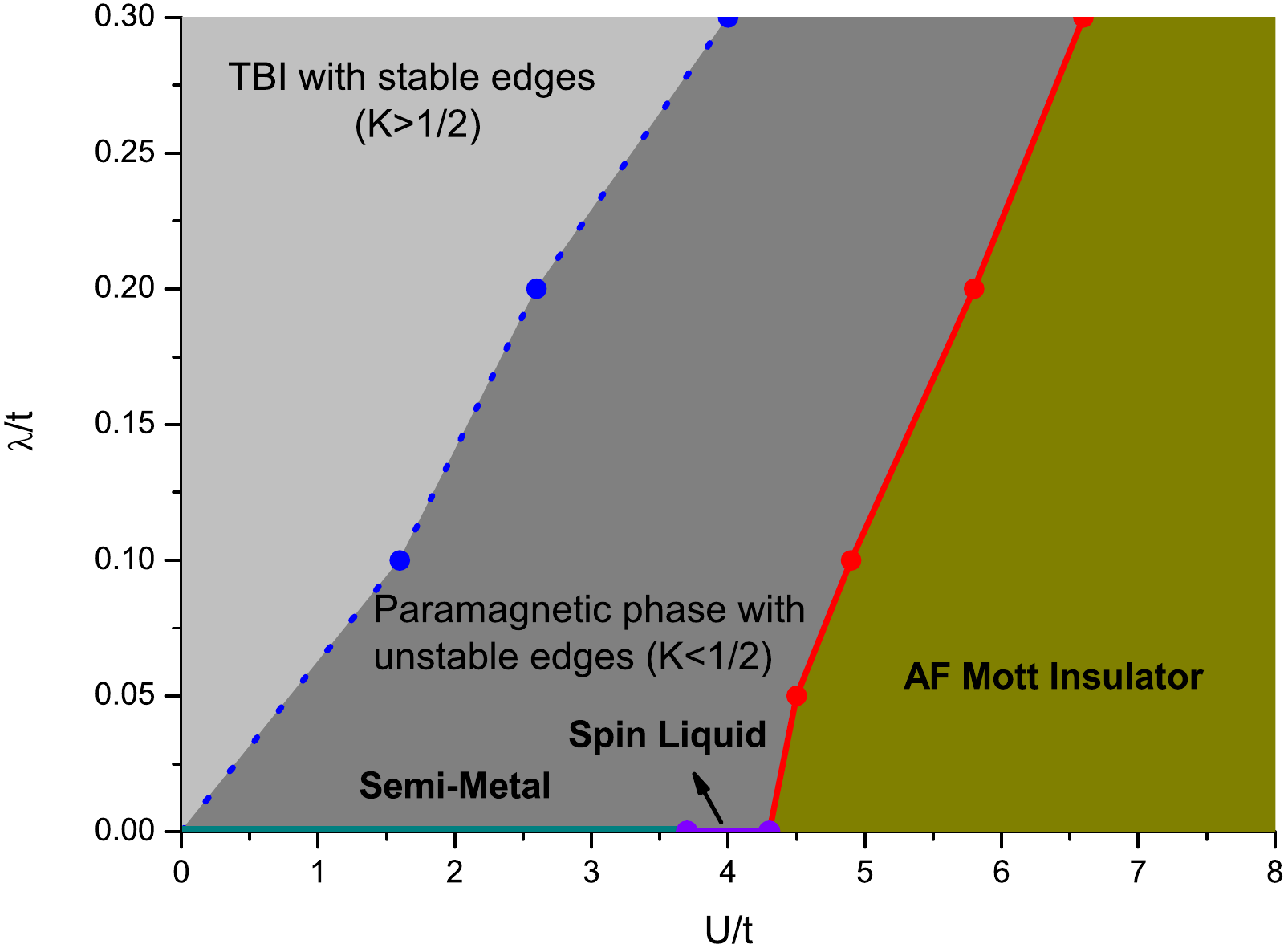}
  \caption{\label{fig:edgephasediagram} Phase diagram of the half-filled KMH
    model with spin-orbit coupling $\lambda\equiv\lso$ and Hubbard
    interaction $U$ from quantum Monte Carlo simulations
    \cite{Zh.Wu.Zh.11}. Also shown are regions where the Luttinger parameter
    $K<1/2$ and $K>1 /2$, respectively.  (Reprinted with permission from
    \cite{Zh.Wu.Zh.11}. Copyright 2011 by the American Physical Society).  }
\end{figure}

The robustness of helical edge states with respect to interactions is a
result of time-reversal symmetry. At the time-reversal invariant momenta
$k=0$, $\pi$, the two edge modes necessarily have the same energy,
$\epsilon_\UP(k)=\epsilon_\DO(k)$, so that the system is gapless. A gap in
the edge states can arise from a breaking of time-reversal symmetry either
globally (for example, due to spontaneous symmetry breaking and the onset of
magnetic order in the bulk, see section~\ref{sec:bulk}), or only at the
edge. In the latter case, the bulk remains a topological insulator, and
the bulk-boundary correspondence breaks down
\cite{PhysRevB.74.195312,PhysRevB.74.045125,PhysRevB.83.085426}. This section
explores the conditions for the occurrence of gapped edge states (such a gap
is assumed to be small compared to the bulk band gap), and reviews
the consequences of strong interactions---compared to the bulk gap---in
systems that remain gapless for symmetry reasons. Quite generally, a gap in the
edge states requires strong interactions, as measured by the Luttinger liquid
parameter $K$.  Evidence for strong interactions ($K\ll1$) for a wide range
of microscopic model parameters comes from numerical simulations
\cite{Ho.As.11,Zh.Wu.Zh.11}, see figures~\ref{fig:luttinger} and
\ref{fig:edgephasediagram}.

The impact of strong interactions can be analyzed by combining bosonization
with the renormalization group method \cite{Cenke06,Wu06}. Whereas forward
scattering does not open a gap, backscattering terms become relevant for
strong enough interactions \cite{Wu06,Cenke06}. The umklapp term, important
at half filling and for systems without spin conservation, becomes a relevant
perturbation for $K<1/2$
\cite{Wu06,Cenke06,PhysRevLett.107.166806}. Backscattering due to quenched
disorder leads to a gap if $K<3/8$ \cite{Wu06,Cenke06}, whereas
backscattering involving magnetic impurities becomes relevant for $K<1/4$
\cite{Maciejko09,PhysRevB.85.245108}. Finally, a gap in the edge states can
also open in the presence of a lattice of magnetic impurities when $K<1/2$
\cite{PhysRevB.85.245108}, and as a result of inter-edge tunneling and/or
umklapp processes
\cite{PhysRevLett.103.166403,PhysRevLett.107.166806,arXiv:1202.3203}.
The helical liquid hence shows remarkable
differences compared to ordinary Luttinger liquids. In the latter, if
$K_\rho<1$, transport is completely blocked at $T=0$ by either disorder
\cite{PhysRevLett.68.1220} or a single magnetic impurity (the Kondo problem)
\cite{PhysRevLett.72.892}. In contrast, a helical liquid is unconditionally
stable with respect to single-particle backscattering \cite{KaMe05b}, and can
avoid the site with the Kondo singlet by penetrating into the bulk provided
$K>1/4$ \cite{Maciejko09,PhysRevB.85.245108}.  

\begin{figure}[t]
  \centering
  \includegraphics[width=0.45\textwidth]{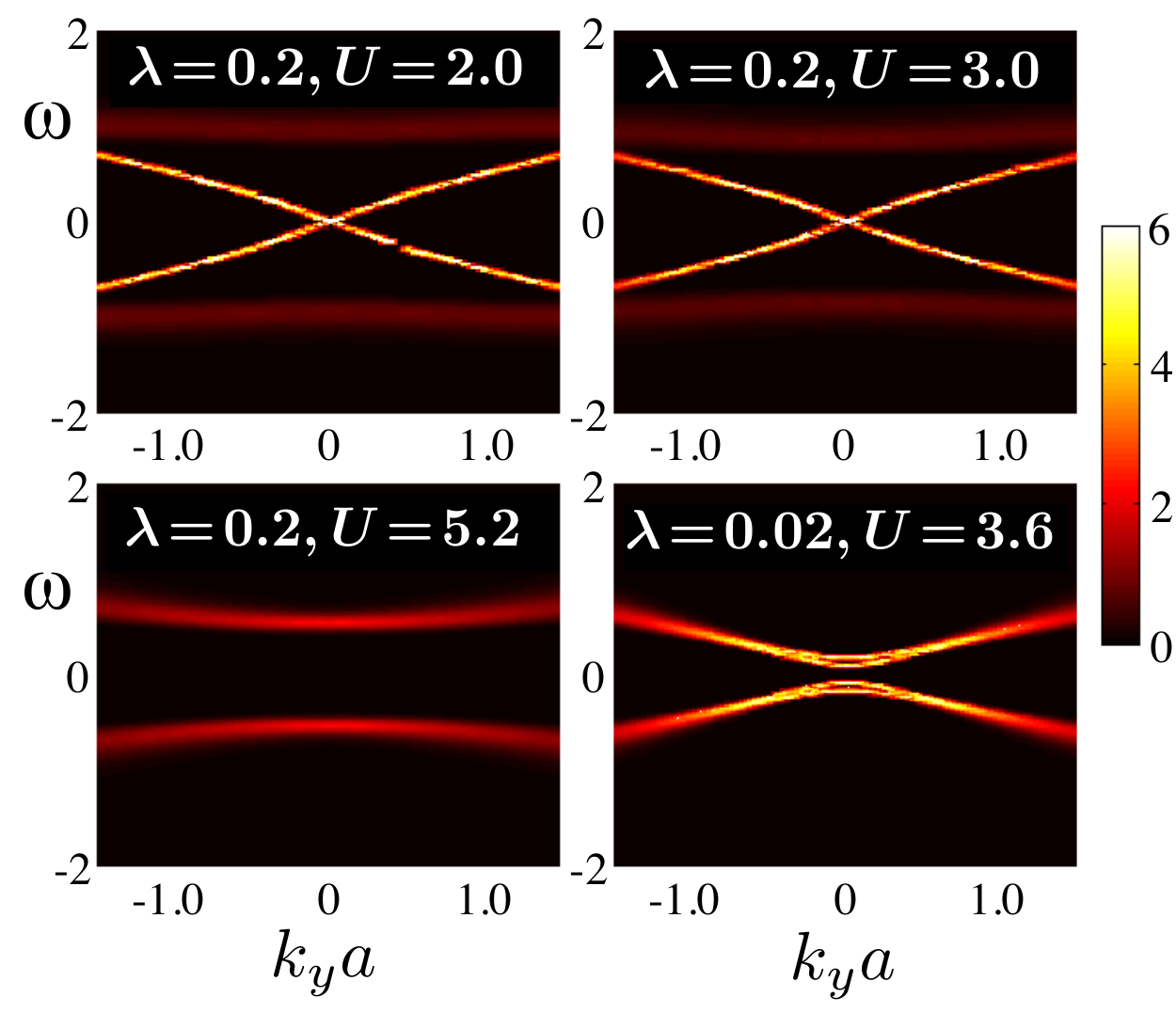}
  \caption{\label{fig:edgestates-armchair}
  Single-particle spectral function of the KMH model on an armchair ribbon
  \cite{Wu.Ra.Li.LH.11} with spin-orbit coupling $\lambda\equiv\lso$ and
  Hubbard interaction $U$. Results were obtained with the cluster dynamical mean-field
  theory. The top row corresponds to the quantum spin Hall phase, whereas the
  bottom row shows the spectrum in the magnetic (left) and the nonmagnetic
  insulator (right), respectively.
  (Reprinted with permission from \cite{Wu.Ra.Li.LH.11}. Copyright 2012
      by the American Physical Society).
  }
\end{figure}

Because of the complexity of the many-body problems with either additional
Rashba interaction, impurities or disorder, the above predictions have not
yet been verified by simulations of microscopic models. Nevertheless,
significant progress in the understanding of correlated helical edges has
been made.  The simplest scenario is the simultaneous breaking of
time-reversal symmetry in the bulk {\it and} at the edge, as realized, for
example, in the KMH model at large $U/t$ (section~\ref{sec:trans}). Because
time-reversal symmetry is broken in the magnetic phase, a topologically
trivial state without gapless edge states is expected. The evolution of the
edge states across the magnetic bulk transition can only be studied in a
model with an interacting bulk. Hence, the effective model of
\cite{Hohenadler10,Ho.As.11} is insufficient. Instead, this problem has
been studied in the framework of the KMH model by Wu \etal
\cite{Wu.Ra.Li.LH.11} using cluster dynamical mean-field theory, and by Yu
\etal \cite{Yu.Xie.Li.11} using the variational cluster approach. Although
these results do not fully capture the Luttinger liquid properties of the
edge states, the results agree qualitatively with expectations. 
 Figure~\ref{fig:edgestates-armchair} shows the
single-particle spectral function for an armchair edge as obtained by Wu
\etal \cite{Wu.Ra.Li.LH.11}. For the value $\lso/t=0.2$ considered, the
critical value for the bulk magnetic transition within their approximation is
$\Uc/t\approx5.2$, which seems consistent with the observed opening of a gap in the edge
states \cite{Wu.Ra.Li.LH.11}. The low-energy model for the gapped edge states
can be written as $H_\text{HL}-\tilde{g} \, m \sin(\sqrt{4\pi}\phi)$
(cf.~(\ref{eq:HLL})), where $m$ is the bulk magnetization that, if nonzero,
leads to a relevant perturbation for repulsive interactions
\cite{Wu.Ra.Li.LH.11}.  A mean-field treatment with additional spin-wave
fluctuations has been given by Lee \cite{PhysRevLett.107.166806}.

To observe a topological insulator with Mott-insulating edges, time-reversal
symmetry has to be broken at the edge. The observed tendency toward strong
magnetic correlations follows already from the simple model~(\ref{eq:HLL}),
which predicts long-ranged but power-law transverse spin correlations for
$K\ll 1$, see (\ref{Luttinger_liquid_corr.eq}). The exponent $K$ has been
calculated numerically for different (spin-conserving) models
\cite{Ho.As.11,Zh.Wu.Zh.11}, and it was found that it can be substantially
smaller than 1, see figures~\ref{fig:luttinger}
and~\ref{fig:edgephasediagram}. However, as shown in
figure~\ref{fig:magneticorder_edge}, and in accordance with
(\ref{Luttinger_liquid_corr.eq}), true long-range order is absent even for
very strong interactions \cite{Ho.As.11}. Assuming that the edge states can
be considered to be strictly one-dimensional, the absence of long-range order is implied by the
Mermin-Wagner theorem, which forbids continuous symmetry breaking in one
dimension. However, the continuous $U(1)$ spin symmetry can be reduced to a
discrete $Z_2$ Ising symmetry in the presence of Rashba spin-orbit coupling
in (\ref{eq:KMH}) or~(\ref{eq:BHZ}),
or in more general models such as (\ref{eq:SI}). In this case, umklapp
processes of the form~(\ref{eq:um}) become possible, and give rise to
Ising long-range magnetic order at $T=0$ if $K<1/2$
\cite{Wu06,Cenke06,PhysRevLett.107.166806}. However, at $T>0$, time-reversal
symmetry will be restored by thermal fluctuations. This mechanism permits the
existence of a topologically nontrivial quantum spin Hall insulator with
gapped edge states and time-reversal symmetry, and causes the bulk-boundary
correspondence to break down
\cite{PhysRevB.74.195312,PhysRevB.74.045125,PhysRevB.83.085426}.
Temperature-dependent edge and bulk magnetism in a quantum spin Hall
insulator has been discussed by Shitade \etal \cite{Irridates-Nagaosa}.  The
possibility of umklapp-driven magnetic order at the edge has been taken into
account by Zheng \etal \cite{Zh.Wu.Zh.11} in their phase diagram of the KMH
model. The latter, shown in figure~\ref{fig:edgephasediagram}, depicts a
rather large region where $K<1/2$, although the simulations were done for
$\lr=0$. (The gap in the edge Green functions in \cite{Zh.Wu.Zh.11} is a
finite-size effect.) To determine $K$ experimentally it may be useful to
exploit the dependence of the single-particle gap induced by a magnetic
field on the electron-electron interaction strength
\cite{PhysRevB.86.165121}.

Based on the above arguments, and for models with conserved spin, the edge
states are expected to become gapped exactly at the point where long-range
magnetic order in the bulk breaks time-reversal symmetry and hence destroys
the topological state. A slight complication is that the helical liquid is
holographic, and hence exists only at the boundary of a two-dimensional
system \cite{Wu06}, whereas the absence of continuous symmetry breaking
applies only to strictly one-dimensional systems. The two-dimensional bulk is
not taken into account in the Luttinger model~(\ref{eq:HLL}), but is included
in the effective model with interactions only at the edge; no evidence for
long-range order has been found (see figure~\ref{fig:magneticorder_edge})
\cite{Ho.As.11}. For graphene edges, it has been argued that the
Mermin-Wagner theorem is invalidated by the momentum dependence of
electron-electron interactions, and numerical results are consistent with
long-range magnetic order \cite{PhysRevB.83.195432,Schmidt2012}.

The difficulties associated with capturing large length scales
(cf. figure~\ref{fig:magneticorder_edge}) and quantum fluctuations, as well
as with the unambiguous detection of topological states, have led to several
claims for the existence of edge Mott insulators (or edge superconductors) in
systems with conserved spin. Paramagnetic, inhomogeneous slave-rotor
calculations for the KMH and the BHZH model suggest the existence of a
paramagnetic edge-Mott state for large values of $\lso$
\cite{PhysRevB.85.235449}. Similar conclusions have been drawn based on
variational Monte Carlo simulations of the KMH model
\cite{PhysRevB.83.205122}. Finally, a symmetry-breaking superconducting state
has been reported for the KMH model with attractive interactions
\cite{Jieetal2012} based on a mean-field calculation. In two dimensions, the
slave-rotor result \cite{PhysRevB.85.235449} is known to be unstable with
respect to gauge fluctuations \cite{RaHu10}. A similar but stable phase (a
topological Mott insulator) has been predicted based on slave-boson
calculations \cite{We.Ka.Va.Fi.11}. The variational Monte Carlo results for
the spin and charge Drude weights \cite{PhysRevB.83.205122} have rather large
error bars, and the authors interpret them either as a transition or a
crossover. The latter possibility agrees with other numerical work
\cite{Hohenadler10,Ho.As.11,Wu.Ra.Li.LH.11}, see discussion below.

The existence of a superconducting edge state can be excluded by exploiting
the exact particle-hole symmetry of the KMH Hamiltonian, and the canonical
transformation $c^\dag_{i\UP}\mapsto (-1)^i c^\nag_{i\DO}$, with
$(-1)^i=\pm1$ depending on the sublattice. Similar to the Hubbard model, this
transformation exchanges the role of spin-$z$ and charge operators in the
Hamiltonian, and flips the sign of the Hubbard term
\cite{PhysRevLett.66.3203}.  The dominant transverse spin correlations
observed in the repulsive model translate into dominant pairing correlations
$P(r)$ in the attractive model. However, all correlation functions decay with
a power law, as can be explicitly confirmed by quantum Monte Carlo
simulations for the attractive case \cite{HoAsunpublished}. The symmetry breaking observed in
\cite{Jieetal2012} at the mean-field level hence has the same origin as
the long-range magnetic order in repulsive model, namely a logarithmically
divergent susceptibility \cite{PhysRevLett.107.166806}.

\begin{figure}[t]
  \centering
  \includegraphics[width=0.45\textwidth]{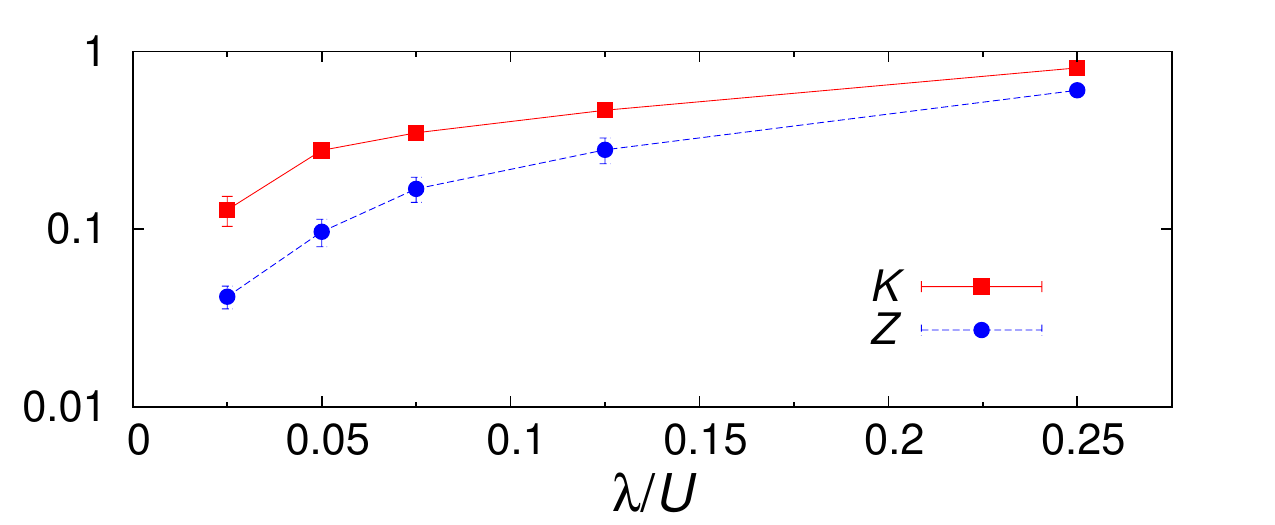}
  \caption{\label{fig:drude}
    Luttinger liquid parameter $K$ and spectral weight of
    long-wavelength charge excitations $Z$ from quantum Monte Carlo simulations
    of an effective model for helical edge states. 
    (Reprinted with permission from \cite{Ho.As.11}. Copyright 2012
      by the American Physical Society).
  }
\end{figure}

Although no gap is opened in quantum spin Hall insulators with conserved
spin, strong interactions lead to important modifications of the edge states.
Apart from a renormalization of the velocity $v$ with increasing $U/\lso$
that follows from the low-energy model~(\ref{eq:HLL}), see
figure~\ref{fig:luttinger}, numerical studies indicate a
suppression of charge transport, visible for example in the dynamical charge
structure factor at long wavelengths and low energies in
figure~\ref{fig:dynamicsU2}(b), or in the Drude weight
\cite{PhysRevB.83.205122}, with a simultaneous enhancement of spectral weight
for transverse spin excitations
\cite{Hohenadler10,Ho.As.11,PhysRevB.83.205122}. Similarly, the spectral
weight of the edge states in the single-particle spectrum is diminished by
strong interactions \cite{Hohenadler10,Ho.As.11,Wu.Ra.Li.LH.11}.

The Drude weight, $D$, can be related to the spectral weight, $Z$, of the
linear mode in the dynamical charge structure factor via $D=Zv$ ($v$ being
the velocity); $Z$ can be measured in quantum Monte Carlo simulations
\cite{Ho.As.11}, and results are shown in figure~\ref{fig:drude}. On the
other hand, for a Luttinger liquid described by (\ref{eq:HLL}), the Drude
weight is given by $D=Kv$ \cite{Ho.As.11}, so that the apparent difference
between $Z$ and $K$ in figure~\ref{fig:drude} indicates that the
model~(\ref{eq:HLL}) is incomplete. While $Z=K$ in the noninteracting limit,
the Drude weight is significantly smaller than predicted theoretically (\ie,
smaller than $D=Kv$) for nonzero interactions.  The discrepancy has been
attributed to inelastic spin-flip scattering mediated by the pronounced
magnetic fluctuations that exist for $K\ll1$, and to scattering processes
that involve bulk states \cite{Ho.As.11}.  Whereas  bulk effects
are expected if the interaction $U$ becomes comparable or larger
than the bulk band gap, the numerical results suggest that deviations occur
even for rather weak interactions \cite{Ho.As.11}.

\section{Topological invariants of correlated systems}\label{sec:index}

The discovery of the remarkable relation between the quantized Hall
conductivity and the first Chern number in the context of the quantum Hall
effect \cite{PhysRevLett.49.405,PhysRevLett.51.51,Kohmoto1985343} has played
a key role for the understanding of topological phases of matter. In
section~\ref{sec:noninteracting}, it has been shown that a topological
invariant can also be defined for quantum spin Hall insulators, and that the
latter is---in the simplest case---determined by the Chern number difference
of two quantum Hall states with opposite Hall conductivities
\cite{PhysRevLett.97.036808}.  Reviews of topological invariants with a focus
on noninteracting systems can be found in
\cite{HaKa10,RevModPhys.83.1057,Bu.Tr.12}.  Here the emphasis is on the
calculation of topological invariants for correlated systems by means of
numerical methods. More general discussions can be found, for example, in
\cite{arXiv:1207.7341,Bu.Tr.12}.

Whereas the direct calculation of the topological invariant of interacting
systems can be rather difficult, its value can in some cases be inferred
indirectly either via the bulk-boundary correspondence, or based on an
adiabatic connection to a noninteracting system.

Because time-reversal symmetry protects topological insulators against weak
interactions, the bulk-boundary correspondence can in principle be used to
determine whether or not a topological state survives upon switching on
electron-electron interactions. However, this approach, followed for example
in \cite{Yu.Xie.Li.11,Wu.Ra.Li.LH.11,Wa.Da.Xi.12}, faces a number of
difficulties. In addition to the inherent difficulty of obtaining numerical
results on large enough systems with open boundaries to demonstrate the
existence of gapless excitations, correlation effects and electron tunnelling
between edges can gap out the edge states despite the existence of
topological order in the bulk (see section~\ref{sec:edge:correl}). Although
gapped edge states can be reconciled with a nonzero topological invariant
theoretically
\cite{PhysRevB.74.195312,PhysRevB.74.045125,PhysRevB.83.085426}, a
distinction of topological and trivial states based on edge states alone will
be challenging in actual numerical calculations for strongly correlated
systems. Nevertheless, the bulk-boundary correspondence is a valuable
theoretical concept. For example, it has been used to formulate criteria for
the stability \cite{PhysRevLett.103.196803,PhysRevB.84.165107} and the
experimental identification \cite{PhysRevLett.108.206804} of fractional
topological insulators. The existence of gapless edge states can also
be inferred from the theory of symmetry-protected topological states
\cite{PhysRevB.82.155138,arXiv:1106.4772,PhysRevB.84.235141,Chen2012,arXiv:1209.4399,arXiv:1212.1726}.

A topological phase may also be identified by exploiting an adiabatic
connection to a noninteracting state (for which the invariant can easily be
calculated).  For example, as discussed in section~\ref{sec:corrTBI}, the KMH
model with $\lr=0$ describes a quantum spin Hall state for $U=0$ and
$\lso\neq0$. For small to intermediate values of $U/t$, the system remains
fully gapped and time-reversal invariant, and can be adiabatically tuned (via
$U/t$) into the noninteracting quantum spin Hall state without closing the
band gap, see Fig.~\ref{fig:tbi-afmi-gaps}. This argument establishes the topological character of the state
at $U>0$. As emphasized in section~\ref{sec:bulk:fqhe}, no such adiabatic
connection to a noninteracting state exists, for example, in the case of
fractional topological insulators or topological Mott insulators.  Finally,
even using exact numerical methods, it is not always possible to
decide if an excitation gap closes or not across a transition
\cite{Ho.Me.La.We.Mu.As.12}.

Given the limitations of these indirect probes for topological states, it is
important to be able to directly calculate the topological invariant.  For
spin-conserved quantum spin Hall insulators, it is sufficient to determine
the Chern number for one spin sector. Because the Chern number is related to
the quantized Hall conductivity, it can in principle be obtained without
approximations from the Kubo formula
\cite{PhysRevB.31.3372,RevModPhys.82.1959}. The $Z_2$ invariant then follows
from (\ref{eq:Z2}). Nevertheless, for correlated electron systems, even the
calculation of the conductivity can be quite demanding.  When spin is not
conserved, the separation into two spin sectors is obviously no longer
possible, and other concepts such as the spin Chern number are required.  In
particular, even an exact calculation of the Hall conductivity via the Kubo
formula may be insufficient because $\sigma^\text{s}_{xy}$ is no longer quantized
\cite{PhysRevLett.97.036808}.

Considering first the integer, spin-conserved case, (\ref{eq:chern:1})
and~(\ref{eq:chern:2}) are not applicable to interacting systems, because they
rely on a representation in terms of Bloch states. Instead, the Chern number
in the presence of interactions and/or disorder can be calculated from
\cite{PhysRevB.31.3372}
\begin{equation}\label{eq:Ctwist}
  C =  \int\int_0^{2\pi}  \frac{\rmd\phi_x \rmd \phi_y}{2\pi\rmi} 
  \left[
    \la \frac{\partial \Psi_0} {\partial \phi_y}\right. \ket{\frac{\partial \Psi_0}{\partial \phi_x}}
    -
    (y \leftrightarrow x)
  \right]
  \,.
\end{equation}
Compared to (\ref{eq:chern:1}), the Bloch states $\ket{u_m}$ are replaced by
the many-body ground state $\ket{\Psi_0}$, and the momenta $k_x$, $k_y$ are
replaced by phase parameters $\phi_x$, $\phi_y$ describing general boundary
conditions $\Psi_0(\dots,\{x_i+L_x,y_i\},\dots)=e^{\rmi
  \phi_x}\Psi_0(\dots,\{x_i,y_i\},\dots)$ (and similar for the $y$ direction)
\cite{PhysRevB.31.3372}. On a torus geometry, the phase shifts $\phi_x$,
$\phi_y$ can be related to magnetic fluxes \cite{RevModPhys.82.1959}.
Assuming the existence of a bulk energy gap and a unique, nondegenerate
ground state, $C$ as defined by (\ref{eq:Ctwist}) is an integer Chern number
directly related to the Hall conductivity \cite{PhysRevB.31.3372}.  An
alternative representation is given by
\cite{PhysRevB.31.3372,RevModPhys.82.1959}
\begin{eqnarray}\label{eq:CHderiv}
  C
  =
  &
  \int\int_0^{2\pi} \frac{d\phi_x d\phi_y}{2\pi\rmi}\\\nonumber
  &\times
  \sum_{n>0} 
  \frac{
    \bra{\Psi_0} \frac{\partial H}{\partial \phi_y}  \ket{\Psi_n} 
    \bra{\Psi_n}  \frac{\partial H}{\partial \phi_x}  \ket{\Psi_0} 
    -
     (y \leftrightarrow x)
  }
  {(E_n - E_{0})^2}\,.
\end{eqnarray}
The derivatives of the Hamiltonian are numerically easier to evaluate than
derivatives of the wave functions \cite{RevModPhys.82.1959},
and~(\ref{eq:CHderiv}) reveals the relation to the Kubo formula
\cite{RevModPhys.82.1959}. For simplicity, the zero-temperature limit has
been taken. At finite temperatures, the Hall and spin Hall conductivities are
in general not exactly quantized, see for example
\cite{Is.Ma.87,arXiv:1111.6250}.

Equations~(\ref{eq:Ctwist}) and~(\ref{eq:CHderiv}) can be extended to the
case of a fractional quantum Hall state with  a $d$-dimensional ground-state
manifold $\{\ket{\Psi_K}\}$, $K=1,\dots,d$. The Chern number reads \cite{PhysRevB.31.3372}
\begin{eqnarray}\label{eq:Ctwistdeg}
  C = &
  \frac{1}{d} \int\int_0^{2\pi} \frac{\rmd\phi_x \rmd \phi_y}{2\pi\rmi}
  \\\nonumber
  &\quad\times \sum_{K=1}^d
  \left[
    \la \frac{\partial \Psi_K} {\partial \phi_y}\right. \ket{\frac{\partial \Psi_K}{\partial \phi_x}}
    -
    \la \frac{\partial \Psi_K} {\partial \phi_x}\right. \ket{\frac{\partial \Psi_K}{\partial \phi_y}}
  \right]
  \,.
\end{eqnarray}
The corresponding generalization of (\ref{eq:CHderiv}) to the fractional case
can be found, for example, in
\cite{Ko.Ve.Da.12}. Equations~(\ref{eq:Ctwist})--(\ref{eq:Ctwistdeg}) are
particularly useful in combination with exact diagonalization methods, and
recent applications in the context of correlated Chern insulators include
\cite{Va.Su.Ri.Ga.11,Wa.Sh.Zh.Wa.Da.Xi.10,PhysRevLett.108.126405,Ko.Ve.Da.12}.
Twisted boundary conditions can also be used in combination with quantum
Monte Carlo simulations \cite{Wa.Sh.Zh.Wa.Da.Xi.10}.

A generalization for integer quantum spin Hall insulators without spin
conservation is based on the Chern number matrix
\cite{PhysRevLett.91.116802,PhysRevLett.97.036808}
\begin{eqnarray}\label{eq:Ctwistspin}
  C^{\sigma\sigma'} = &
  \int\int_0^{2\pi} \frac{\rmd \phi^\sigma_x \rmd \phi^{\sigma'}_y }{2\pi\rmi} \\\nonumber
  &\qquad\times
  \left[
    \la \frac{\partial \Psi_0} {\partial \phi^{\sigma'}_y}\right. \ket{\frac{\partial \Psi_0}{\partial \phi^\sigma_x}}
    -
    \la \frac{\partial \Psi_0} {\partial \phi^\sigma_x}\right. \ket{\frac{\partial \Psi_0}{\partial \phi^{\sigma'}_y}}
  \right]
  \,,
\end{eqnarray}
with spin-dependent boundary twists $\phi^\sigma_x$, $\phi^{\sigma'}_y$. If
spin is conserved, $C^{\UP\DO}=C^{\DO\UP}=0$ whereas
$C^{\UP\UP}=C^{\DO\DO}=\pm1$.  Sheng \etal \cite{PhysRevLett.97.036808}
defined a total charge Chern number
$C^\text{c}=\sum_{\sigma\sigma'}C^{\sigma\sigma'}$ and a total spin Chern
number $C^\text{s}=\sum_{\sigma\sigma'}\sigma C^{\sigma\sigma'}$. In contrast
to $\sigma^\text{s}_{xy}$, which is only quantized if spin is conserved (in
this case, the calculation of $\sigma^\text{s}_{xy}$ and $C^\text{s}$ are
completely equivalent, see section~\ref{sec:noninteracting}), the spin Chern
number $C^\text{s}$ as defined in \cite{PhysRevLett.97.036808} remains
quantized throughout the topological phase of the KM model with $\lr\neq0$
and/or disorder \cite{PhysRevLett.97.036808}. Whereas Sheng \etal
\cite{PhysRevLett.97.036808} suggest that (\ref{eq:Ctwistspin}) permits one to
distinguish two different quantum spin Hall phases (with the sign of
$\sigma^\text{s}_{xy}$ depending on the sign of the spin-orbit coupling), Fu
and Kane \cite{PhysRevB.74.195312} as well as Fukui and Hatsugai
\cite{PhysRevB.75.121403} argue that this distinction is merely a boundary
effect, and that the state is fully characterized by a $Z_2$ index $\nu=1$.
The flux insertion related to the twisted boundary conditions can cause a
spurious closing of the band gap on small systems
\cite{PhysRevB.75.121403,PhysRevLett.100.186807}. Although~(\ref{eq:Ctwist})
and~(\ref{eq:Chern:GF}) are equivalent for the spin-conserved case, it is not
clear if the above definition of the spin Chern number can also be applied
(and remains quantized) to fractional topological insulators without spin
conservation \cite{PhysRevLett.106.236804}.

Equation~(\ref{eq:Ctwistspin}) provides a rather general way to calculate the
$Z_2$ index in correlated topological phases with time-reversal symmetry.
However, its numerical evaluation can be quite involved or even technically
impossible in the context of numerical approaches based on quantum Monte
Carlo or quantum cluster methods. Consequently, a lot of effort has been
devoted to developing simplified ways of calculating topological invariants.

The starting point for one of the most fruitful recent developments is to
express the Chern number in terms of the single-particle Green function
\cite{springerlink:10.1007/BF01410451,So85,Is.Ma.87,Volovik,PhysRevLett.105.256803,PhysRevB.83.085426},
\begin{eqnarray}\label{eq:Chern:GF}
  C = & 
  \frac{\epsilon^{\mu\nu\rho}}{6} \int \rmd k_0 \int \frac{\rmd k_x \rmd k_y}{(2\pi)^2} \\\nonumber
  &\qquad\times\tr \left[
  G \partial_{\mu} G^{-1} G \partial_{\nu} G^{-1} G \partial_{\rho} G^{-1}\right]\,,
\end{eqnarray}
with $k_0=\om$ (and $\om$ real), $\bm{k}=(k_x,k_y)$, and
$\partial_\mu\equiv \partial /\partial k_\mu$ etc. $G=G(\rmi\omega,\bm{k})$
is the frequency and momentum-dependent single-particle Green function (in
general a matrix in spin and orbital space), and $G^{-1}$ its matrix inverse
\cite{PhysRevB.83.085426}.  The trace is over the matrix indices of $G$ and
$G^{-1}$, and a summation over $\mu$, $\nu$, and $\rho$ is implied.
Equation~(\ref{eq:Chern:GF}) can be applied to spin-conserving quantum spin
Hall systems by calculating $C$ in each spin sector, see for example
\cite{arXiv:1111.6250,arXiv:1207.4547v1}. In this context, stochastic
integration methods have been employed to sample the strongly varying
integrands \cite{Wa.Ji.Da.Xi.11}, and an algorithm that preserves gauge
invariance and the quantization of the Chern number even on a discretized
Brillouin zone has been given \cite{Fukui_Chern}. A simplification of~(\ref{eq:Chern:GF}),
valid within the local approximation of dynamical mean-field theory, has been
used to calculate $\sigma^\text{s}_{xy}$ for the BHZH model
\cite{arXiv:1111.6250}, see figure~\ref{fig:sigma_bhz}.  The $Z_2$ invariant
of general, time-reversal invariant insulators can also be expressed in terms
of the Green function; it is related to the electromagnetic polarization, and
involves a five-dimensional integral \cite{PhysRevLett.105.256803}.

Equation~(\ref{eq:Chern:GF}) can in principle be used directly to calculate
the Chern number. Moreover, the formulation in terms of the single-particle
Green function provides valuable insights into the bulk-boundary
correspondence and interaction effects by analyzing the properties of $G$
itself \cite{Volovik,PhysRevB.83.085426,PhysRevB.84.125132}.  Most
importantly, (\ref{eq:Chern:GF}) implies that the Chern number can only
change if a singularity appears in either $G$ or $G^{-1}$
\cite{PhysRevB.83.085426,PhysRevB.84.125132}. Considering first a
noninteracting system, the relation $G(\rmi\om,\mathbf{k})=[\rmi \om -
H(\mathbf{k})]^{-1}$ reveals that, for $\om$ on the real axis, poles can only
occur at $\om=0$, where they are related to zero eigenvalues of $H$ and hence
to zero-energy states. The change of $C$ in this case is hence related to
either a closing of the bulk band gap, or to an interface with a
topologically different system where the zero-energy states correspond to
edge states. In this way, the Green function can be used to derive the
bulk-boundary correspondence principle \cite{PhysRevB.84.125132}.

In interacting systems, poles in the self-energy and hence in $G^{-1}$
provide another mechanism for a change of the Chern number
\cite{PhysRevB.83.085426,arXiv:1203.2928}, see~(\ref{eq:Chern:GF}). However,
as pointed out in \cite{PhysRevB.83.085426,PhysRevB.84.125132}, such {\it
  Green function zeros} are in general not related to a gap closing or to the
existence of edge states. Consequently, the bulk-boundary correspondence can
break down in the presence of interactions
\cite{PhysRevB.83.085426,PhysRevB.84.125132}. Although intuitive, this
picture based on the single-particle Green function may not capture
interaction-generated collective excitations \cite{PhysRevB.83.085426}, such
as gapless spinon excitations in a topological Mott insulator \cite{PeBa10}.

Whereas~(\ref{eq:Chern:GF}) is exact for noninteracting systems, its physical
meaning is not yet fully understood in the interacting case.  As discussed in
\cite{PhysRevB.83.085426,Bu.Tr.12}, for interacting systems,
(\ref{eq:Chern:GF}) relies on an adiabatic connection between the interacting
state and a noninteracting, topological band insulator. It is therefore in
general not applicable to interaction-driven states such as topological Mott
insulators and fractional topological insulators. Although
(\ref{eq:Chern:GF}) remains a quantized, topological quantity even in the
presence of interactions \cite{PhysRevB.83.085426}, its relation to physical
observables such as $\sigma_{xy}^\text{s}$ for general interacting states is
an interesting open question
\cite{PhysRevB.83.085426,PhysRevB.84.125132,Bu.Tr.12,Gurarie2013}. On the other hand,
(\ref{eq:Chern:GF}) can be extended to disordered systems by replacing the
physical momenta with phases representing twisted boundary conditions
\cite{Bu.Tr.12}.  The Green function has also been used to study topological
superconductors \cite{PhysRevB.86.165116,arXiv:1207.1104}, and correlated
topological states in one \cite{arXiv:1205.5095} and three dimensions
\cite{PhysRevLett.109.066401}.

The multiple integrals, and the required knowledge of the Green function and
its derivative, make the numerical evaluation of~(\ref{eq:Chern:GF}) rather
difficult. Starting from (\ref{eq:Chern:GF}), further simplifications have
been achieved by Wang and Zhang \cite{arXiv:1203.1028} and by Wang \etal
\cite{PhysRevB.85.165126}, and the single-particle Green function again plays
a key role. Wang and Zhang \cite{arXiv:1203.1028} showed that given an
adiabatic connection to a noninteracting state (see also \cite{Bu.Tr.12}),
the topological invariant can be obtained from the Green function at $\om=0$
only \cite{arXiv:1203.1028}, thereby simplifying practical calculations
significantly.  As shown in \cite{arXiv:1203.1028}, the Chern number can be
written in the form (\ref{eq:chern:1}), with the Berry flux defined over the
space of eigenvalues of the zero-frequency Green function. A related,
physically more intuitive concept is the so-called {\it topological
  Hamiltonian} defined as $\tilde{h}(\bm{k})=-G^{-1}(0,\bm{k})=h_0(\bm{k}) +
\Sigma(0,\bm{k})$ \cite{arXiv:1207.7341}, where $h_0(\bm{k})$ is the
noninteracting part of a many-body Hamiltonian, and $\Sigma(0,\bm{k})$ is the
zero-frequency part of the self-energy. Remarkably, provided there is an
adiabatic connection to a noninteracting system, the topological Hamiltonian
captures the topological invariant of the full, interacting problem
\cite{arXiv:1203.1028,arXiv:1203.1028}; see also
\cite{arXiv:1207.7341,Bu.Tr.12}. It therefore provides a highly nontrivial
extension of the fact that for a noninteracting system, where
$-G^{-1}(0,\bm{k})=H(\bm{k})$, the Green function inherits the symmetries
(time-reversal, particle-hole, and chiral symmetry) of the Hamiltonian, and
can hence be used for a topological classification \cite{PhysRevB.84.125132}.
An illuminating discussion of why the zero-frequency self-energy is
sufficient to calculate the topological invariant can be found in
\cite{arXiv:1207.7341}.  For practical applications, the power of the concept
of the topological Hamiltonian is related to the possibility of calculating
the Chern number by any method that is applicable to noninteracting
systems. For example, given inversion symmetry, the Green function has to be
determined only at the four time-reversal invariant momenta
\cite{PhysRevB.85.165126}. This method has been applied to the KMH model
\cite{arXiv:1203.2928} and the BHZH model \cite{arXiv:1211.3059}, and is
quite similar in spirit to the work of Fu and Kane \cite{FuKa07}, in which
the eigenvalues of the parity operator at the time-reversal invariant points
in the Brillouin zone determine the topological invariant.

In addition to this rather general approach, several other simplifications
have been proposed and used in combination with numerical methods.  The
calculation of topological invariants becomes even simpler if the self-energy
is local, \ie, independent of $\bm{k}$. This approximation underlies the
dynamical mean-field theory, and is expected to yield quite accurate results
in high dimensions \cite{Georges96}. Wang \etal \cite{PhysRevB.84.205116}
showed that a local self-energy that is diagonal in orbital space permits one
to write the Chern number as a product of the {\it frequency-domain winding
  number} and a Chern number of a mean-field Hamiltonian. The former is
related to fluctuation-driven topological phase transitions, whereas the
latter captures band structure effects. This idea has been successfully
applied to the KMH model \cite{arXiv:1203.2928}. For local self-energies that
are not diagonal in orbital space, a simplification can be achieved using a
pole expansion of the self-energy, and leads to an effective, noninteracting
Hamiltonian whose topological invariant can be shown to be the same as that
of the original, correlated model \cite{Wa.Ji.Da.Xi.11}. Applications of this
method can be found in
\cite{Wa.Ji.Da.Xi.11,Wa.Da.Xi.12,arXiv:1203.2928}. Interestingly, these
calculations predict interaction-driven transitions to topological phases
that are not captured by mean-field theory and hence related to quantum
fluctuations \cite{Wa.Da.Xi.12,arXiv:1203.2928}. A drawback for
applications to two-dimensional systems is the complete neglect of spatial
correlations.

\begin{figure}[t]
  \centering
  \includegraphics[width=0.5\textwidth]{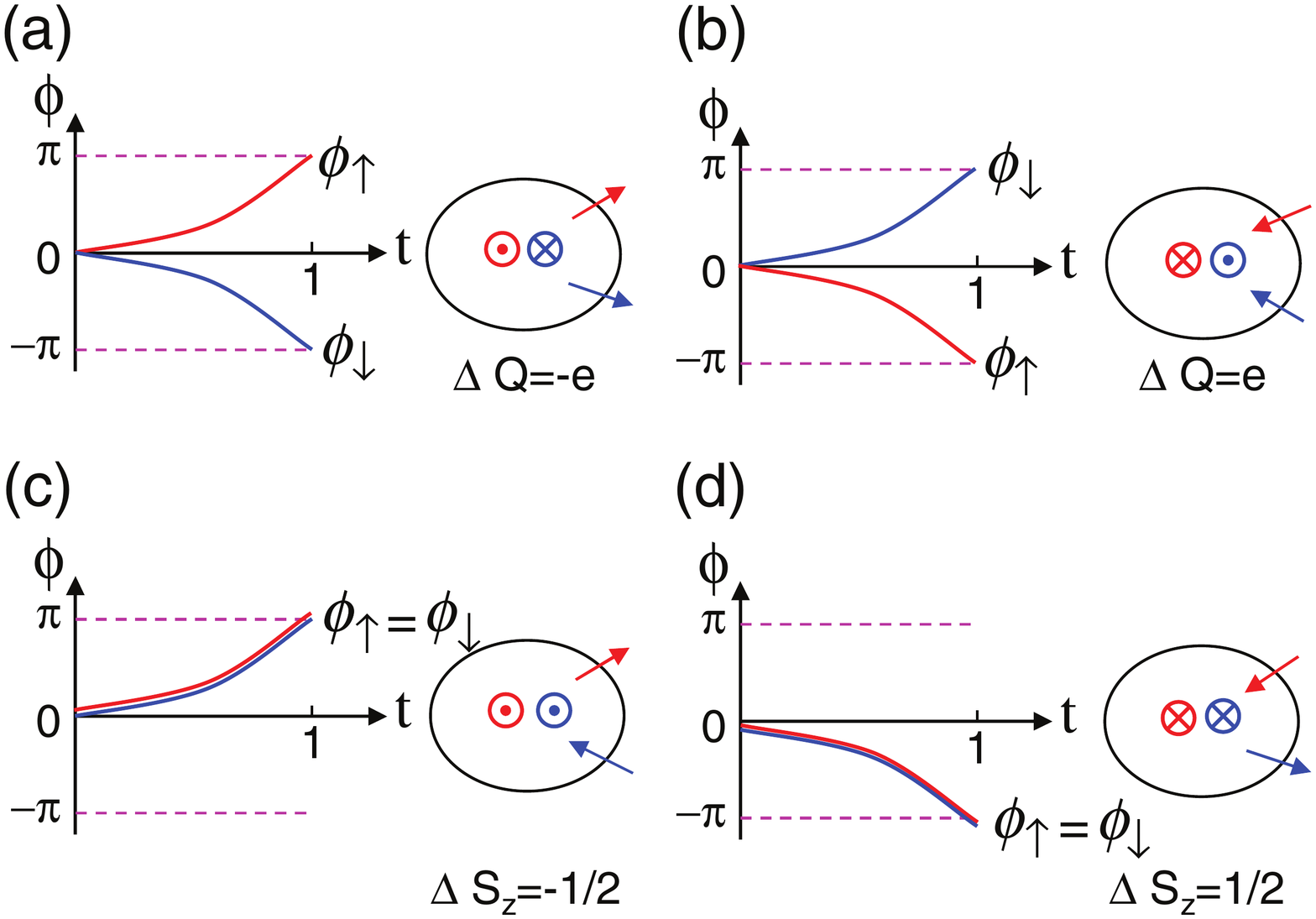}
  \caption{\label{fig:spincharge} The four possible adiabatic processes
    $\phi_\UP(t)$, $\phi_\DO(t)$ for inserting a $\pi$ flux into a quantum
    spin Hall insulator \cite{Qi08}. Considering a closed loop around the
    flux, Faraday's law implies the existence of charge fluxons with charge
    $\pm e$ and spin $0$, and spin fluxons with charge $0$ and spin $\pm1/2$.
    (Reprinted with permission from \cite{Qi08}. Copyright 2008 by the
    American Physical Society).  }
\end{figure}

Topological insulators can also be identified by their response to
topological defects such as $\pi$ fluxes \cite{Qi08,PhysRevLett.101.086801},
twisted boundary conditions \cite{PhysRevLett.99.196805}, dislocations
\cite{Zaanen12}, vacancies \cite{arXiv:1210.0266}, or disclinations
\cite{RueggLin12}. Such defects, near which the topological properties are
perturbed compared to the rest of the system, typically give rise to new,
low-energy excitations with well-defined quantum numbers. Similar to the
metallic edge states arising from open boundary conditions, the existence of
these states is protected by time-reversal symmetry.  They also have a close
conceptual relation to the familiar soliton excitations in polyacetylene
first proposed by Su \etal \cite{SuShHe79}.  For numerical methods, $\pi$
fluxes are of particular interest as they can be easily implemented, and will
be the focus here.

The existence of low-energy excitations as a result of a $\pi$ flux becomes
quite apparent from the flux insertion argument by Qi and Zhang \cite{Qi08};
see also \cite{RevModPhys.83.1057}.  Consider a disk-shaped, noninteracting
quantum spin Hall insulator that conserves spin \cite{Qi08}. Through a hole
at the centre, a magnetic flux of magnitude $hc/2e$ (equal to $\pi$ in units
where $\hbar=c=e=1$) is inserted adiabatically. This insertion can be
described by the time-dependent fluxes $\phi_\UP(t)$, $\phi_\DO(t)$, with
$\phi_\sigma(t=0)=0$ and $\phi_\sigma(t=1)=\pm\pi$. In total, there are four
different possibilities to switch on the fluxes that differ by the signs of
the fluxes $\phi_\UP(t)$ and $\phi_\DO(t)$, as shown in
figure~\ref{fig:spincharge}. Importantly, because the resulting Berry phases
$e^{\rmi \pi}=e^{-\rmi \pi}=-1$, all four choices lead to the same final
state \cite{Qi08}. Considering a loop around the flux tube, Faraday's law
states that the flux gives rise to a tangential electric field, which for a
quantum spin Hall insulator in turn causes a flow of charge perpendicular to
the loop. The four different processes give rise to a doublet of states with
charge $Q=\pm e$ and spin $S^z=0$ (the so-called {\it charge fluxons}
\cite{PhysRevLett.101.086801},
figure~\ref{fig:spincharge}(a) and (b)), and a pair of states with charge
$Q=0$ and spin $S^z=\pm1/2$ ({\it spin fluxons}
\cite{PhysRevLett.101.086801}, figure~\ref{fig:spincharge}(c) and (d)).
Given time-reversal and particle-hole symmetry, these states can be shown to lie in
the middle of the band gap \cite{Qi08,PhysRevLett.101.086801}. Moreover, they
are exponentially localized near the $\pi$ flux, and the two spin fluxon states
form a Kramers doublet related by time reversal, just like the two states of
a spin $1/2$. The whole argument can also be generalized to systems with only
a $Z_2$ spin symmetry \cite{Qi08}.

\begin{figure}[t]
  \centering
  \includegraphics[width=0.45\textwidth]{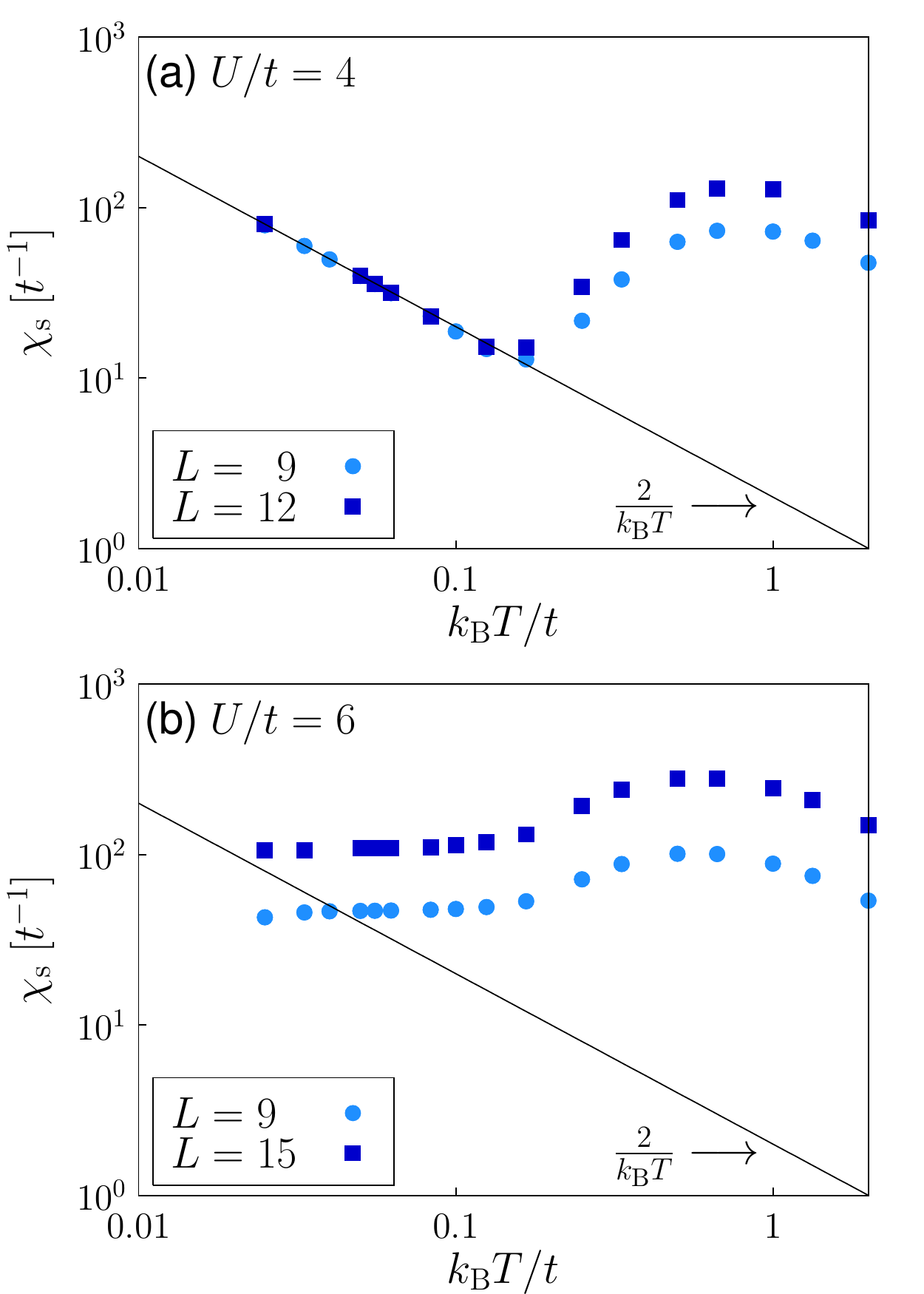}
  \caption{\label{fig:spinons} Magnetic susceptibility of the KMH model with
    one pair of $\pi$ fluxes from quantum Monte Carlo simulations on $L\times
    L$ lattices. (a) In the quantum spin Hall phase, midgap spin fluxon states
    give rise to a Curie law at low temperatures. In contrast, such a Curie
    law is absent in the results for the magnetic insulating state shown in
    (b). Data taken from \cite{As.Be.Ho.2012}.  }
\end{figure}

Fluxon states as a response to a $\pi$ flux provide an alternative way to
define the $Z_2$ topological invariant, and were suggested to be a useful
tool to study correlated topological states in combination with numerical
methods \cite{Qi08,PhysRevLett.101.086801}. In particular, there is no need
to study systems with open boundaries, or to perform high-dimensional
integrals over Green functions. On periodic lattices, $\pi$ fluxes can be
inserted in pairs, and are implemented in terms of Berry phase factors for
hopping processes that cross the {\it string} or {\it branch cut} that
connects two fluxes \cite{As.Be.Ho.2012,Qi08,PhysRevLett.99.196805}. A first
study in this direction was recently carried out based on exact quantum Monte
Carlo simulations of the KMH model \cite{As.Be.Ho.2012}. In that work, it was
established that the spin fluxon states can be identified from a characteristic
Curie law (given by $N_{\pi} / k_\text{B}T$ for $N_\pi$ fluxes) in the
temperature-dependent magnetic susceptibility. This Curie law is related
to the low-energy spin fluxon doublet (in the presence of repulsive electronic
interactions, charge fluxon states become gapped
\cite{Qi08,PhysRevLett.101.086801}). As illustrated in
figure~\ref{fig:spinons}, the presence or absence
of a Curie law at low temperatures provides a clear distinction between a
correlated quantum spin Hall state and an antiferromagnet with spontaneously
broken time-reversal symmetry \cite{As.Be.Ho.2012}. The presence of localized
spin fluxon excitations can also be visualized by means of the lattice-site
dependent spin excitation spectrum \cite{As.Be.Ho.2012}. Based on the flux
insertion argument (figure~\ref{fig:spincharge}), a generalization to
fractional topological insulators seems possible.

Finally, quantum entanglement \cite{PhysRevB.82.155138} has in recent years emerged as one of the most
useful tools to detect topological order in numerical calculations, most
notably in the framework of the density matrix renormalization group and
related approaches \cite{Schollwoeck05_rev}. The {\it entanglement entropy}
\cite{PhysRevLett.90.227902,PhysRevLett.96.110404,PhysRevLett.96.110405} is
calculated as the partial trace over the density matrix of the ground state
and provides a measure for the quantum correlations between different parts
of the quantum system under consideration. It directly reveals the presence
of excitations with exotic quantum numbers and can be used to identify, for
example, quantum spin liquids \cite{PhysRevLett.109.067201,arXiv:1205.4289}
and the exotic QSH* phase of the SI model \cite{PhysRevLett.108.046401}. Even
more information is contained in the {\it entanglement spectrum}
\cite{PhysRevLett.101.010504}, corresponding to the spectrum of the reduced
density matrix.  In the context of topological insulators, it has been shown
that topological phases possess protected gapless modes in their entanglement
spectrum, which remain gapless even if the physical edge states are gapped
out by interaction effects or perturbations that break time-reversal
symmetry \cite{PhysRevB.82.241102,PhysRevLett.108.196402}; for more details
see \cite{Fi.Ch.Hu.Ka.Lu.Ru.Zy.11,PhysRevB.84.155116} and references therein.
Whereas the two-dimensional setting of quantum spin Hall insulators is
typically difficult to study with the density matrix renormalization group
method, important progress has been made using tensor networks, and a
suitable ansatz for fractional topological insulators has been proposed in
\cite{PhysRevLett.106.156401}. For recent advances in the
application to quantum Hall states see \cite{arXiv:1211.3733}.

\section{Conclusions and outlook}\label{sec:conclusions}

Despite their rather short history, time-reversal invariant topological
insulators have already evolved into a multi-faceted research field 
driven by interests ranging from future technological applications of the
quantum spin Hall effect to the intricate physics of interaction-driven
topological order. This review has aimed at providing an overview of
the fascinating topic of electronic correlation effects, which on their own
have been at the focus of attention in the condensed matter community for
decades. Similar to the fractional quantum Hall effect, the combination of
electron-electron interactions and topological aspects in strongly-correlated
topological insulators provides an ideal setting to explore beautiful new
physics.

Among the two-dimensional systems discussed here, the currently best
understood correlated topological insulators are those with strong, intrinsic
spin-orbit coupling that gives rise to a topological band insulator in the
absence of electronic correlations. If spin is conserved, such states have
close conceptual relations to the integer quantum Hall effect.  The impact of
electronic interactions can in many cases be explored by means of
sophisticated numerical methods.  Because time-reversal symmetry is at the
heart of topological insulators, magnetic correlations induced by strong
electron-electron interactions play a crucial role. For several of the
available theoretical models, it is known that an extended topological
insulator phase exists up to rather strong electronic interactions, and is
adiabatically connected to a topological band insulator.
At even stronger interactions, time-reversal symmetry can be spontaneously
broken by the onset of long-range magnetic order. Generically, such a
transition destroys the topological state but under special conditions,
antiferromagnetic topological insulators with robust edge states exist.

Correlation effects also play an important role for the helical edge
states. Time-reversal symmetry provides protection against single-particle
backscattering processes, but two-particle backscattering can give rise to
long-range magnetic order at the edge, and to a breakdown of the
bulk-boundary correspondence. However, whereas the relevant Luttinger
parameter has been shown to become small enough for such a transition to
occur, the necessary umklapp scattering is not allowed if spin is conserved.
In the absence of magnetic impurities, the edges therefore retain their
metallic character even for strong interactions. Helical edge states are also
remarkably different from ordinary one-dimensional liquids concerning their
stability with respect to magnetic impurities or disorder.

Topological states that are not adiabatically connected to band insulators
have been explored less thoroughly. Examples include topological Mott
insulators, in which the topological state is generated from electronic
correlations rather than from spin-orbit coupling, and fractional topological
insulators that can, in the simplest case, be constructed by combining two
fractional quantum Hall states in a time-reversal invariant way. In these
cases, the resulting many-body state cannot be understood in the
noninteracting limit, and intriguing phenomena such as quasiparticles with
fractional charge and statistics appear. Moreover, while Rashba spin-orbit
coupling---violating the conservation of total spin---is often considered as
an experimental nuisance, theoretical results suggest that it may actually
lead to exciting and not yet fully explored new physics. Topological Mott
insulators and fractional topological insulators, as well as the general
class of topologically ordered states with time-reversal symmetry, have so
far been studied mostly using mean-field or topological field theory, and
pose challenging problems such as finding suitable ways to determine whether
or not a given state is topological. The identification and comprehensive
numerical investigation of microscopic models is one of the future
goals. An important recent development is the use of quantum entanglement
and local unitary transformations to study and classify phases with
topological order as well as symmetry-protected topological states such as
topological insulators.
Finally, the current search for experimental realizations, for example
in cold-atom systems and transition-metal oxide heterostructures, will
hopefully soon lead to a fruitful interplay of theory and experiment.

The authors are grateful to 
M. Bercx,
B. B\'eri,
J. Budich,
N. Cooper,
M. Daghofer,
V. Gurarie,
M. Kharitonov,
T. Lang,
M. Laubach,
A. Liebsch,
Z. Meng,
J. Moore,
A. Muramatsu,
S. Rachel,
M. Schmidt,
R. Thomale,
B. Trauzettel,
C. Varney,
S. Wessel, and
C. Xu for valuable discussions. Financial support from the DFG grants
As~120/4-3 and Ho~4489/2-1 is gratefully acknowledged.

\section*{References}


\begin{thebibliography}{100}

\bibitem{GinzburgLandau}
V.~L. Ginzburg and L.~D. Landau, Zh. Eksp. Teor. Fiz. {\bf 20},  1064  (1950).

\bibitem{PhysRevLett.45.494}
K.~v. Klitzing, G. Dorda, and M. Pepper, Phys. Rev. Lett. {\bf 45},  494
  (1980).

\bibitem{RevModPhys.80.1083}
C. Nayak, S.~H. Simon, A. Stern, M. Freedman, and S. Das~Sarma, Rev. Mod. Phys.
  {\bf 80},  1083  (2008).

\bibitem{PhysRevLett.49.405}
D.~J. Thouless, M. Kohmoto, M.~P. Nightingale, and M. den Nijs, Phys. Rev.
  Lett. {\bf 49},  405  (1982).

\bibitem{PhysRevLett.51.51}
J.~E. Avron, R. Seiler, and B. Simon, Phys. Rev. Lett. {\bf 51},  51  (1983).

\bibitem{Kohmoto1985343}
M. Kohmoto, Annals of Physics {\bf 160},  343   (1985).

\bibitem{PhysRevB.25.2185}
B.~I. Halperin, Phys. Rev. B {\bf 25},  2185  (1982).

\bibitem{FuKa07}
L. Fu and C.~L. Kane, Phys. Rev. B {\bf 76},  045302  (2007).

\bibitem{Haldane98}
F.~D.~M. Haldane, Phys. Rev. Lett. {\bf 61},  2015  (1988).

\bibitem{PhysRevB.75.121306}
J.~E. Moore and L. Balents, Phys. Rev. B {\bf 75},  121306  (2007).

\bibitem{KaMe05a}
C.~L. Kane and E.~J. Mele, Phys. Rev. Lett. {\bf 95},  146802  (2005).

\bibitem{KaMe05b}
C.~L. Kane and E.~J. Mele, Phys. Rev. Lett. {\bf 95},  226801  (2005).

\bibitem{PhysRevLett.98.106803}
L. Fu, C.~L. Kane, and E.~J. Mele, Phys. Rev. Lett. {\bf 98},  106803  (2007).

\bibitem{RevModPhys.83.1057}
X.-L. Qi and S.-C. Zhang, Rev. Mod. Phys. {\bf 83},  1057  (2011).

\bibitem{PhysRevB.74.165310}
H. Min, J.~E. Hill, N.~A. Sinitsyn, B.~R. Sahu, L. Kleinman, and A.~H.
  MacDonald, Phys. Rev. B {\bf 74},  165310  (2006).

\bibitem{PhysRevLett.107.076802}
C.-C. Liu, W. Feng, and Y. Yao, Phys. Rev. Lett. {\bf 107},  076802  (2011).

\bibitem{PhysRevLett.108.155501}
P. Vogt, P. De~Padova, C. Quaresima, J. Avila, E. Frantzeskakis, M.~C. Asensio,
  A. Resta, B. Ealet, and G. Le~Lay, Phys. Rev. Lett. {\bf 108},  155501
  (2012).

\bibitem{PhysRevLett.108.245501}
A. Fleurence, R. Friedlein, T. Ozaki, H. Kawai, Y. Wang, and Y.
  Yamada-Takamura, Phys. Rev. Lett. {\bf 108},  245501  (2012).

\bibitem{PhysRevLett.109.056804}
L. Chen, C.-C. Liu, B. Feng, X. He, P. Cheng, Z. Ding, S. Meng, Y. Yao, and K.
  Wu, Phys. Rev. Lett. {\bf 109},  056804  (2012).

\bibitem{Koenig07}
M. K\"onig, S. Wiedmann, C. Br\"une, A. Roth, H. Buhmann, L.~W. Molenkamp,
  X.-L. Qi, and S.-C. Zhang, Science {\bf 318},  766  (2007).

\bibitem{Roth09}
A. Roth, C. Br\"une, H. Buhmann, L.~W. Molenkamp, J. Maciejko, X.-L. Qi, and
  S.-C. Zhang, Science {\bf 325},  294  (2009).

\bibitem{Br.Ro.Bu.Ha.Mo.Ma.Qi.Zh.12}
C. Br\"une, A. Roth, H. Buhmann, E.~M. Hankiewicz, L.~W. Molenkamp, J.
  Maciejko, X.-L. Qi, and S.-C. Zhang, Nat. Phys. {\bf 8},  486  (2012).

\bibitem{BeHuZh06}
B.~A. Bernevig, T.~L. Hughes, and S. Zhang, Science {\bf 314},  1757  (2006).

\bibitem{PhysRevB.78.195125}
A.~P. Schnyder, S. Ryu, A. Furusaki, and A.~W.~W. Ludwig, Phys. Rev. B {\bf
  78},  195125  (2008).

\bibitem{Kitaev.09}
A. Kitaev, AIP Conf. Proc. {\bf 1134},  22  (2009).

\bibitem{Bu.Tr.12}
J.~C. Budich and B. Trauzettel, physica status solidi (RRL) {\bf 7},  109
  (2013).

\bibitem{Sl.Me.Ju.Za.12}
R.-J. Slager, A. Mesaros, V. Juricic, and J. Zaanen, Nat. Phys. {\bf 9},  98
  (2013).

\bibitem{Irridates-Nagaosa}
A. Shitade, H. Katsura, J. Kunes, X.-L. Qi, S.-C. Zhang, and N. Nagaosa, Phys.
  Rev. Lett. {\bf 102},  256403  (2009).

\bibitem{PhysRevB.84.100406}
J. Reuther, R. Thomale, and S. Trebst, Phys. Rev. B {\bf 84},  100406  (2011).

\bibitem{Chaloupka10}
J. Chaloupka, G. Jackeli, and G. Khaliullin, Phys. Rev. Lett. {\bf 105},
  027204  (2010).

\bibitem{Re.Th.Ra.12}
J. Reuther, R. Thomale, and S. Rachel, Phys. Rev. B {\bf 86},  155127  (2012).

\bibitem{Ka.La.Fi.12}
M. Kargarian, A. Langari, and G.~A. Fiete, Phys. Rev. B {\bf 86},  205124
  (2012).

\bibitem{arXiv:1210.2290}
S. Okamoto, Phys. Rev. Lett. {\bf 110},  066403  (2013).

\bibitem{Balents10}
L. Balents, Nature {\bf 464},  199  (2010).

\bibitem{RaQiHo08}
S. Raghu, X. Qi, C. Honerkamp, and S. Zhang, Phys. Rev. Lett. {\bf 100},
  156401  (2008).

\bibitem{PeBa10}
D. Pesin and L. Balents, Nat. Phys. {\bf 6},  376  (2010).

\bibitem{PhysRevLett.101.076402}
B.~J. Kim, H. Jin, S.~J. Moon, J.-Y. Kim, B.-G. Park, C.~S. Leem, J. Yu, T.~W.
  Noh, C. Kim, S.-J. Oh, J.-H. Park, V. Durairaj, G. Cao, and E. Rotenberg,
  Phys. Rev. Lett. {\bf 101},  076402  (2008).

\bibitem{Kimetal2009}
B.~J. Kim, H. Ohsumi, T. Komesu, S. Sakai, T. Morita, H. Takagi, and T. Arima,
  Science {\bf 323},  1329  (2009).

\bibitem{FQHE_sheng2011}
D.~N. Sheng, Z.-C. Gu, K. Sun, and L. Sheng, Nat. Commun. {\bf 2},  389
  (2011).

\bibitem{PhysRevLett.108.126405}
J.~W.~F. Venderbos, S. Kourtis, J. van~den Brink, and M. Daghofer, Phys. Rev.
  Lett. {\bf 108},  126405  (2012).

\bibitem{PhysRevLett.103.196803}
M. Levin and A. Stern, Phys. Rev. Lett. {\bf 103},  196803  (2009).

\bibitem{PhysRevB.86.125119}
Y.-M. Lu and A. Vishwanath, Phys. Rev. B {\bf 86},  125119  (2012).

\bibitem{Wen_book}
X.-G. Wen, {\em Quantum Field Theory of Many-body Systems: From the Origin of
  Sound to an Origin of Light and Electrons} (Oxford University Press, Oxford,
  2004).

\bibitem{PhysRevB.82.155138}
X. Chen, Z.-C. Gu, and X.-G. Wen, Phys. Rev. B {\bf 82},  155138  (2010).

\bibitem{PhysRevB.84.235141}
X. Chen, Z.-X. Liu, and X.-G. Wen, Phys. Rev. B {\bf 84},  235141  (2011).

\bibitem{arXiv:1106.4772}
X. Chen, Z.-C. Gu, Z.-X. Liu, and X.-G. Wen, arXiv:1106.4772  (2011).

\bibitem{Chen2012}
X. Chen, Z.-C. Gu, Z.-X. Liu, and X.-G. Wen, Science {\bf 21},  1604  (2012).

\bibitem{arXiv:1209.4399}
C. Xu, arXiv:1209.4399  (2012).

\bibitem{arXiv:1212.1726}
J. Oon, G.~Y. Cho, and C. Xu, arXiv:1212.1726  (2012).

\bibitem{HaKa10}
M.~Z. Hasan and C.~L. Kane, Rev. Mod. Phys. {\bf 82},  3045  (2010).

\bibitem{doi:10.1146/annurev-conmatphys-062910-140432}
M.~Z. Hasan and J.~E. Moore, Annu. Rev. Cond. Mat. Phys. {\bf 2},  55  (2011).

\bibitem{arXiv:1211.5104}
S. Wolgast, C. Kurdak, K. Sun, J.~W. Allen, D.-J. Kim, and Z. Fisk,
  arXiv:1211.5104  (2012).

\bibitem{arXiv:1211.6769}
J. Botimer, D.~J. Kim, S. Thomas, T. Grant, Z. Fisk, and J. Xia,
  arXiv:1211.6769  (2012).

\bibitem{Zhang23032012}
X. Zhang, H. Zhang, J. Wang, C. Felser, and S.-C. Zhang, Science {\bf 335},
  1464  (2012).

\bibitem{Xi.Zh.Ra.Na.Ok.11}
D. Xiao, W. Zhu, Y. Ran, N. Nagaosa, and S. Okamoto, Nat. Commun. {\bf 2},  596
   (2011).

\bibitem{PhysRevB.86.235141}
X. Hu, A. R\"uegg, and G.~A. Fiete, Phys. Rev. B {\bf 86},  235141  (2012).

\bibitem{PhysRevB.84.201104}
K.-Y. Yang, W. Zhu, D. Xiao, S. Okamoto, Z. Wang, and Y. Ran, Phys. Rev. B {\bf
  84},  201104  (2011).

\bibitem{Ru.Fi.11}
A. R\"uegg and G.~A. Fiete, Phys. Rev. B {\bf 84},  201103  (2011).

\bibitem{PhysRevB.85.245131}
A. R\"uegg, C. Mitra, A.~A. Demkov, and G.~A. Fiete, Phys. Rev. B {\bf 85},
  245131  (2012).

\bibitem{PhysRevB.84.241103}
F. Wang and Y. Ran, Phys. Rev. B {\bf 84},  241103  (2011).

\bibitem{PhysRevB.82.115124}
R. Nandkishore and L. Levitov, Phys. Rev. B {\bf 82},  115124  (2010).

\bibitem{PhysRevLett.106.156801}
F. Zhang, J. Jung, G.~A. Fiete, Q. Niu, and A.~H. MacDonald, Phys. Rev. Lett.
  {\bf 106},  156801  (2011).

\bibitem{PhysRevX.1.021001}
C. Weeks, J. Hu, J. Alicea, M. Franz, and R. Wu, Phys. Rev. X {\bf 1},  021001
  (2011).

\bibitem{Go.Ma.Ko.Gu.Ma.12}
K.~K. Gomes, W. Mar, W. Ko, F. Guinea, and H.~C. Manoharan, Nature {\bf 483},
  306  (2012).

\bibitem{RevModPhys.83.1523}
J. Dalibard, F. Gerbier, G. Juzeli\ifmmode~\bar{u}\else \={u}\fi{}nas, and P.
  \"Ohberg, Rev. Mod. Phys. {\bf 83},  1523  (2011).

\bibitem{PhysRevLett.107.145301}
B. B\'eri and N.~R. Cooper, Phys. Rev. Lett. {\bf 107},  145301  (2011).

\bibitem{PhysRevLett.105.255302}
N. Goldman, I. Satija, P. Nikoli\'{c}, A. Bermudez, M.~A. Martin-Delgado, M.
  Lewenstein, and I.~B. Spielman, Phys. Rev. Lett. {\bf 105},  255302  (2010).

\bibitem{PhysRevA.86.053618}
A. Dauphin, M. M\"uller, and M.~A. Martin-Delgado, Phys. Rev. A {\bf 86},
  053618  (2012).

\bibitem{arXiv:1207.3716}
F. Grusdt and M. Fleischhauer, arXiv:1207.3716  .

\bibitem{PhysRevA.82.043811}
J. Koch, A.~A. Houck, K.~L. Hur, and S.~M. Girvin, Phys. Rev. A {\bf 82},
  043811  (2010).

\bibitem{arXiv:1206.1539}
A. Petrescu, A.~A. Houck, and K. Le~Hur, Phys. Rev. A {\bf 86},  053804
  (2012).

\bibitem{Moore10}
J.~E. Moore, Nature {\bf 464},  194  (2010).

\bibitem{DzSuGa10}
M. Dzero, K. Sun, V. Galitski, and P. Coleman, Phys. Rev. Lett. {\bf 104},
  106408  (2010).

\bibitem{PhysRevB.85.045130}
M. Dzero, K. Sun, P. Coleman, and V. Galitski, Phys. Rev. B {\bf 85},  045130
  (2012).

\bibitem{PhysRevB.74.085308}
X.-L. Qi, Y.-S. Wu, and S.-C. Zhang, Phys. Rev. B {\bf 74},  085308  (2006).

\bibitem{Fi.Ch.Hu.Ka.Lu.Ru.Zy.11}
G.~A. Fiete, V. Chua, X. Hu, M. Kargarian, R. Lundgren, A. R\"uegg, J. Wen, and
  V. Zyuzin, Physica E {\bf 44},  845  (2012).

\bibitem{QiHuZh08}
X. Qi, T.~L. Hughes, and S. Zhang, Phys. Rev. B {\bf 78},  195424  (2008).

\bibitem{Wu06}
C. Wu, B.~A. Bernevig, and S.-C. Zhang, Phys. Rev. Lett. {\bf 96},  106401
  (2006).

\bibitem{Cenke06}
C. Xu and J.~E. Moore, Phys. Rev. B {\bf 73},  045322  (2006).

\bibitem{PhysRevB.85.195116}
A.~M. Essin and V. Gurarie, Phys. Rev. B {\bf 85},  195116  (2012).

\bibitem{Meng10}
Z.~Y. Meng, T.~C. Lang, S. Wessel, F.~F. Assaad, and A. Muramatsu, Nature {\bf
  464},  847  (2010).

\bibitem{Novoselov22102004}
K.~S. Novoselov, A.~K. Geim, S.~V. Morozov, D. Jiang, Y. Zhang, S.~V. Dubonos,
  I.~V. Grigorieva, and A.~A. Firsov, Science {\bf 306},  666  (2004).

\bibitem{Neto_rev}
A.~H. {Castro Neto}, F. Guinea, N.~M.~R. Peres, K.~S. Novoselov, and A.~K.
  Geim, Rev. Mod. Phys. {\bf 81},  109  (2009).

\bibitem{Hu63}
J. Hubbard, Proc. R. Soc. London {\bf 276},  238  (1963).

\bibitem{Wallace47}
P.~R. Wallace, Phys. Rev. {\bf 71},  622  (1947).

\bibitem{PhysRevLett.97.036808}
D.~N. Sheng, Z.~Y. Weng, L. Sheng, and F.~D.~M. Haldane, Phys. Rev. Lett. {\bf
  97},  036808  (2006).

\bibitem{RaHu10}
S. Rachel and K. {Le Hur}, Phys. Rev. B {\bf 82},  075106  (2010).

\bibitem{PhysRevLett.103.046811}
K. Sun, H. Yao, E. Fradkin, and S.~A. Kivelson, Phys. Rev. Lett. {\bf 103},
  046811  (2009).

\bibitem{Haldane88}
F.~D.~M. Haldane, Phys. Rev. Lett. {\bf 61},  1029  (1988).

\bibitem{Wa.Sh.Zh.Wa.Da.Xi.10}
L. Wang, H. Shi, S. Zhang, X. Wang, X. Dai, and X.~C. Xie, arXiv:1012.5163
  (2010).

\bibitem{VaSuRi10}
C.~N. Varney, K. Sun, M. Rigol, and V. Galitski, Phys. Rev. B {\bf 82},  115125
   (2010).

\bibitem{Va.Su.Ri.Ga.11}
C.~N. Varney, K. Sun, M. Rigol, and V. Galitski, Phys. Rev. B {\bf 84},  241105
   (2011).

\bibitem{Zaanen12}
V. Juricic, A. Mesaros, R.-J. Slager, and J. Zaanen, Phys. Rev. Lett. {\bf
  108},  106403  (2012).

\bibitem{PhysRevB.85.235449}
A. Medhi, V.~B. Shenoy, and H.~R. Krishnamurthy, Phys. Rev. B {\bf 85},  235449
   (2012).

\bibitem{Ro.Re.Li.Mo.Zh.Ha.10}
D.~G. Rothe, R.~W. Reinthaler, C.-X. Liu, L.~W. Molenkamp, S.-C. Zhang, and
  E.~M. Hankiewicz, New J. Phys. {\bf 12},  065012  (2012).

\bibitem{PhysRevLett.108.156402}
T.~L. Schmidt, S. Rachel, F. von Oppen, and L.~I. Glazman, Phys. Rev. Lett.
  {\bf 108},  156402  (2012).

\bibitem{arXiv:1111.6250}
T. Yoshida, S. Fujimoto, and N. Kawakami, Phys. Rev. B {\bf 85},  125113
  (2012).

\bibitem{Wa.Da.Xi.12}
L. Wang, X. Dai, and X.~C. Xie, Eur. Phys. Lett. {\bf 98},  57001  (2012).

\bibitem{arXiv:1202.3203}
Y. Tada, R. Peters, M. Oshikawa, A. Koga, N. Kawakami, and S. Fujimoto, Phys.
  Rev. B {\bf 85},  165138  (2012).

\bibitem{arXiv:1211.3059}
J.~C. Budich, B. Trauzettel, and G. Sangiovanni, arXiv:1211.3059  (2012).

\bibitem{arXiv:1207.4547v1}
T. Yoshida, R. Peters, S. Fujimoto, and N. Kawakami, Phys. Rev. B {\bf 87},
  085134  (2013).

\bibitem{PhysRevLett.108.046401}
A. R\"uegg and G.~A. Fiete, Phys. Rev. Lett. {\bf 108},  046401  (2012).

\bibitem{PhysRevB.80.113102}
H.-M. Guo and M. Franz, Phys. Rev. B {\bf 80},  113102  (2009).

\bibitem{PhysRevB.81.205115}
A. R\"uegg, J. Wen, and G.~A. Fiete, Phys. Rev. B {\bf 81},  205115  (2010).

\bibitem{PhysRevB.82.085310}
C. Weeks and M. Franz, Phys. Rev. B {\bf 82},  085310  (2010).

\bibitem{PhysRevB.82.085106}
M. Kargarian and G.~A. Fiete, Phys. Rev. B {\bf 82},  085106  (2010).

\bibitem{PhysRevB.84.155116}
X. Hu, M. Kargarian, and G.~A. Fiete, Phys. Rev. B {\bf 84},  155116  (2011).

\bibitem{Kitaev20032}
A. Kitaev, Annals of Physics {\bf 303},  2   (2003).

\bibitem{Kitaev20062}
A. Kitaev, Annals of Physics {\bf 321},  2   (2006).

\bibitem{Ar.Ca.Sa.12}
M.~A.~N. Ara\'ujo, E.~V. Castro, and P.~D. Sacramento, Phys. Rev. B {\bf 87},
  085109  (2013).

\bibitem{JPSJ.80.044707}
J. Goryo and N. Maeda, J. Phys. Soc. Jpn. {\bf 80},  044707  (2011).

\bibitem{Cocks.12}
D. Cocks, P.~P. Orth, S. Rachel, M. Buchhold, K. Le~Hur, and W. Hofstetter,
  Phys. Rev. Lett. {\bf 109},  205303  (2012).

\bibitem{PhysRevB.31.3372}
Q. Niu, D.~J. Thouless, and Y.-S. Wu, Phys. Rev. B {\bf 31},  3372  (1985).

\bibitem{RevModPhys.82.1959}
D. Xiao, M.-C. Chang, and Q. Niu, Rev. Mod. Phys. {\bf 82},  1959  (2010).

\bibitem{PhysRevA.64.052101}
W.-Y. Hsiang and D.-H. Lee, Phys. Rev. A {\bf 64},  052101  (2001).

\bibitem{Volovik}
G.~E. Volovik, {\em The Universe in a Helium Droplet} (Clarendon Press, Oxford,
  2003).

\bibitem{PhysRevLett.91.116802}
D.~N. Sheng, L. Balents, and Z. Wang, Phys. Rev. Lett. {\bf 91},  116802
  (2003).

\bibitem{PhysRevLett.53.2449}
G.~W. Semenoff, Phys. Rev. Lett. {\bf 53},  2449  (1984).

\bibitem{PhysRevB.74.195312}
L. Fu and C.~L. Kane, Phys. Rev. B {\bf 74},  195312  (2006).

\bibitem{PhysRevB.41.12838}
X.~G. Wen, Phys. Rev. B {\bf 41},  12838  (1990).

\bibitem{Wen92}
X.~G. Wen, Int. J. Mod. Phys. B {\bf 6},  1711  (1992).

\bibitem{PhysRevLett.68.1220}
C.~L. Kane and M.~P.~A. Fisher, Phys. Rev. Lett. {\bf 68},  1220  (1992).

\bibitem{PhysRevLett.98.076802}
M. Onoda, Y. Avishai, and N. Nagaosa, Phys. Rev. Lett. {\bf 98},  076802
  (2007).

\bibitem{PhysRevB.84.165107}
T. Neupert, L. Santos, S. Ryu, C. Chamon, and C. Mudry, Phys. Rev. B {\bf 84},
  165107  (2011).

\bibitem{Hohenadler10}
M. Hohenadler, T.~C. Lang, and F.~F. Assaad, Phys. Rev. Lett. {\bf 106},
  100403  (2011), erratum {\bf 109}, 229902(E) (2012).

\bibitem{Ho.Me.La.We.Mu.As.12}
M. Hohenadler, Z.~Y. Meng, T.~C. Lang, S. Wessel, A. Muramatsu, and F.~F.
  Assaad, Phys. Rev. B {\bf 85},  115132  (2012).

\bibitem{As.Be.Ho.2012}
F.~F. Assaad, M. Bercx, and M. Hohenadler, Phys. Rev. X {\bf 3},  011015
  (2013).

\bibitem{Gr.Xu.11}
C. Griset and C. Xu, Phys. Rev. B {\bf 85},  045123  (2012).

\bibitem{Zh.Wu.Zh.11}
D. Zheng, G.-M. Zhang, and C. Wu, Phys. Rev. B {\bf 84},  205121  (2011).

\bibitem{PhysRevB.83.205122}
Y. Yamaji and M. Imada, Phys. Rev. B {\bf 83},  205122  (2011).

\bibitem{Wu.Ra.Li.LH.11}
W. Wu, S. Rachel, W.-M. Liu, and K. Le~Hur, Phys. Rev. B {\bf 85},  205102
  (2012).

\bibitem{Yu.Xie.Li.11}
S.-L. Yu, X.~C. Xie, and J.-X. Li, Phys. Rev. Lett. {\bf 107},  010401  (2011).

\bibitem{arXiv:1203.2928}
J.~C. Budich, R. Thomale, G. Li, M. Laubach, and S.-C. Zhang, Phys. Rev. B {\bf
  86},  201407  (2012).

\bibitem{PhysRevB.83.035113}
A. Liebsch, Phys. Rev. B {\bf 83},  035113  (2011).

\bibitem{We.Ka.Va.Fi.11}
J. Wen, M. Kargarian, A. Vaezi, and G.~A. Fiete, Phys. Rev. B {\bf 84},  235149
   (2011).

\bibitem{Ma.Va.Va.11}
M. Mardani, M.-S. Vaezi, and A. Vaezi, arXiv:1111.5980  (2011).

\bibitem{PhysRevB.85.195126}
A. Vaezi, M. Mashkoori, and M. Hosseini, Phys. Rev. B {\bf 85},  195126
  (2012).

\bibitem{PhysRevLett.107.166806}
D.-H. Lee, Phys. Rev. Lett. {\bf 107},  166806  (2011).

\bibitem{So.Ot.Yu.}
S. Sorella, Y. Otsuka, and S. Yunoki, Sci. Rep. {\bf 2},  992  (2012).

\bibitem{Ho.As.11}
M. Hohenadler and F.~F. Assaad, Phys. Rev. B {\bf 85},  081106  (2012), erratum
  {\bf 86}, 199901(E) (2012).

\bibitem{PhysRevB.81.245209}
R.~S.~K. Mong, A.~M. Essin, and J.~E. Moore, Phys. Rev. B {\bf 81},  245209
  (2010).

\bibitem{PhysRevB.74.144506}
M. Campostrini, M. Hasenbusch, A. Pelissetto, and E. Vicari, Phys. Rev. B {\bf
  74},  144506  (2006).

\bibitem{GrVi12}
T. Grover and A. Vishwanath, arXiv:1206.1332  (2012).

\bibitem{e20020021}
V. Anisimov, I. Nekrasov, D. Kondakov, T. Rice, and M. Sigrist, Eur. Phys. J. B
  {\bf 25},  191  (2002).

\bibitem{PhysRevLett.91.226401}
A. Liebsch, Phys. Rev. Lett. {\bf 91},  226401  (2003).

\bibitem{PhysRevLett.92.216402}
A. Koga, N. Kawakami, T.~M. Rice, and M. Sigrist, Phys. Rev. Lett. {\bf 92},
  216402  (2004).

\bibitem{PhysRevB.72.201102}
R. Arita and K. Held, Phys. Rev. B {\bf 72},  201102  (2005).

\bibitem{PhysRevB.83.045114}
H. Guo, S. Feng, and S.-Q. Shen, Phys. Rev. B {\bf 83},  045114  (2011).

\bibitem{PhysRevB.84.035127}
J. He, Y.-H. Zong, S.-P. Kou, Y. Liang, and S. Feng, Phys. Rev. B {\bf 84},
  035127  (2011).

\bibitem{He.Wa.Ko.12}
J. He, B. Wang, and S.-P. Kou, Phys. Rev. B {\bf 86},  235146  (2012).

\bibitem{PhysRevLett.108.046806}
T. Neupert, L. Santos, S. Ryu, C. Chamon, and C. Mudry, Phys. Rev. Lett. {\bf
  108},  046806  (2012).

\bibitem{PhysRevLett.97.146401}
I.~F. Herbut, Phys. Rev. Lett. {\bf 97},  146401  (2006).

\bibitem{PhysRevLett.100.146404}
C. Honerkamp, Phys. Rev. Lett. {\bf 100},  146404  (2008).

\bibitem{PhysRevLett.101.086801}
Y. Ran, A. Vishwanath, and D.-H. Lee, Phys. Rev. Lett. {\bf 101},  086801
  (2008).

\bibitem{PhysRevB.86.045128}
R. Nandkishore, M.~A. Metlitski, and T. Senthil, Phys. Rev. B {\bf 86},  045128
   (2012).

\bibitem{PhysRevLett.93.036403}
C. Wu and S.-C. Zhang, Phys. Rev. Lett. {\bf 93},  036403  (2004).

\bibitem{PhysRevB.81.085105}
C. Weeks and M. Franz, Phys. Rev. B {\bf 81},  085105  (2010).

\bibitem{PhysRevB.82.045102}
Q. Liu, H. Yao, and T. Ma, Phys. Rev. B {\bf 82},  045102  (2010).

\bibitem{PhysRevB.82.075125}
J. Wen, A. R\"uegg, C.-C.~J. Wang, and G.~A. Fiete, Phys. Rev. B {\bf 82},
  075125  (2010).

\bibitem{PhysRevB.79.245331}
Y. Zhang, Y. Ran, and A. Vishwanath, Phys. Rev. B {\bf 79},  245331  (2009).

\bibitem{PhysRevB.78.125316}
M.~W. Young, S.-S. Lee, and C. Kallin, Phys. Rev. B {\bf 78},  125316  (2008).

\bibitem{PhysRevLett.96.106802}
B.~A. Bernevig and S.-C. Zhang, Phys. Rev. Lett. {\bf 96},  106802  (2006).

\bibitem{PhysRevLett.106.236802}
E. Tang, J.-W. Mei, and X.-G. Wen, Phys. Rev. Lett. {\bf 106},  236802  (2011).

\bibitem{PhysRevLett.106.236803}
K. Sun, Z. Gu, H. Katsura, and S. Das~Sarma, Phys. Rev. Lett. {\bf 106},
  236803  (2011).

\bibitem{PhysRevLett.106.236804}
T. Neupert, L. Santos, C. Chamon, and C. Mudry, Phys. Rev. Lett. {\bf 106},
  236804  (2011).

\bibitem{Le.Bu.KJ.St.11}
M. Levin, F.~J. Burnell, M. Koch-Janusz, and A. Stern, Phys. Rev. B {\bf 84},
  235145  (2011).

\bibitem{La.Li.Be.Mo.12}
A.~M. L\"auchli, Z. Liu, E.~J. Berholtz, and R. Moessner, arXiv:1207.6094
  (2012).

\bibitem{PhysRevX.1.021014}
N. Regnault and B.~A. Bernevig, Phys. Rev. X {\bf 1},  021014  (2011).

\bibitem{Wu.Ja.Su.12}
Y.-H. Wu, J.~K. Jain, and K. Sun, Phys. Rev. B {\bf 86},  165129  (2012).

\bibitem{PhysRevLett.109.246805}
T. Scaffidi and G. M\"oller, Phys. Rev. Lett. {\bf 109},  246805  (2012).

\bibitem{Cho20111515}
G.~Y. Cho and J.~E. Moore, Annals of Physics {\bf 326},  1515   (2011).

\bibitem{Le.St.12}
M. Levin and A. Stern, Phys. Rev. B {\bf 86},  115131  (2012).

\bibitem{PhysRevB.85.165134}
Y.-M. Lu and Y. Ran, Phys. Rev. B {\bf 85},  165134  (2012).

\bibitem{Fe.Vi.11}
D. Ferraro and G. Viola, arXiv:1112.5399  (2011).

\bibitem{Roy.12}
R. Roy, arXiv:1208.2055  (2012).

\bibitem{Ni.11}
P. Nikoli\'{c}, J. Phys.: Condensed Matt. {\bf 25},  025602  (2013).

\bibitem{Ni.12}
P. Nikoli\'{c}, arXiv:1206.1055  (2012).

\bibitem{Gr.Ne.Ch.Mu.12}
A.~G. Grushin, T. Neupert, C. Chamon, and C. Mudry, Phys. Rev. B {\bf 86},
  205125  (2012).

\bibitem{PhysRevLett.108.206804}
B. B\'eri and N.~R. Cooper, Phys. Rev. Lett. {\bf 108},  206804  (2012).

\bibitem{arXiv:1206.2626}
T. Liu, C. Repellin, B.~A. Bernevig, and N. Regnault, arXiv:1206.2626  (2012).

\bibitem{PhysRevLett.106.156401}
B. B\'eri and N.~R. Cooper, Phys. Rev. Lett. {\bf 106},  156401  (2011).

\bibitem{Voit94}
J. Voit, Rep. Prog. Phy. {\bf 57},  977  (1995).

\bibitem{Tk.Ha.12}
G. Tkachov and E.~M. Hankiewicz, Phys. Status Solidi {\bf 250},  215  (2013).

\bibitem{PhysRevLett.103.166403}
Y. Tanaka and N. Nagaosa, Phys. Rev. Lett. {\bf 103},  166403  (2009).

\bibitem{1210.4818}
G. Aut\`{e}s and O.~V. Yazyev, physica status solid (RRL) {\bf 7},  151
  (2013).

\bibitem{QiHuZh.08}
X.-L. Qi, T.~L. Hughes, and S.-C. Zhang, Nat. Phys. {\bf 4},  273  (2008).

\bibitem{PhysRevLett.108.086602}
J.~C. Budich, F. Dolcini, P. Recher, and B. Trauzettel, Phys. Rev. Lett. {\bf
  108},  086602  (2012).

\bibitem{0953-8984-24-35-355001}
A. Medhi and V.~B. Shenoy, J. Phys: Condens. Matter {\bf 24},  355001  (2012).

\bibitem{PhysRevLett.84.4164}
A.~V. Moroz, K.~V. Samokhin, and C.~H.~W. Barnes, Phys. Rev. Lett. {\bf 84},
  4164  (2000).

\bibitem{PhysRevB.68.075107}
A. Iucci, Phys. Rev. B {\bf 68},  075107  (2003).

\bibitem{Rasha}
Y.~A. Bychkov and E.~I. Rashba, J. Phys. C: Solid State Phys. {\bf 17},  6039
  (1984).

\bibitem{Maciejko09}
J. Maciejko, C. Liu, Y. Oreg, X.-L. Qi, C. Wu, and S.-C. Zhang, Phys. Rev.
  Lett. {\bf 102},  256803  (2009).

\bibitem{PhysRevB.85.245108}
J. Maciejko, Phys. Rev. B {\bf 85},  245108  (2012).

\bibitem{PhysRevB.86.121106}
F. Cr\'epin, J.~C. Budich, F. Dolcini, P. Recher, and B. Trauzettel, Phys. Rev.
  B {\bf 86},  121106  (2012).

\bibitem{PhysRevB.86.165121}
M. Kharitonov, Phys. Rev. B {\bf 86},  165121  (2012).

\bibitem{Schulz93}
H.~J. Schulz, Phys. Rev. Lett. {\bf 71},  1864  (1993).

\bibitem{PhysRevB.82.161302}
D. Soriano and J. Fern\'andez-Rossier, Phys. Rev. B {\bf 82},  161302  (2010).

\bibitem{PhysRevLett.106.226401}
H. Feldner, Z.~Y. Meng, T.~C. Lang, F.~F. Assaad, S. Wessel, and A. Honecker,
  Phys. Rev. Lett. {\bf 106},  226401  (2011).

\bibitem{PhysRevB.83.195432}
D.~J. Luitz, F.~F. Assaad, and M.~J. Schmidt, Phys. Rev. B {\bf 83},  195432
  (2011).

\bibitem{Schmidt2012}
M.~J. Schmidt, Phys. Rev. B {\bf 86},  075458  (2012).

\bibitem{PhysRevB.85.035136}
B. Braunecker, C. Bena, and P. Simon, Phys. Rev. B {\bf 85},  035136  (2012).

\bibitem{PhysRevB.74.045125}
X.-L. Qi, Y.-S. Wu, and S.-C. Zhang, Phys. Rev. B {\bf 74},  045125  (2006).

\bibitem{PhysRevB.83.085426}
V. Gurarie, Phys. Rev. B {\bf 83},  085426  (2011).

\bibitem{PhysRevLett.72.892}
A. Furusaki and N. Nagaosa, Phys. Rev. Lett. {\bf 72},  892  (1994).

\bibitem{Jieetal2012}
J. Yuan, J.-H. Gao, W.-Q. Chen, F. Ye, Y. Zhou, and F.-C. Zhang, Phys. Rev. B
  {\bf 86},  104505  (2012).

\bibitem{PhysRevLett.66.3203}
R.~R.~P. Singh and R.~T. Scalettar, Phys. Rev. Lett. {\bf 66},  3203  (1991).

\bibitem{HoAsunpublished}
M. Hohenadler and F.~F. Assaad,   (unpublished).

\bibitem{arXiv:1207.7341}
Z. Wang and B. Yan, arXiv:1207.7341  (2012).

\bibitem{Is.Ma.87}
K. Ishikawa and T. Matsuyama, Nucl. Phys. B {\bf 280},  523  (1987).

\bibitem{Ko.Ve.Da.12}
S. Kourtis, J.~W.~F. Venderbos, and M. Daghofer, Phys. Rev. B {\bf 86},  235118
   (2012).

\bibitem{PhysRevB.75.121403}
T. Fukui and Y. Hatsugai, Phys. Rev. B {\bf 75},  121403  (2007).

\bibitem{PhysRevLett.100.186807}
S.-S. Lee and S. Ryu, Phys. Rev. Lett. {\bf 100},  186807  (2008).

\bibitem{springerlink:10.1007/BF01410451}
K. Ishikawa and T. Matsuyama, Z. Physik C {\bf 33},  41  (1986).

\bibitem{So85}
H. So, Prog. Theor. Phys. {\bf 74},  585  (1985).

\bibitem{PhysRevLett.105.256803}
Z. Wang, X.-L. Qi, and S.-C. Zhang, Phys. Rev. Lett. {\bf 105},  256803
  (2010).

\bibitem{Wa.Ji.Da.Xi.11}
L. Wang, H. Jiang, X. Dai, and X.~C. Xie, Phys. Rev. B {\bf 85},  235135
  (2012).

\bibitem{Fukui_Chern}
T. Fukui, Y. Hatsugai, and H. Suzuki, J. Phys. Soc. Jpn. {\bf 74},  1674
  (2005).

\bibitem{PhysRevB.84.125132}
A.~M. Essin and V. Gurarie, Phys. Rev. B {\bf 84},  125132  (2011).

\bibitem{Gurarie2013}
V. Gurarie and A.~M. Essin, arXiv:1301.3941  (2013).

\bibitem{PhysRevB.86.165116}
Z. Wang and S.-C. Zhang, Phys. Rev. B {\bf 86},  165116  (2012).

\bibitem{arXiv:1207.1104}
J.~C. Budich and B. Trauzettel, arXiv:1207.1104  (2012).

\bibitem{arXiv:1205.5095}
S.~R. Manmana, A.~M. Essin, R.~M. Noack, and V. Gurarie, Phys. Rev. B {\bf 86},
   205119  (2012).

\bibitem{PhysRevLett.109.066401}
A. Go, W. Witczak-Krempa, G.~S. Jeon, K. Park, and Y.~B. Kim, Phys. Rev. Lett.
  {\bf 109},  066401  (2012).

\bibitem{arXiv:1203.1028}
Z. Wang and S.-C. Zhang, Phys. Rev. X {\bf 2},  031008  (2012).

\bibitem{PhysRevB.85.165126}
Z. Wang, X.-L. Qi, and S.-C. Zhang, Phys. Rev. B {\bf 85},  165126  (2012).

\bibitem{Georges96}
A. Georges, G. Kotliar, W. Krauth, and M.~J. Rozenberg, Rev. Mod. Phys. {\bf
  68},  13  (1996).

\bibitem{PhysRevB.84.205116}
L. Wang, X. Dai, and X.~C. Xie, Phys. Rev. B {\bf 84},  205116  (2011).

\bibitem{Qi08}
X.-L. Qi and S.-C. Zhang, Phys. Rev. Lett. {\bf 101},  086802  (2008).

\bibitem{PhysRevLett.99.196805}
D.-H. Lee, G.-M. Zhang, and T. Xiang, Phys. Rev. Lett. {\bf 99},  196805
  (2007).

\bibitem{arXiv:1210.0266}
J. He, Y.-X. Zhu, Y.-J. Wu, L.-F. Liu, Y. Liang, and S.-P. Kou, arXiv:1210.0266
   (2012).

\bibitem{RueggLin12}
A. R\"uegg and C. Lin, Phys. Rev. Lett. {\bf 110},  046401  (2013).

\bibitem{SuShHe79}
W.~P. Su, J.~R. Schrieffer, and A.~J. Heeger, Phys. Rev. Lett. {\bf 42},  1698
  (1979).

\bibitem{Schollwoeck05_rev}
U. Schollw\"ock, Rev. Mod. Phys. {\bf 77},  259  (2005).

\bibitem{PhysRevLett.90.227902}
G. Vidal, J.~I. Latorre, E. Rico, and A. Kitaev, Phys. Rev. Lett. {\bf 90},
  227902  (2003).

\bibitem{PhysRevLett.96.110404}
A. Kitaev and J. Preskill, Phys. Rev. Lett. {\bf 96},  110404  (2006).

\bibitem{PhysRevLett.96.110405}
M. Levin and X.-G. Wen, Phys. Rev. Lett. {\bf 96},  110405  (2006).

\bibitem{PhysRevLett.109.067201}
S. Depenbrock, I.~P. McCulloch, and U. Schollw\"ock, Phys. Rev. Lett. {\bf
  109},  067201  (2012).

\bibitem{arXiv:1205.4289}
H.-C. Jiang, Z. Wang, and L. Balents, Nat. Phys. {\bf 8},  902  (2012).

\bibitem{PhysRevLett.101.010504}
H. Li and F.~D.~M. Haldane, Phys. Rev. Lett. {\bf 101},  010504  (2008).

\bibitem{PhysRevB.82.241102}
A.~M. Turner, Y. Zhang, and A. Vishwanath, Phys. Rev. B {\bf 82},  241102
  (2010).

\bibitem{PhysRevLett.108.196402}
X.-L. Qi, H. Katsura, and A.~W.~W. Ludwig, Phys. Rev. Lett. {\bf 108},  196402
  (2012).

\bibitem{arXiv:1211.3733}
M. P.Zaletel, R.~S.~K. Mong, and F. Pollmann, arXiv:1211.3733  (2012).

\end{thebibliography}

\end{document}